\documentclass[floatfix,aps,prb,twocolumn,superscriptaddress]{revtex4-2}
\usepackage{float}
\restylefloat{table}
\usepackage{adjustbox}
\usepackage{tikz}
\usetikzlibrary{calc,positioning}
\usepackage{amsmath,amssymb}
\usepackage{subfig}
\usepackage{graphicx}
\usepackage{braket}
\usepackage{epstopdf}
\usepackage{hyperref}
\usepackage{float}
\usepackage{xcolor}
\usepackage{tikz}
\usetikzlibrary{positioning}
\usepackage{graphicx}% Include figure files
\usepackage{dcolumn}
\usepackage[font=small,labelfont=bf,
justification=justified,
format=plain]{caption}
\usepackage{bm}% bold math
\usepackage[normalem]{ulem}
\usepackage{amsmath,amssymb}
\usepackage{subfig}
\usepackage{graphicx}
\usepackage{braket}
\usepackage{epstopdf}
\usepackage{hyperref}
\usepackage{float}
\usepackage{xcolor}
\usepackage{tikz}
\usetikzlibrary{positioning}
\usepackage{amsmath,mleftright}
\usepackage{xparse}
\usepackage{ragged2e}
\usepackage{caption}
\usepackage{amsmath}
\usepackage{hyperref}
\usepackage{graphicx}
\usepackage{float}
\usepackage{soul}
\usepackage{color}
\usepackage[normalem]{ulem}
\usepackage{braket}
\usepackage{tikz}
\newcommand{\comment}[1]{}

\begin{document}

%\preprint{}

\title{Metal to insulator crossover in the repulsive Fermi Hubbard model \\ probed by static correlations}
%Signatures of weak to strong coupling crossover in the repulsive Fermi Hubbard model through static correlations} 
%\\
%Comparison between simulations and optical lattice emulators}
%Entropy and Moment Correlations in 2D Fermi-Hubbard model}



%\thanks{}
\author{Sayantan Roy}
\affiliation{Department of Physics, The Ohio State University, Columbus OH 43210, USA}
%\author{Abhisek Samanta}
%\affiliation{Department of Physics, The Ohio State University, Columbus OH 43210, USA}
%\author{Ningyuan Jia}
%\affiliation{MIT-Harvard Center for Ultracold Atoms, Research Laboratory of Electronics, and Department of Physics, Massachusetts Institute of Technology, Cambridge, Massachusetts 02139, USA}
\author{Sameed Pervaiz}
\affiliation{Department of Physics, The Ohio State University, Columbus OH 43210, USA}

\author{Thereza Paiva}
\affiliation{Instituto de Fisica, Universidade Federal do Rio de Janeiro, 21941-972 Rio de Janeiro RJ, Brazil}
%\author{Martin Zwierlein}
%\affiliation{MIT-Harvard Center for Ultracold Atoms, Research Laboratory of Electronics, and Department of Physics, Massachusetts Institute of Technology, Cambridge, Massachusetts 02139, USA}
\author{Nandini Trivedi}
\affiliation{Department of Physics, The Ohio State University, Columbus OH 43210, USA}


%%%%%%%%%%%%% %%%%%%%%%%%%%%%%%% TODO: %%%%%%%%%%%%%%%%%%%%%%%%%%%%%%%%%
% Latest batch of changes that should finish up my (Sameed's) part:
% Plan: finish by Friday, Jan 28
% - Plot Structure Factors
%     [ ] Heatmap below half filling for multiple temperatures
%     [ ] Heatmap at half filling for multiple temperatures
%     [ ] Heatmap above half filling for multiple temperatures
%     [ ] AF cf vs density and temperature
%     [ ] Ferro cf vs density and temperature
% - Replot graphs with hot/cold color scheme and/or no grid lines
%     [x] Cs/Cm vs distance
%     [ ] Entropy
%     [ ] Decomposition plots
%     [ ] Structure Factors
%%%%%%%%%%%%%%%%%%%%%%%%%%%%%%%%%%%%%%%%%%%%%%%%%%%%%%%%%%%%%%%%%%%%%%

\date{\today}

\begin{abstract}
Recent progress in fluorescence imaging technique has enabled identification of nontrivial doublon, singlon, and holon correlation functions in cold atom experiments. We show that these correlators can be used to identify an extended crossover between metallic to insulating phase, at intermediate to high temperatures. Toward this end, we report a Determinantal Quantum Monte Carlo (DQMC) study of such correlation functions in the two-dimensional repulsive Fermi Hubbard model on a square lattice as a function of doping, interaction strength $U$ and temperature $T$. We find definite signatures of an extended crossover from small $U$ (metallic regime) to large $U$ (Mott Insulator regime). Our key findings are: (1) The extended crossover at half filling is marked by suppression of weight in the thermodynamic density of states $\tilde{\kappa} = \frac{\partial n}{\partial \mu}$ and local density of states $N(\omega)$, that occur at different interaction strengths and can be tracked separately by different correlators. (2) At a given ($U,T$), we define $n_{cr}(U,T)$ to be the density at which $\frac{\partial \tilde{ \kappa}}{\partial T}$ changes sign. With strong interactions, local density fluctuations lead to an unexpected non monotonic temperature dependence of $n_{cr}(T;U)$. (3) Correlations between local moments on neighboring sites arises from a competition between doublon-holon attraction and density-density repulsion. Around half filling, local moment correlations grow with $U$ in the metallic regime. On crossover to the Mott Insulator regime, local moment correlations decrease with $U$, although individual moment size increases. Our results allow comparisons of different correlation functions with recent experimental findings and guide further experimental investigations.
\end{abstract}

\maketitle

\section{Introduction}
\label{sec_introduction}

Of particular interest in strongly correlated system is the metal-insulator transition, where a system with a partially filled band can turn into an insulator when subjected to large interactions~\cite{imada1998metal}. The energy cost due to Coulomb repulsion $U$ between electrons can split a half filled band into a completely occupied lower band and an empty upper band, forming a ``Mott" insulator~\cite{mott1961transition}.
The repulsive Fermi-Hubbard model is a prototype for understanding 
such transitions,
 proposed independently by Gutzwiller~\cite{gutzwiller1963effect}, Kanamori~\cite{kanamori1963electron} and Hubbard~\cite{hubbard1964electron}, and thought to be a model for explaining high $T_c$ superconductivity by Anderson~\cite{anderson1992experimental}.

 On a square lattice, the half filled Hubbard model with hopping $t$ and interaction $U$ exhibits an excitation gap for all $U$ at temperature $T=0$~\cite{schafer2015fate}. For small interactions, magnetic instabilities of the Fermi surface~\cite{schrieffer1989dynamic} create an excitation gap $\Delta_c \propto e^{-2\pi \sqrt{\frac{t}{U}}}$, turning the system into a Slater insulator. For large interactions, local moments are well formed that turn the system into a Mott Insulator with an excitation gap $\Delta_c \propto U$~\cite{paiva2010fermions,borejsza2004antiferromagnetism}. The crossover from a Slater insulator to Mott insulator in the repulsive Hubbard model is analogous to the BCS-BEC crossover in the attractive Hubbard model. However,  thermal effects can lead to a metallic (gapless) state even at a finite interaction strength at non zero temperatures~\cite{paiva2010fermions}. In this paper, we will determine the criteria by which a metallic state can be distinguised from an insulating state at a finite temperature.

 Cold atomic systems provide the advantage of creating specific hamiltonians~\cite{bloch2008many,bloch2005ultracold,bloch2012quantum,bakr2009quantum,aidelsburger2013realization,gross2017quantum,weitenberg2011single}, where measurements of quantum mechanical and thermodynamic quantities can be compared directly with theoretical modeling~\cite{sherson2010single,greiner2002quantum,aidelsburger2015measuring,cocchi2016equation,greif2016site,cheuk2016observation,drewes2016thermodynamics,parsons2016site,drewes2017antiferromagnetic,brown2017spin,nichols2019spin,brown2019bad,guardado2020subdiffusion,mazurenko2017cold,koepsell2019imaging}.
  Early quantum gas microscope experiments were very successful in measuring spin-spin correlation functions~(\cite{mazurenko2017cold,boll2016spin,cheuk2016observation}). 
  Developments in quantum gas microscope technology have furthermore allowed experiments to produce  site-resolved measurements of local moment~\cite{cheuk2016observation} and site resolved particle number, doublon, and holon densities~\cite{hartke2020doublon}, as well as correlation functions between these quantities~\cite{koepsell2019imaging,koepsell2020robust,hirthe2023magnetically,salomon2019direct,vijayan2020time,koepsell2021microscopic,sompet2022realizing,chiu2018quantum,kale2022schrieffer,ji2021coupling,ji2021coupling,chiu2018quantum,chiu2019string,greif2016site,parsons2016site,parsons2015site,bakr2009quantum,simon2011quantum,simon2011quantum,greiner2004detection,cheuk2015quantum,sommer2011universal,nichols2019spin,cheuk2016observation,zwierlein2005formation,hertkorn2021density,hartke2023direct}.
 
 A new development was the construction of a bilayer Fermi gas microscope~\cite{hartke2020doublon,koepsell2020robust,drewes2016thermodynamics}, in which each lattice site is split into a double well, separating atom pairs. This allows for measuring all possible correlation functions between doubly occupied, singly occupied, and empty sites. In light of these recent advances, 
it is natural to ask whether cold atom experiments can track the metal-to-insulator crossover at intermediate temperatures and whether there are definite signatures of this crossover, at and away from half-filling in the absence of any spontaneous symmetry breaking.  We seek to answer these questions in this work.

\begin{figure*}[t]
\begin{tikzpicture}
\node (img1) 
{\includegraphics[width=6.4cm,height=4.8cm]{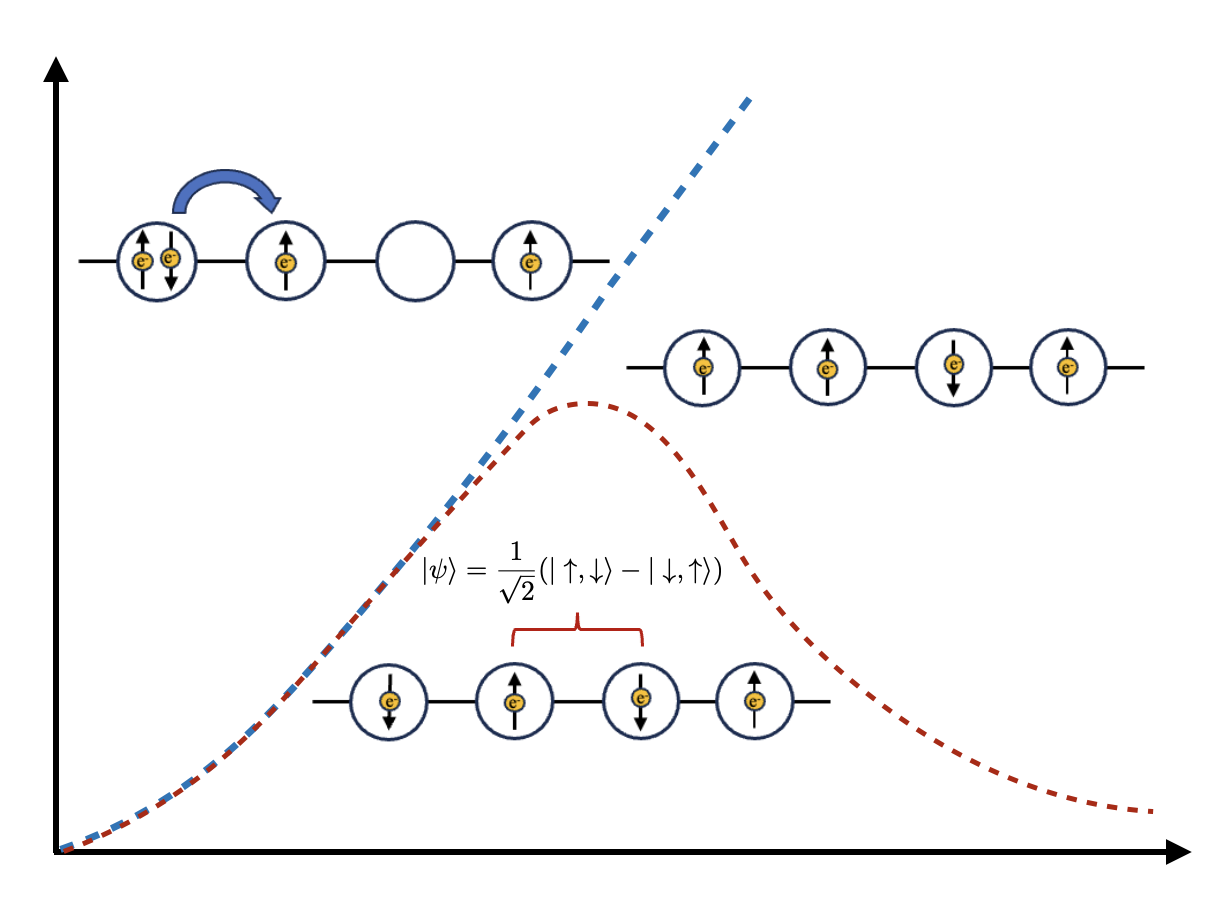}};
%\node[left=of img1,node distance=0cm,yshift=1.7cm,xshift=5.0cm]{\normalsize{\textcolor{blue}{T}$_{\text{\textcolor{blue}{charge}}}$}};
\node[left=of img1,node distance=0cm,yshift=1.70cm,xshift=2.2cm]{\large{(a)}};
\node[left=of img1,node distance=0cm,rotate=0,anchor=center,yshift=0cm,xshift=0.9cm]{\large{ $T$}};
\node[below=of img1,node distance=0cm,yshift=1.3cm,xshift=0.0cm]{\large{$U$}};
%%%%%%%%%%%%%Labelling the phase diagram%%%%
\node[left=of img1,node distance=0cm,rotate=0,anchor=center,yshift=0.3cm,xshift=2.8cm]{\scriptsize{ Paramagnetic}};
\node[left=of img1,node distance=0cm,rotate=0,anchor=center,yshift=0.0cm,xshift=2.5cm]{\scriptsize{Metal}};
\node[left=of img1,node distance=0cm,rotate=0,anchor=center,yshift=0.3cm,xshift=1.8cm]{\scriptsize{(A)}};
\node[left=of img1,node distance=0cm,rotate=0,anchor=center,yshift=-0.1cm,xshift=6.1cm]{\scriptsize{ Paramagnetic}};
\node[left=of img1,node distance=0cm,rotate=0,anchor=center,yshift=-0.5cm,xshift=6.1cm]{\scriptsize{Insulator}};
\node[left=of img1,node distance=0cm,rotate=0,anchor=center,yshift=1.6cm,xshift=5.5cm]{\normalsize{$T_{\rm charge}$}};
\node[left=of img1,node distance=0cm,rotate=0,anchor=center,yshift=-1.6cm,xshift=6.6cm]{\normalsize{$T_{\rm spin}$}};

\node[left=of img1,node distance=0cm,rotate=0,anchor=center,yshift=-0.1cm,xshift=5.1cm]{\scriptsize{(B)}};
\node[left=of img1,node distance=0cm,rotate=0,anchor=center,yshift=-1.9cm,xshift=4.4cm]{\scriptsize{ (C) Quasi-LRO AFM}};
%\node[left=of img1,node distance=0cm,rotate=0,anchor=center,yshift=-1.9cm,xshift=3.8cm]{\scriptsize{(C)}};
%%%%%%%%%%%%%%%%%%%%%%%%%%%%%%%%%%%%%%%%%%%%
\node (img2)[right=of img1,yshift=1.00cm,xshift = 0.0cm] 
{\includegraphics[width=2.6cm,height=1.8cm]{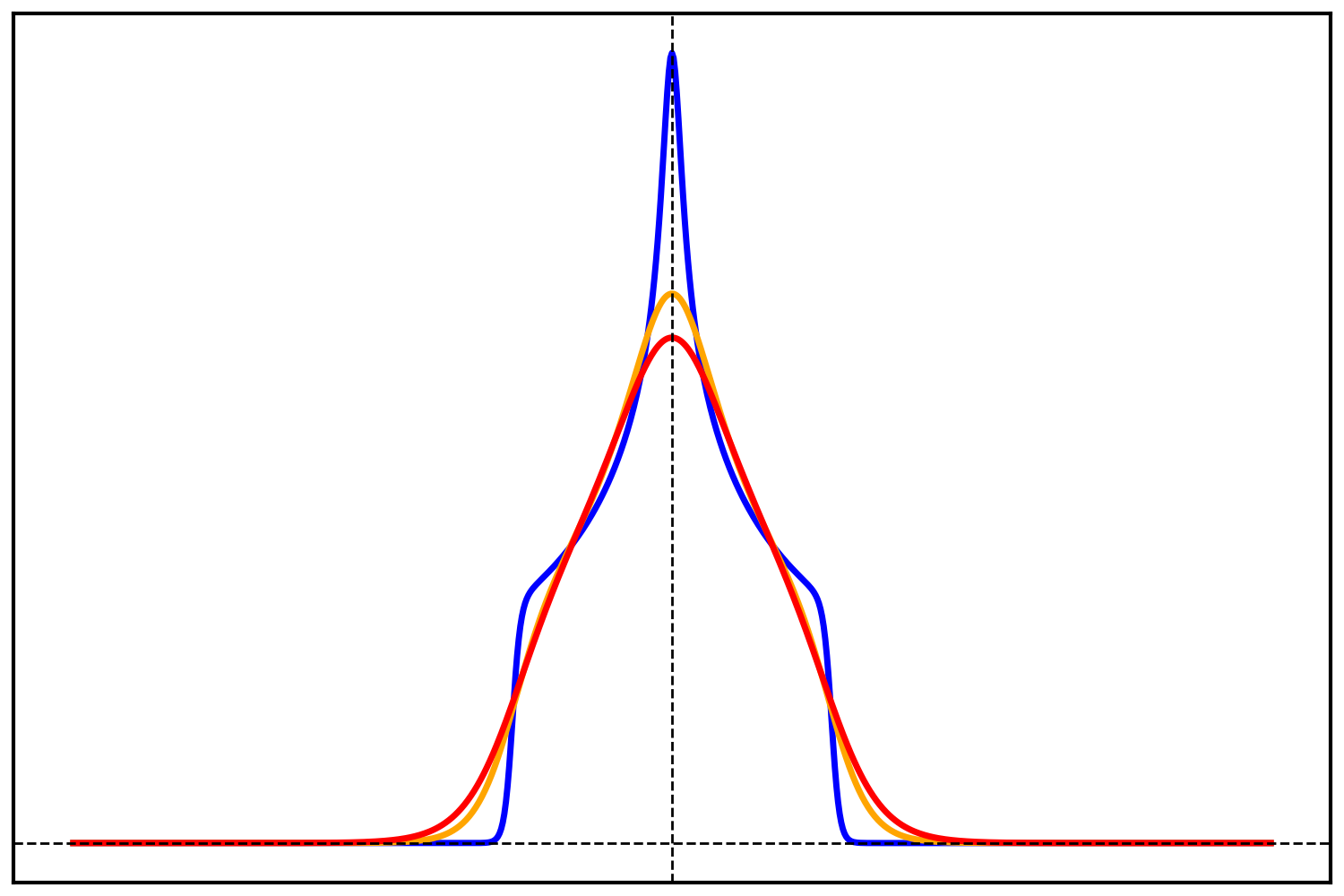}};
\node[left=of img2,node distance=0cm,rotate=0,anchor=center,yshift=0.0cm,xshift=0.8cm]{\normalsize{ $\frac{\partial n}{\partial \mu}$}};
\node[left=of img2,node distance=0cm,rotate=0,anchor=center,yshift=0.65cm,xshift=1.5cm]{\scriptsize{(b)}};
\node[left=of img2,node distance=0cm,rotate=0,anchor=center,yshift=0.65cm,xshift=3.2cm]{\scriptsize{$\Delta=0$}};
\node[below=of img2,node distance=0cm,yshift=1.2cm,xshift=0.0cm]{\small{$\mu$}};
%%%%%%%%%%%%%%%%%%%%%%%%%%%%%%%%%%%%%%%
\node (img3) [right=of img2,xshift = -0.65cm] 
{\includegraphics[width=2.6cm,height=1.8cm]{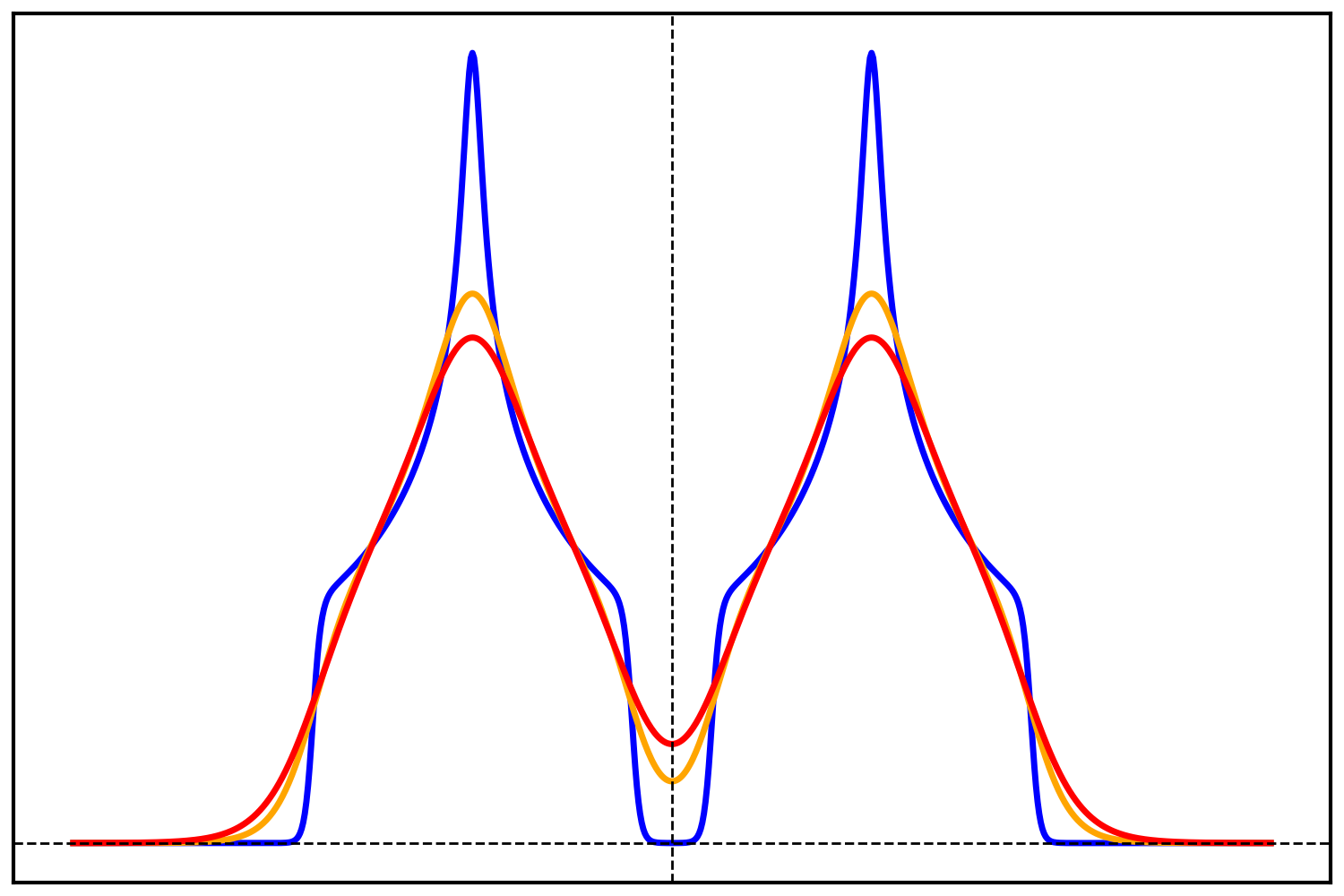}};
\node[left=of img3,node distance=0cm,rotate=0,anchor=center,yshift=0.0cm,xshift=0.8cm]{\normalsize{ $\frac{\partial n}{\partial \mu}$}};
\node[left=of img3,node distance=0cm,rotate=0,anchor=center,yshift=0.65cm,xshift=1.5cm]{\scriptsize{(c)}};
\node[left=of img3,node distance=0cm,rotate=0,anchor=center,yshift=0.65cm,xshift=3.2cm]{\scriptsize{$\Delta\neq0$}};
\node[below=of img3,node distance=0cm,yshift=1.2cm,xshift=0.0cm]{\small{$\mu$}};
%%%%%%%%%%%%%%%%%%%%%%%%%%%%%%%%%%%%%%%%%
%%%%%%%%%%%%%%%%%%%%%%%%%%%%%%%%%%%%%%%
\node (img4) [below=of img2,yshift =0.65cm] 
{\includegraphics[width=2.6cm,height=1.8cm]{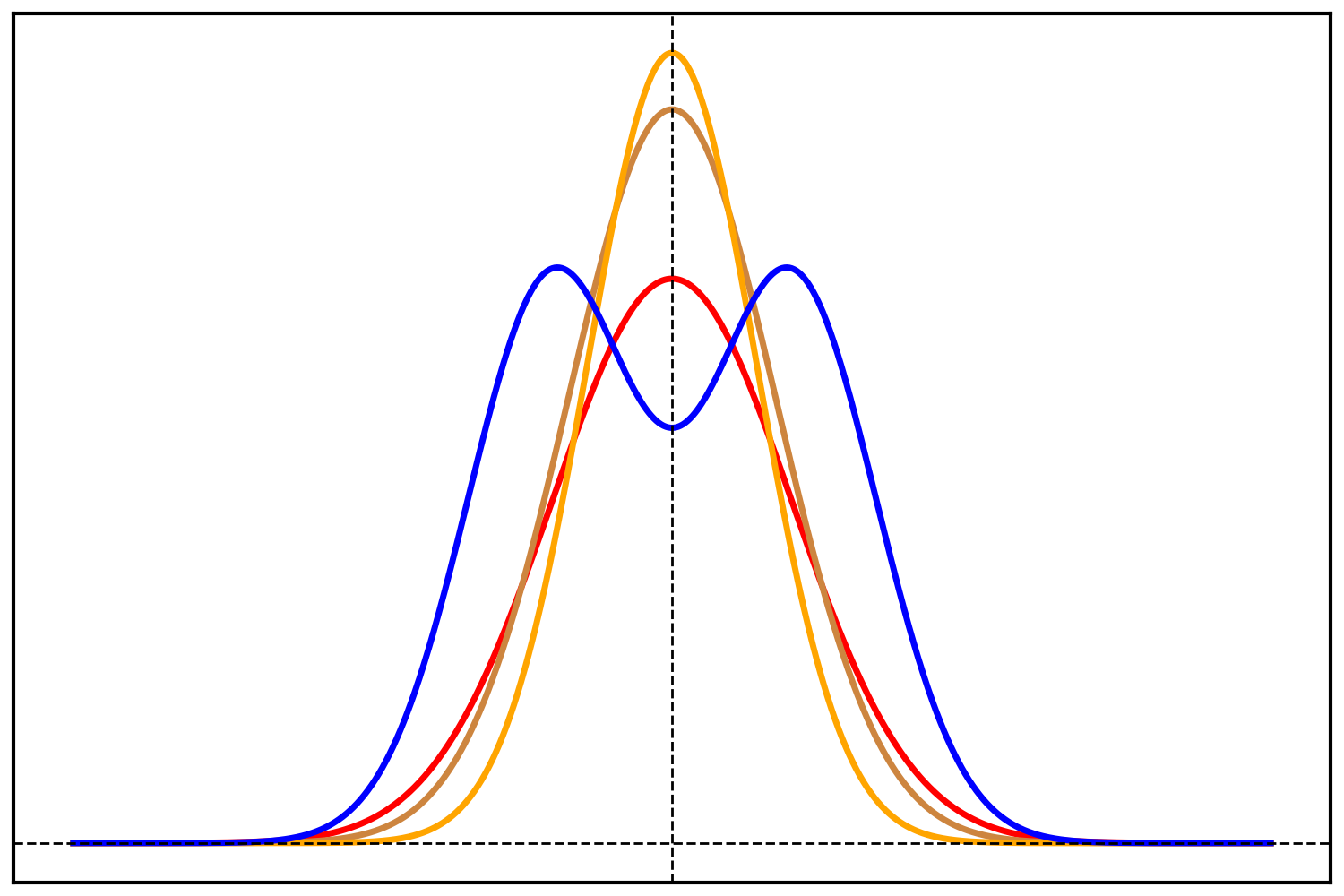}};
\node[left=of img4,node distance=0cm,rotate=90,anchor=center,yshift=-0.8cm,xshift=0.0cm]{\small{ $N(\omega)$}};
\node[left=of img4,node distance=0cm,rotate=0,anchor=center,yshift=0.65cm,xshift=1.5cm]{\scriptsize{(d)}};
\node[below=of img4,node distance=0cm,yshift=1.2cm,xshift=0.0cm]{\small{$\omega$}};
%%%%%%%%%%%%%%%%%%%%%%%%%%%%%%%%%%%%%%%%%
%%%%%%%%%%%%%%%%%%%%%%%%%%%%%%%%%%%%%%%
\node (img5) [right=of img4,xshift = -0.65cm] 
{\includegraphics[width=2.6cm,height=1.8cm]{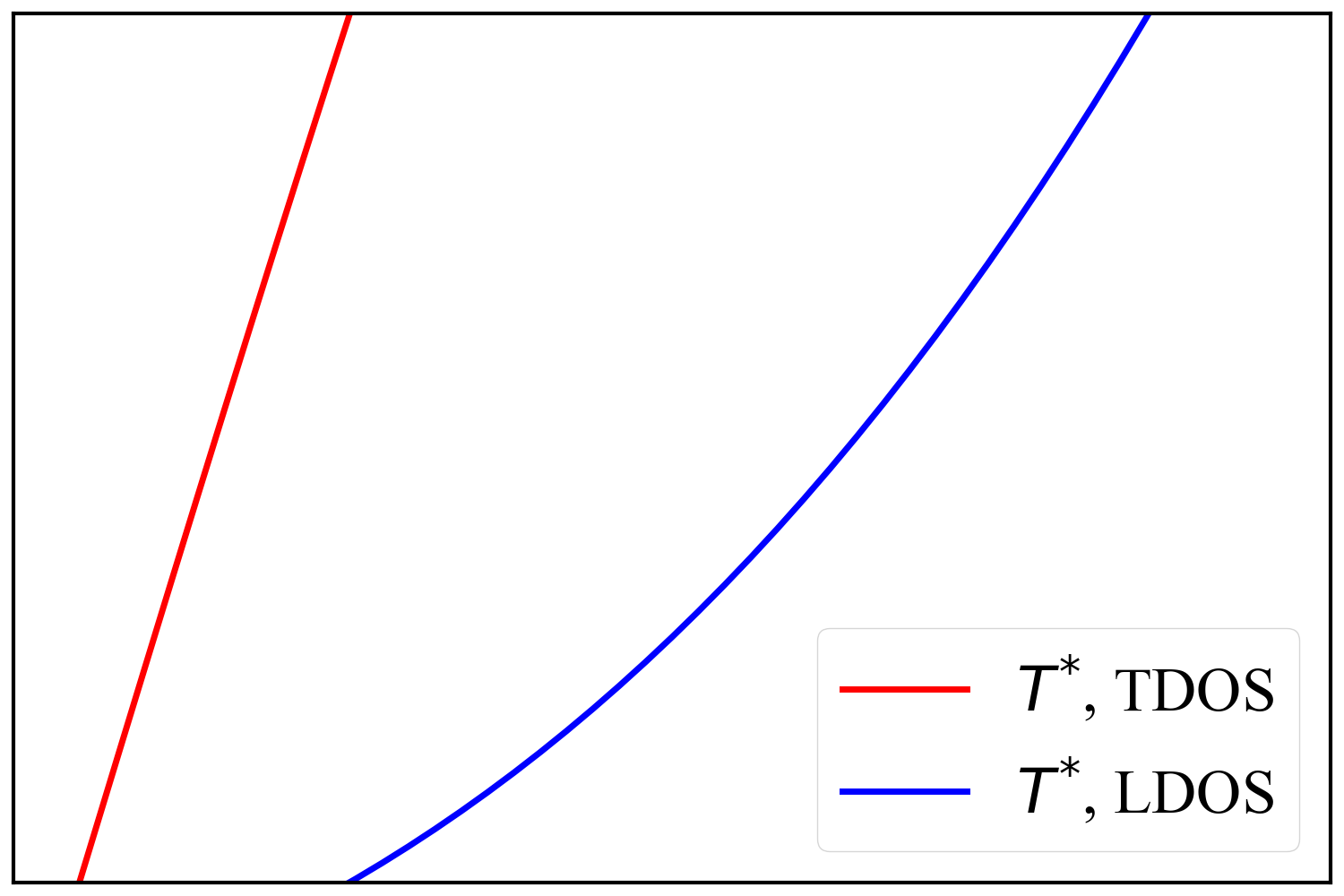}};
\node[left=of img5,node distance=0cm,rotate=0,anchor=center,yshift=0.0cm,xshift=0.9cm]{\small{ $T$}};
\node[left=of img5,node distance=0cm,rotate=0,anchor=center,yshift=0.65cm,xshift=1.5cm]{\scriptsize{(e)}};
\node[below=of img5,node distance=0cm,yshift=1.2cm,xshift=0.0cm]{\small{$U$}};
%%%%%%%%%%%%%%%%%%%%%%%%%%%%%%%%%%%%%%%%%
\end{tikzpicture}
\caption{\textbf{(a)} Temperature and interaction scales in the 2D square lattice Hubbard model at half filling. There are two energy scales, the charge gap $\Delta_{ch}$, which sets the energy scale for charge fluctuation, and spin gap $\Delta_S$, which sets energy scales for spin fluctuations. The corresponding temperature scales separate the $U-T$ plane at half filling into three regions\cite{paiva2010fermions}: \textbf{(A)} Metallic -- for $T>T_{\text{charge}}(U)$, the system is a  paramagnetic metal. \textbf{(B)} Mott Insulator -- below the temperature scale $T_{\text{charge}}(U)$, the system is a paramagnetic insulator, driven by strong onsite repulsion. Itinerant fermions get pinned down to form local moments, however, the moments are not yet correlated due to thermal fluctuations. \textbf{(C)} Short range ordered state with strong AF correlations -- on lowering the temperature below $T_{\text{spin}}(U)$, the Mott-insulator develops strong antiferromagnetic correlations that are short ranged. %An ordering is prohibited in the thermodynamic limit by Mermin Wagner theorem \cite{mermin1966absence}. However in numerical simulations, systems can appear long range ordered due to finite sized effects. 
\textbf{(b)-(e)} Different criteria for isolating a metal-to-insulator crossover, which forms the main ideas explored in this work. \textbf{(b),(c)} Thermodynamic density of states $\tilde{\kappa} = \frac{\partial n}{\partial \mu}$ for a toy model $E_{\pm}(k) = \pm(\varepsilon_k+\Delta)$ at different temperatures (T increases from blue to red). $\Delta = 0$ corresponds to a metal, $\Delta >W$ corresponds to a band insulator($W$ being the bandwidth). For an insulator, compressibility is finite at $T>0$; hence metallic vs insulating systems should be distinguished by sign of $\frac{\partial \tilde{\kappa}}{\partial T}$. The temperature at which $\frac{\partial \tilde{\kappa}}{\partial T}$ changes sign defines $T^{*},\rm TDOS$. \textbf{(d)} The temperature variation of the local density of states (LDOS) $N(\omega) = -\frac{1}{\pi} \text{Im} G(r=0,\omega)$ for the half-filled repulsive Hubbard model. Low energy weight increases with lowering temperature until a gap onsets at $T^{*},\rm LDOS$. \textbf{(e)} The two temperature scales mentioned before are not concurrent, giving rise to an extended crossover in $T-U$ plane that can be probed by static correlation functions in cold atom experiments.}
\label{schematic}
\end{figure*}
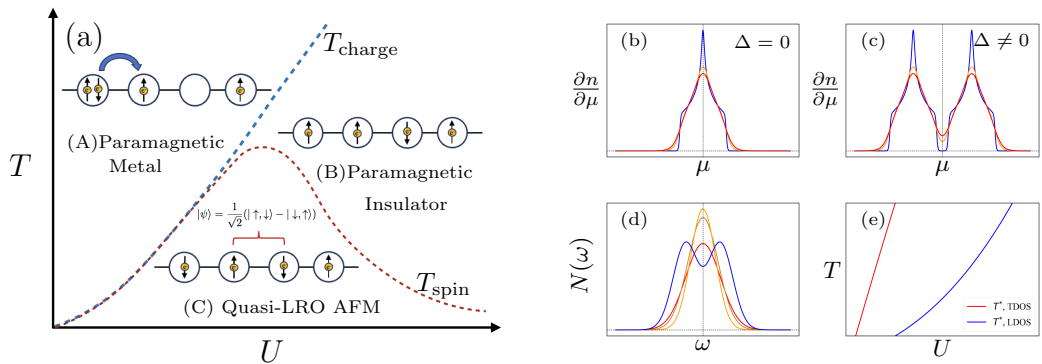

In the intermediate temperature range, Determinant Quantum Monte Carlo (DQMC) \cite{blankenbecler1981monte,white1989numerical,hirsch1985two} is not plagued by the infamous ``sign" problem and is a powerful tool to perform unbiased numerical simulations. 
 We have investigated both thermodynamic observables as well as equal time correlation functions (density, doublons, holons and local moments) as functions of doping, temperature and interaction strength. Our key findings are as follows:

(1) At half-filling, the temperature dependence of the thermodynamic density of states (TDOS), $\tilde{\kappa}=\frac{\partial n}{\partial \mu}$ shows a density fluctuation driven opening of a charge gap and consequently a metal to insulator crossover at finite $U$ and $T$. It is followed by a second metal-to-insulator crossover due to ``pseudogap" onset in the local density of states (LDOS), $N(\omega) = (1/N_s)\sum_{k}A(k,\omega)$ at a larger $U$. Different equal-time correlators can track the onset of gaps in both TDOS and LDOS in experiments.

(2) The temperature dependence of $\tilde{\kappa}$ defines densities $n_{cr}(U,T)$ at which there is a crossover between an insulator and a metal. With intermediate to strong interactions, local density fluctuations lead to a non-monotonic temperature dependence of $n_{cr}(U,T)$. The crossover density grows up to a temperature scale set by interaction strength; increasing temperature further causes $n_{cr}(U,T)$ to decrease monotonically as the gap in $\tilde{\kappa}$ closes.

(3) Moment-moment correlation between neighboring sites arises because of competition between nearest neighbor holon-doublon attraction and density-density repulsion. Close to half-filling, moment-moment correlations increase with increasing $U$ in the metallic regime. On crossover to the Mott Insulator regime, moment-moment correlations decrease with $U$ although individual local moments are better formed.

(4) Next nearest neighbor correlations (density, doublon, holon, and local moments) show a marked difference between small and large $U$ regimes, both at and away from half-filling (shown in Appendix \ref{app_e}).

\section{Model and Method}
\label{sec_model_method}

We consider the single-band Fermi Hubbard model with nearest neighbor hopping and onsite repulsive interaction, defined by the hamiltonian,
\begin{align}
    \mathcal{H} &= -t\sum_{\langle i,j \rangle,\sigma}{ \left( \hat{c}^\dag_{i,\sigma} \hat{c}_{j,\sigma} + h.c. \right) }-\mu \sum_{i}\hat{n}_{i} \nonumber \\
    &+U\sum_{i}{ \left( \hat{n}_{i\uparrow} - \frac{1}{2} \right) \left( \hat{n}_{i\downarrow} - \frac{1}{2} \right)}
\label{Hamiltonian}
\end{align}

 We write the Hamiltonian in a symmetric particle-hole form so that the system is half-filled when the chemical potential $\mu = 0$. The hopping amplitude $t$ sets the energy scale, unless otherwise mentioned.  
The spatial index $i$ labels a site on a 2D square lattice, and 
 $\hat{c}_{i,\sigma}$ and $\hat{c}^\dag_{i,\sigma}$ are fermionic annihilation and creation operators respectively.
The number operator is defined as $\hat{n}_{i,\sigma} \equiv \hat{c}^\dag_{i,\sigma} \hat{c}_{i,\sigma}$,  $\hat{n}_i = \hat{n}_{i, \uparrow} + \hat{n}_{i, \downarrow}$, and the particle density per site ${n} = \sum_i{\langle\hat{n}_i\rangle}/{N_s}$, where $N_s$ is the total number of sites.

We perform Determinant Quantum Monte Carlo (DQMC)
simulations that maps a many-particle interacting fermionic system
into a single-particle Hamiltonian (quadratic form), with the
aid of bosonic auxiliary fields. For the analyses conducted in this paper, ensembles of 10x10 lattices are simulated over a finite set of $\mu$ and different $U$ from weak to strong coupling. We look at evolution of the equation of state ($n$ vs $\mu$), thermodynamic density of states $\tilde{\kappa} = \frac{\partial n }{\partial \mu}$ %(a proxy for the charge gap)
, local density of states, $N(\omega) = \frac{-1}{\pi}\text{Im}G(r,\omega)$, doublon occupancy $d = \langle n_{\uparrow}n_{\downarrow} \rangle$, and thermodynamic entropy $s$ across the metal-insulator crossover in the square lattice, by varying $T$ and $U$. We also look at the interplay between various charge degrees of freedom, from the weak coupling to the strong coupling limit, by investigating equal time correlation functions between density, holons, doublons, and local moments. We have restricted our analysis to parameter regimes where the sign problem 
is not an issue, and at temperatures where the coherence length is smaller than the system size.

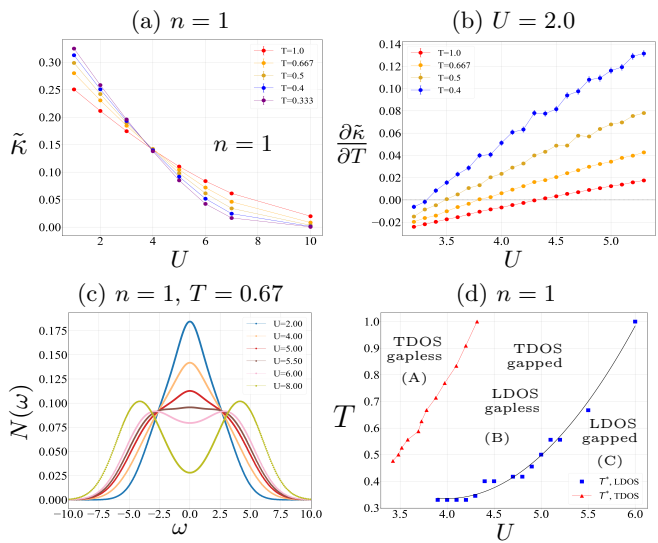
\begin{figure}[t]
\begin{tikzpicture}
%%%%%%%%%%%%%%%%%%%%%%%%%%%%%%%%%%%%%%%
\node (img1)
{\includegraphics[width=3.8cm,height=2.8cm]{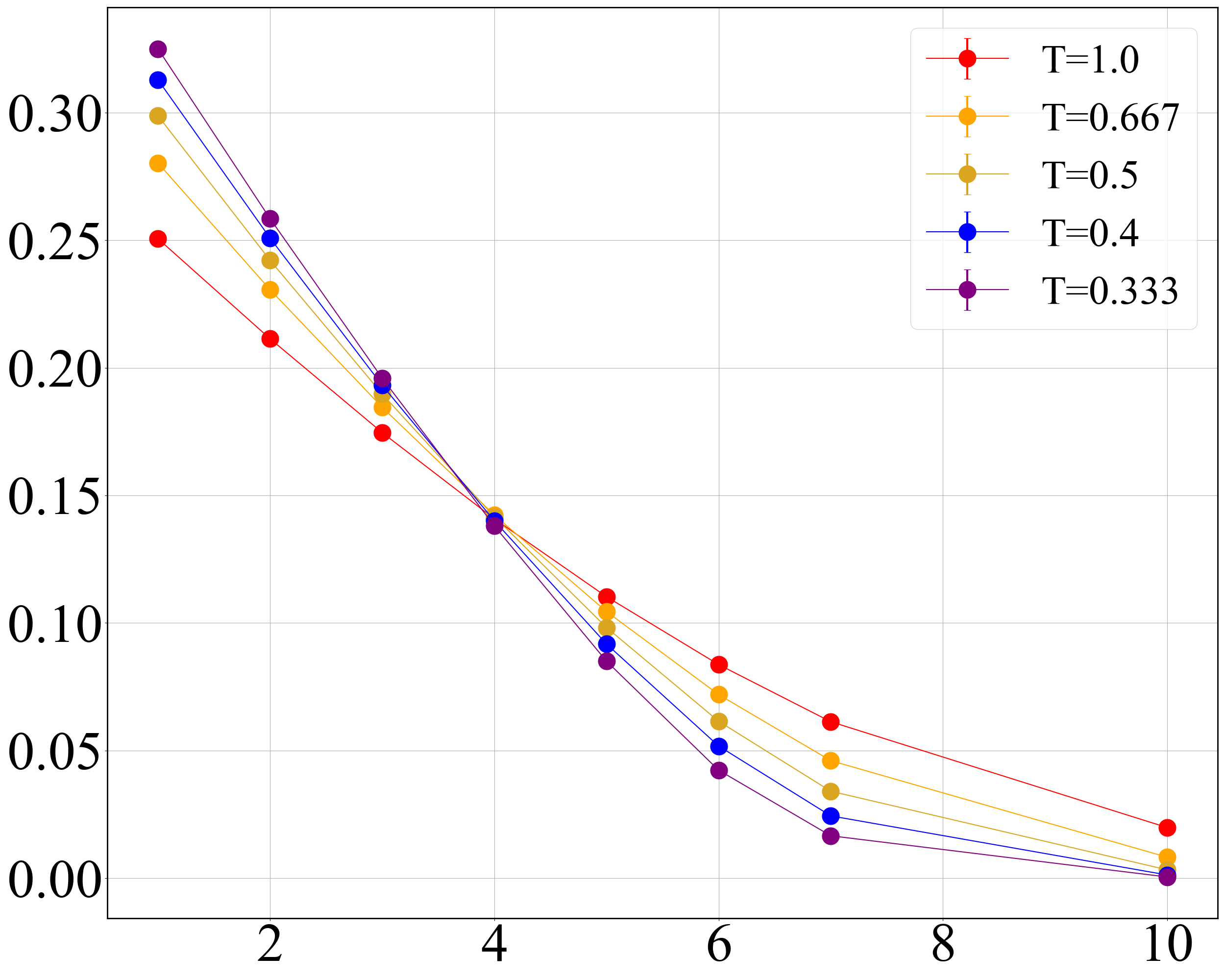}};
%\node[left=of img3,node distance=0cm,yshift=2.0cm,xshift=1.6cm]{\large{(b)}};
\node[left=of img1,node distance=0cm,rotate=0,anchor=center,yshift=0.0cm,xshift=0.8cm]{\large{ $\tilde{\kappa}$}};
\node[above=of img1,node distance=0cm,rotate=0,anchor=center,yshift=-0.9cm,xshift=0.0cm]{\small{(a) $n =1$}};
\node[left=of img1,node distance=0cm,rotate=0,anchor=center,yshift=0.0cm,xshift=3.8cm]{\small{ $n=1$}};
\node[below=of img1,node distance=0cm,yshift=1.2cm,xshift=0.0cm]{\small{$U$}};
%\node[above=of img3,node distance=0cm,yshift=-1.3cm,xshift=0.0cm]{\large{(a)}\Large{ $\frac{\partial \tilde{\kappa}}{\partial T}$}\large{, $T = 0.67$}};
%%%%%%%%%%%%%%%%%%%%%%%%%%%%%%%%%%%%%%%
\node (img2)[right=of img1, xshift = -0.6cm]
{\includegraphics[width=3.8cm,height=2.8cm]{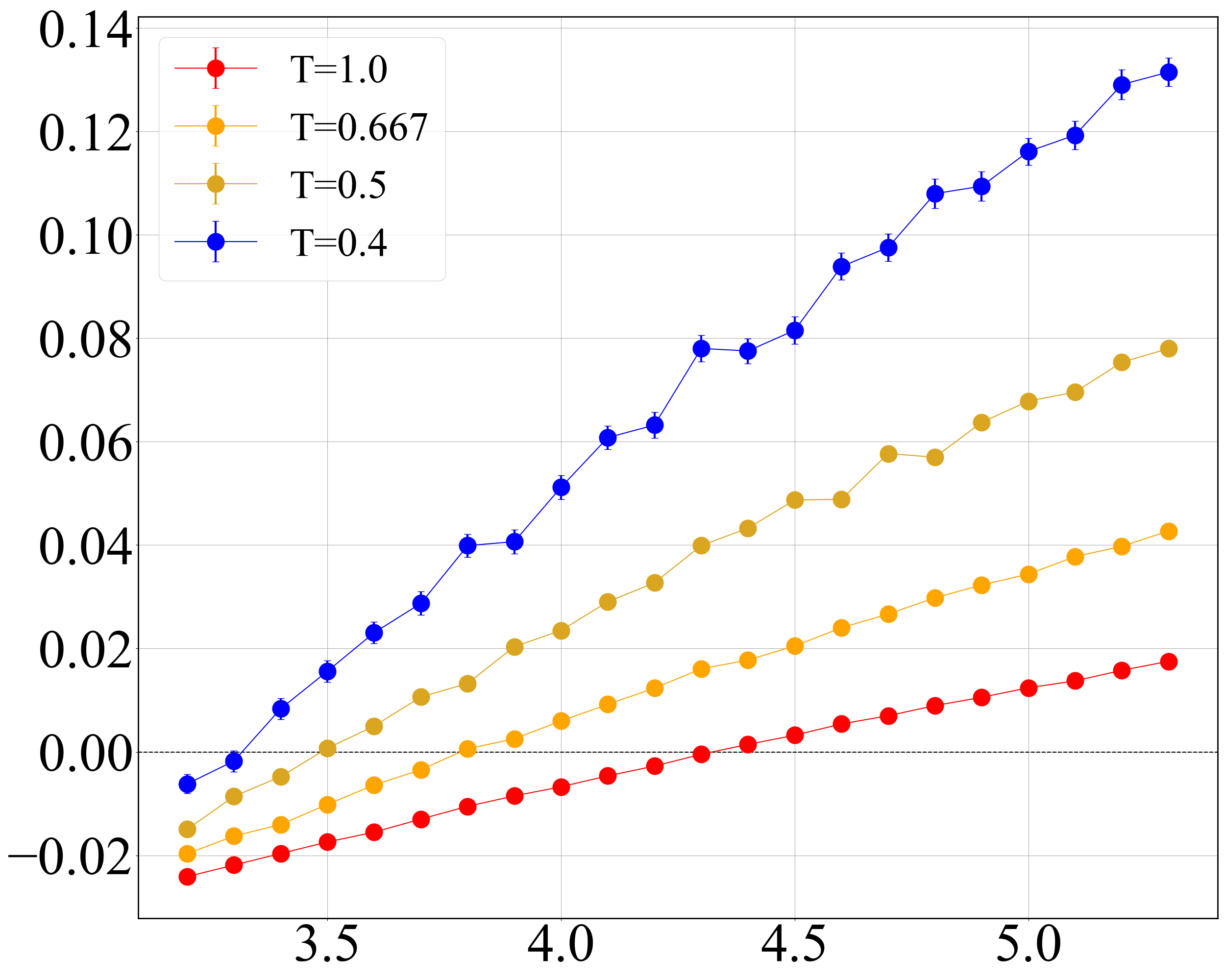}};
%\node[left=of img3,node distance=0cm,yshift=2.0cm,xshift=1.6cm]{\large{(b)}};
\node[left=of img2,node distance=0cm,rotate=0,anchor=center,yshift=0.0cm,xshift=0.8cm]{\large{ $\frac{\partial \tilde{\kappa}}{\partial T}$}};
\node[below=of img2,node distance=0cm,yshift=1.2cm,xshift=0.0cm]{\small{$U$}};
%\node[left=of img2,node distance=0cm,rotate=0,anchor=center,yshift=0.0cm,xshift=3.8cm]{\small{ $U=2.0$}};
\node[above=of img2,node distance=0cm,rotate=0,anchor=center,yshift=-0.9cm,xshift=0.0cm]{\small{(b) $U=2.0$}};
%\node[above=of img3,node distance=0cm,yshift=-1.3cm,xshift=0.0cm]{\large{(a)}\Large{ $\frac{\partial \tilde{\kappa}}{\partial T}$}\large{, $T = 0.67$}};
%%%%%%%%%%%%%%%%%%%%%%%%%%%%%%%%%%%%%%%
\node (img3)[below=of img1, yshift = 0.4cm]
{\includegraphics[width=3.8cm,height=2.8cm]{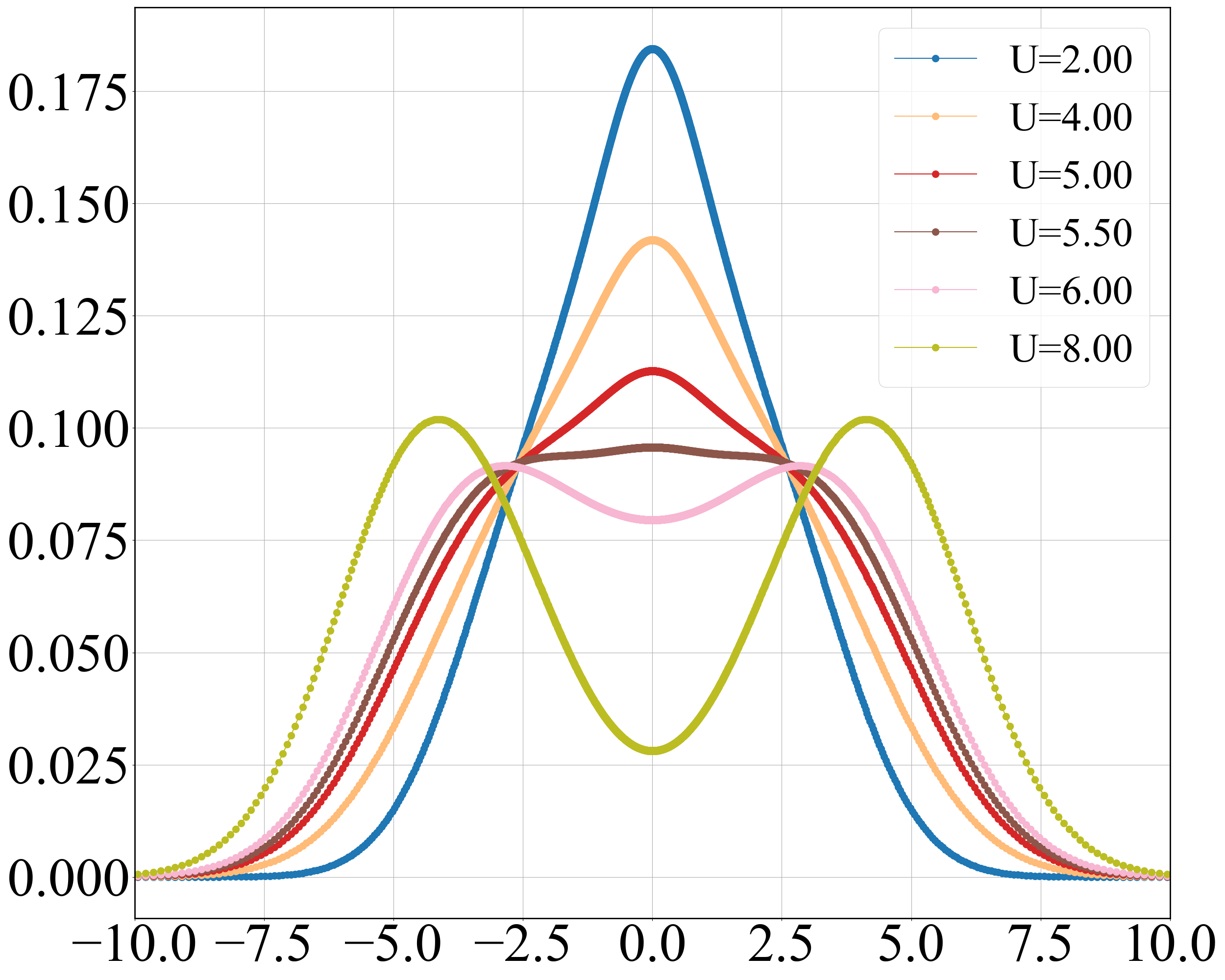}};
%\node[left=of img3,node distance=0cm,yshift=2.0cm,xshift=1.6cm]{\large{(b)}};
\node[left=of img3,node distance=0cm,rotate=90,anchor=center,yshift=-0.9cm,xshift=0.0cm]{\small{ $N(\omega)$}};
\node[below=of img3,node distance=0cm,yshift=1.2cm,xshift=0.0cm]{\small{$\omega$}};
%\node[left=of img3,node distance=0cm,rotate=0,anchor=center,yshift=0.0cm,xshift=3.8cm]{\small{ $U=5.0$}};
\node[above=of img3,node distance=0cm,rotate=0,anchor=center,yshift=-0.9cm,xshift=0.0cm]{\small{(c) $n=1$, $T=0.67$}};
%\node[above=of img3,node distance=0cm,yshift=-1.3cm,xshift=0.0cm]{\large{(a)}\Large{ $\frac{\partial \tilde{\kappa}}{\partial T}$}\large{, $T = 0.67$}};
%%%%%%%%%%%%%%%%%%%%%%%%%%%%%%%%%%%%%%%
\node (img4)[right=of img3, xshift = -0.7cm]
{\includegraphics[width=3.8cm,height=2.8cm]{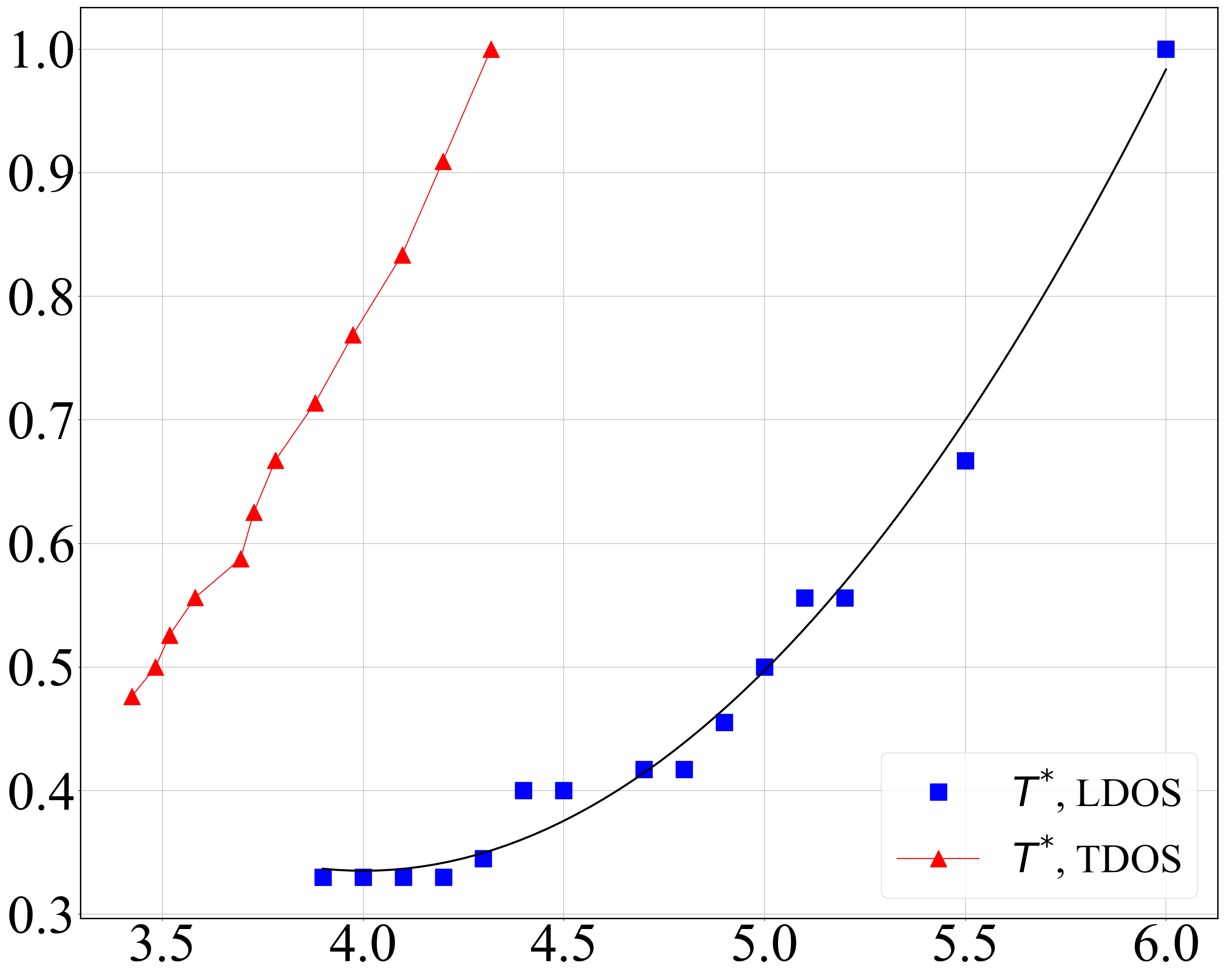}};
%\node[left=of img3,node distance=0cm,yshift=2.0cm,xshift=1.6cm]{\large{(b)}};
\node[left=of img4,node distance=0cm,rotate=0,anchor=center,yshift=0.0cm,xshift=0.8cm]{\large{ $T$}};
\node[below=of img4,node distance=0cm,yshift=1.2cm,xshift=0.0cm]{\small{$U$}};
%\node[left=of img4,node distance=0cm,rotate=0,anchor=center,yshift=0.0cm,xshift=3.0cm]{\small{ $U=8.0$}};
\node[above=of img4,node distance=0cm,rotate=0,anchor=center,yshift=-0.9cm,xshift=0.0cm]{\small{(d) $n = 1$}};
%%%%%%%%%%%%%%%%%%%%%%%%%%%%%%%%%%%%%%%
\node[left=of img4,node distance=0cm,rotate=0,anchor=center,yshift=1.0cm,xshift=1.8cm]{\tiny{ TDOS}};
\node[left=of img4,node distance=0cm,rotate=0,anchor=center,yshift=0.8cm,xshift=1.8cm]{\tiny{gapless}};
\node[left=of img4,node distance=0cm,rotate=0,anchor=center,yshift=0.5cm,xshift=1.8cm]{\tiny{(A)}};
\node[left=of img4,node distance=0cm,rotate=0,anchor=center,yshift=0.9cm,xshift=3.4cm]{\tiny{ TDOS}};
\node[left=of img4,node distance=0cm,rotate=0,anchor=center,yshift=0.7cm,xshift=3.4cm]{\tiny{gapped}};

\node[left=of img4,node distance=0cm,rotate=0,anchor=center,yshift=0.3cm,xshift=3.1cm]{\tiny{ LDOS}};
\node[left=of img4,node distance=0cm,rotate=0,anchor=center,yshift=0.1cm,xshift=3.1cm]{\tiny{gapless}};
\node[left=of img4,node distance=0cm,rotate=0,anchor=center,yshift=-0.3cm,xshift=2.9cm]{\tiny{(B)}};
\node[left=of img4,node distance=0cm,rotate=0,anchor=center,yshift=-0.1cm,xshift=4.4cm]{\tiny{ LDOS}};
\node[left=of img4,node distance=0cm,rotate=0,anchor=center,yshift=-0.3cm,xshift=4.4cm]{\tiny{gapped}};
\node[left=of img4,node distance=0cm,rotate=0,anchor=center,yshift=-0.6cm,xshift=4.4cm]{\tiny{(C)}};
%%%%%%%%%%%%%%%%%%%%%%%%%%%%%%%%%%%%%%%%%
\end{tikzpicture}
\caption{Metal-to-insulator crossover in the half-filled Hubbard model at intermediate to high temperature. \textbf{(a)} TDOS $\tilde{\kappa} = \frac{\partial n}{\partial \mu}$ at half filling. $\tilde{\kappa}$ at different temperatures cross around $U \sim 4$. This signifies the opening of a charge gap in TDOS near half-filling.\textbf{(b)} Zero crossings of $\frac{\partial \tilde{\kappa}}{\partial T}$ at half-filling determine the crossover strength $U_c(T)$ at the respective temperature. At $U_c(0.67) \sim 3.8$, a Mott insulating state with $\frac{\partial \tilde{\kappa}}{\partial T}>0$ is formed around half-filling. \textbf{(c)} Gap opening in LDOS at $T=0.67$ at half filling. Beyond $U=5.5$, a gap develops at $\omega=0$ in LDOS. \textbf{(d)} Comparison of gap onset temperature $T^{*}$ for TDOS contrasted with LDOS. At a fixed temperature, the gap in TDOS opens up at a smaller $U$ than the gap in LDOS, giving rise to the phases (A), (B), (C).}
\label{gap_scales}
\end{figure}

\section{Metal insulator crossover at half filling: TDOS vs LDOS}
\label{sec_mottness}

In this section, we look at two distinct criteria for characterizing a metal vs an insulator: thermodynamic density of states (TDOS), and the local density of states (LDOS). We establish how the difference between the two criteria gives rise to an extended crossover in the $T-U$ plane at half-filling.

(i) The fate of an interacting system at a finite $T$ is decided by temperature variation of TDOS, $\tilde{\kappa} = \frac{\partial n}{\partial \mu}$, shown in Fig \ref{gap_scales}(a). For a metallic state, $\tilde{\kappa}$ is dominated by density fluctuations from Van Hove singularity in the single-particle density of states\cite{kim2020spin}, which are suppressed with temperature (seen for $U <4$). For an insulator, density fluctuations inside the TDOS gap are instead thermally activated, so $\tilde{\kappa}$ increases with temperature. The zero crossings of $\frac{\partial \tilde{\kappa}}{\partial T}$ at a fixed temperature determines the crossover strength $U_{cr}^{\rm TDOS}(T)$, at which the system goes from a metal ($\frac{\partial \tilde{\kappa}}{\partial T} <0$) to an insulator ($\frac{\partial \tilde{\kappa}}{\partial T}>0$), shown in Fig \ref{gap_scales}(b). Opening of a charge gap in TDOS due to density fluctuations is accompanied by an insulating phase around half-filling, and ``isosbestic" points where $n$ vs $\mu$ curves show little temperature variation at specific $\mu$ values (shown in Appendix \ref{app_a}).

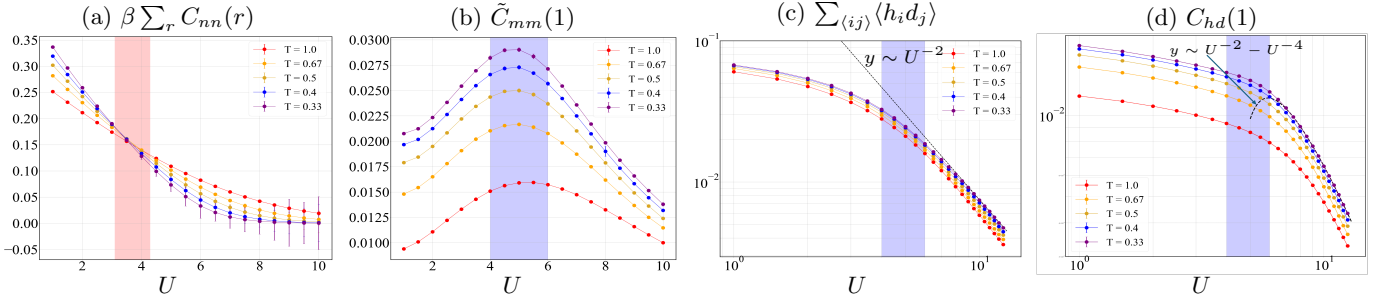
\begin{figure*}[t]
\begin{tikzpicture}

\node (img1) {\includegraphics[width=4.4cm,height=3.2cm]{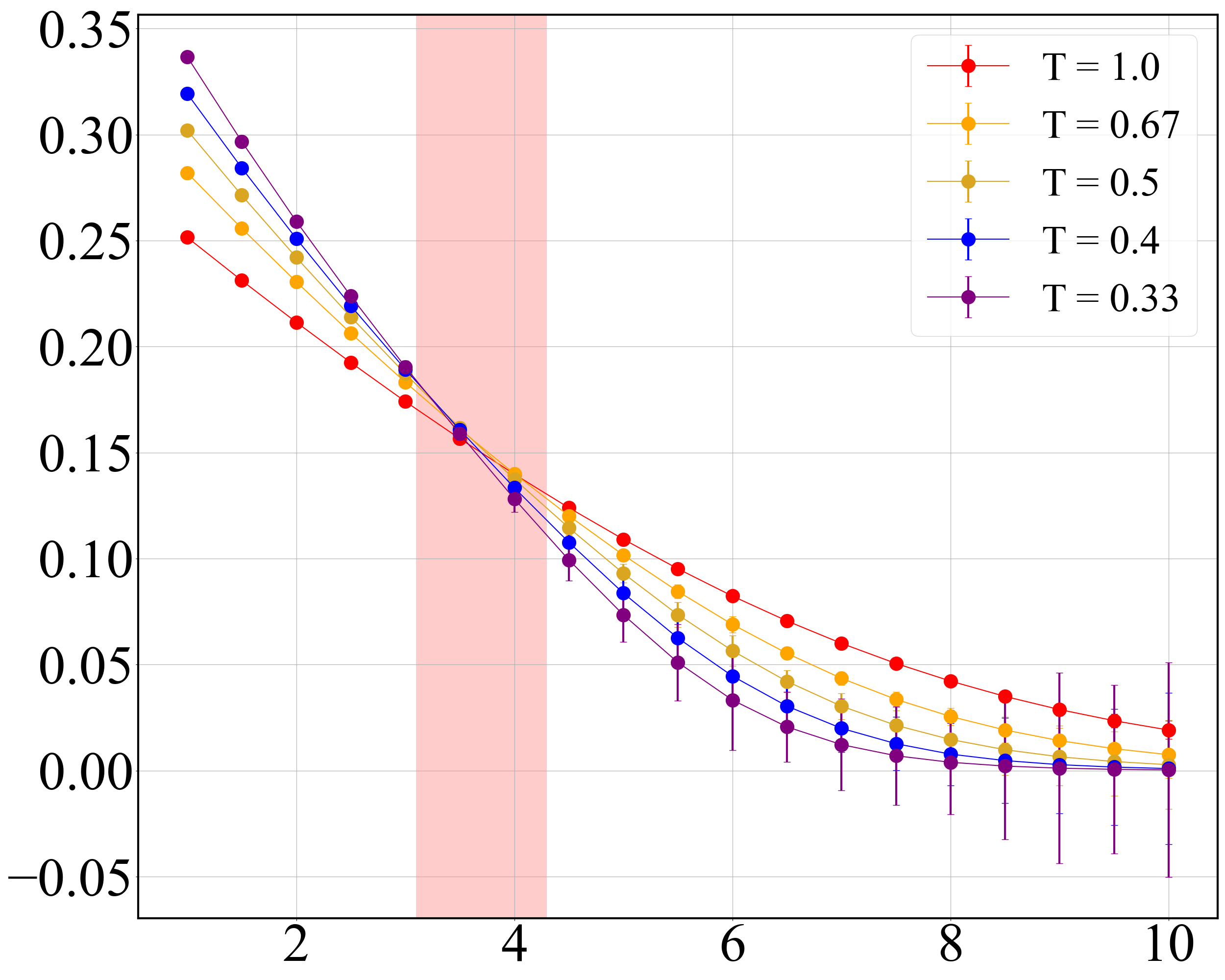}};
\node[above=of img1,node distance=0cm,yshift=-1.2
cm,xshift=0.0cm]{\small{(a) $\beta \sum_{r}C_{nn}(r)$}};
%\node[left=of img1,node distance=0cm,rotate=90,anchor=center,yshift=-0.8cm,xshift=0.0cm]{\small{ $\sum_{r}C_{nn}(r)$}};
\node[below=of img1,node distance=0cm,yshift=1.2cm,xshift=0.0cm]{\small{$U$}};
%%%%%%%%%%%%%%%%%%%%%%%%%%%%%%%%%%%%%%%%%%
\node (img2) [right=of img1,xshift=-1.05cm]{\includegraphics[width=4.4cm,height=3.2cm]{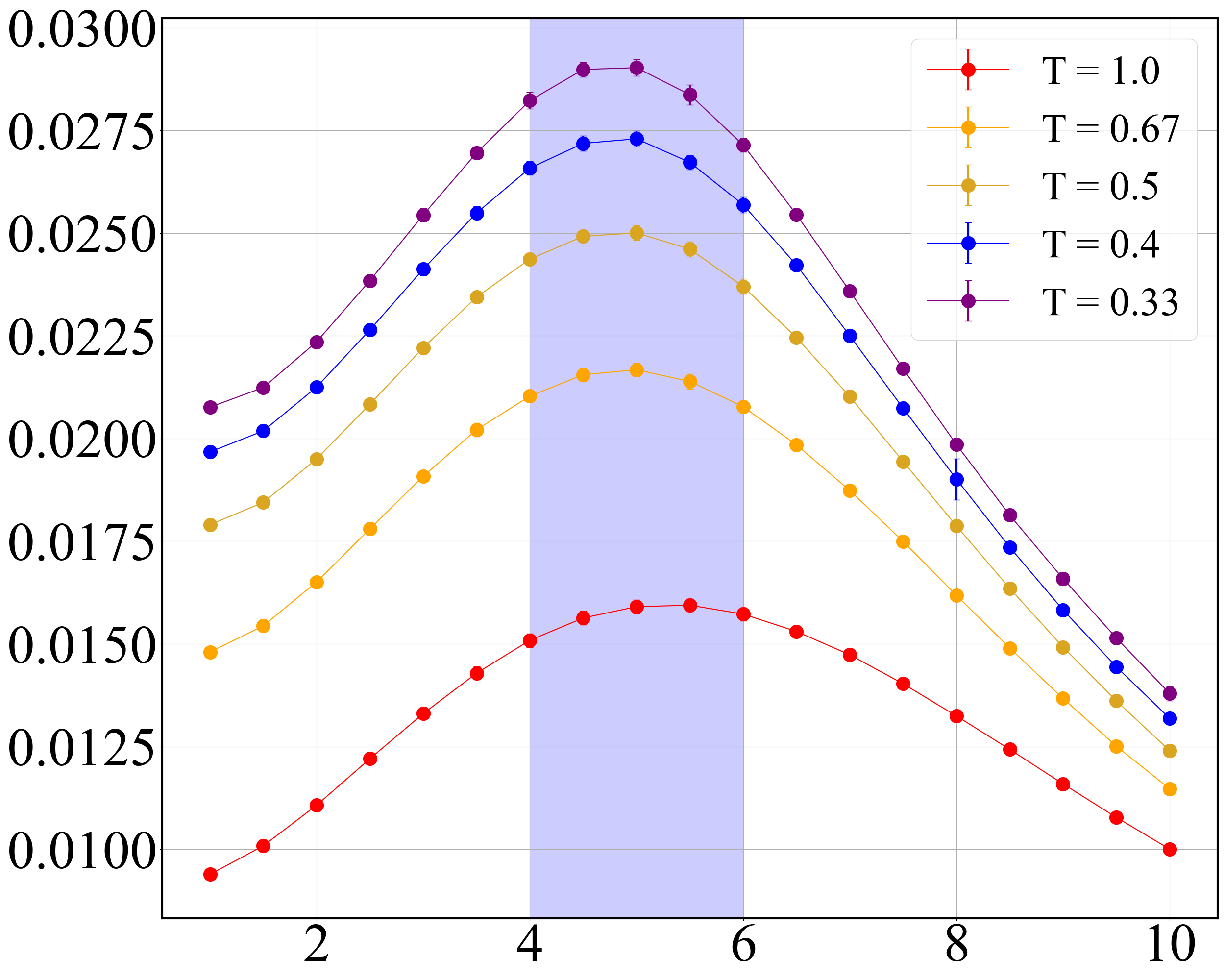}};
\node[above=of img2,node distance=0cm,yshift=-1.2
cm,xshift=0.0cm]{\small{(b) $\tilde{C}_{mm}(1)$}};
%\node[left=of img2,node distance=0cm,rotate=90,anchor=center,yshift=-0.9cm,xshift=0.0cm]{\scriptsize{ $\sum_{\langle ij \rangle}\langle h_id_j\rangle$}};
\node[below=of img2,node distance=0cm,yshift=1.2cm,xshift=0.0cm]{\small{$U$}};
%%%%%%%%%%%%%%%%%%%%%%%%%%%%%%%%%%%%%%%%%%%
\node (img3) [right=of img2,xshift=-1.05cm]{\includegraphics[width=4.4cm,height=3.2cm]{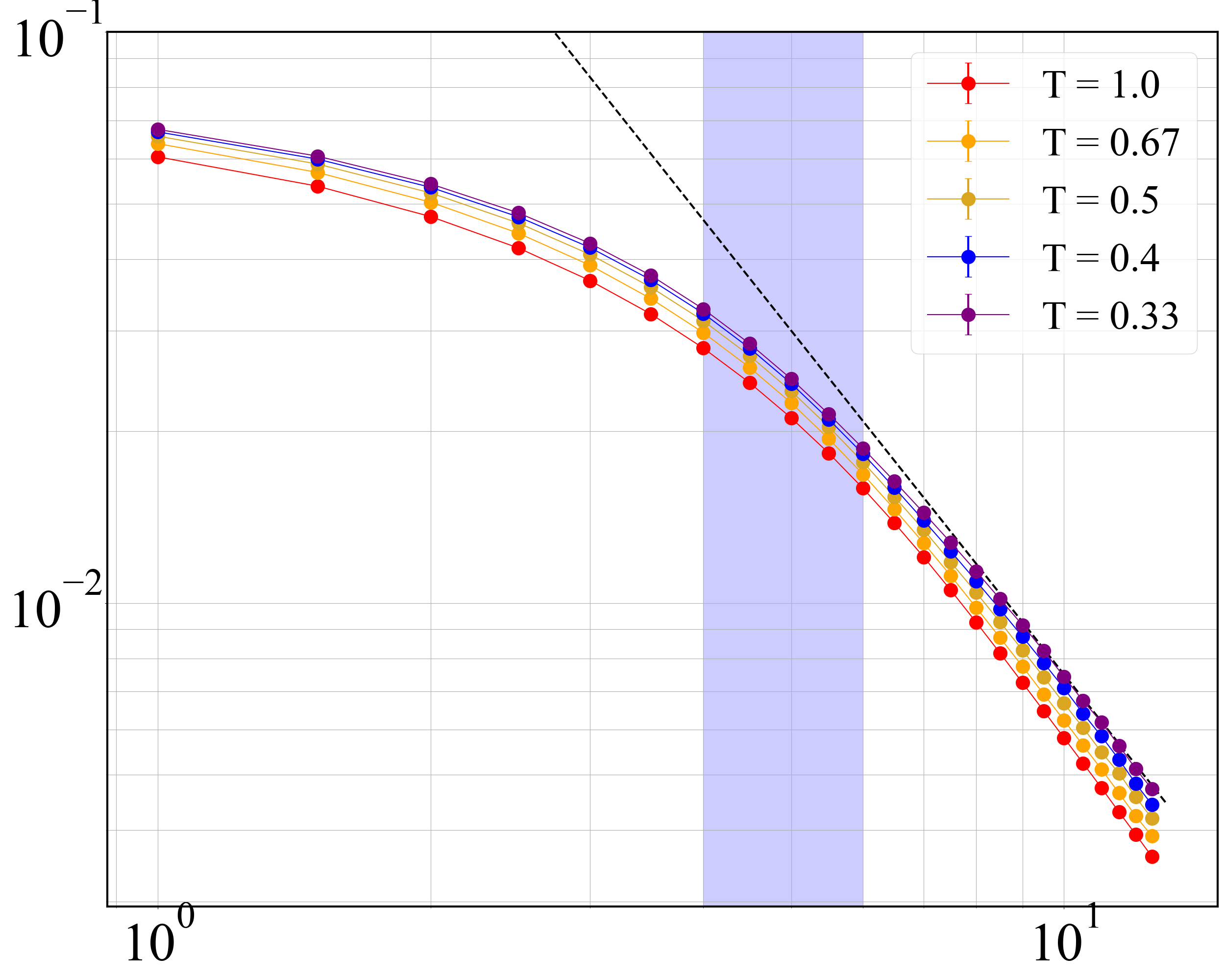}};
%\node[left=of img4,node distance=0cm,rotate=90,anchor=center,yshift=-0.8cm,xshift=0.0cm]{\scriptsize{ $\tilde{C}_{mm}(1)$}};
\node[above=of img3,node distance=0cm,yshift=-1.2
cm,xshift=0.0cm]{\small{(c) $\sum_{\langle ij \rangle}\langle h_i d_j\rangle$}};
\node[left=of img3,node distance=0cm,yshift=1.3cm,xshift=4.6cm]{\scriptsize{$y \sim U^{-2}$}};
\node[below=of img3,node distance=0cm,yshift=1.2cm,xshift=0.0cm]{\small{$U$}};
%%%%%%%%%%%%%%%%%%%%%%%%%%%%%%%%%%%%%%%%%
\node (img4) [right=of img3,xshift=-1.05cm]{\includegraphics[width=4.4cm,height=3.2cm]{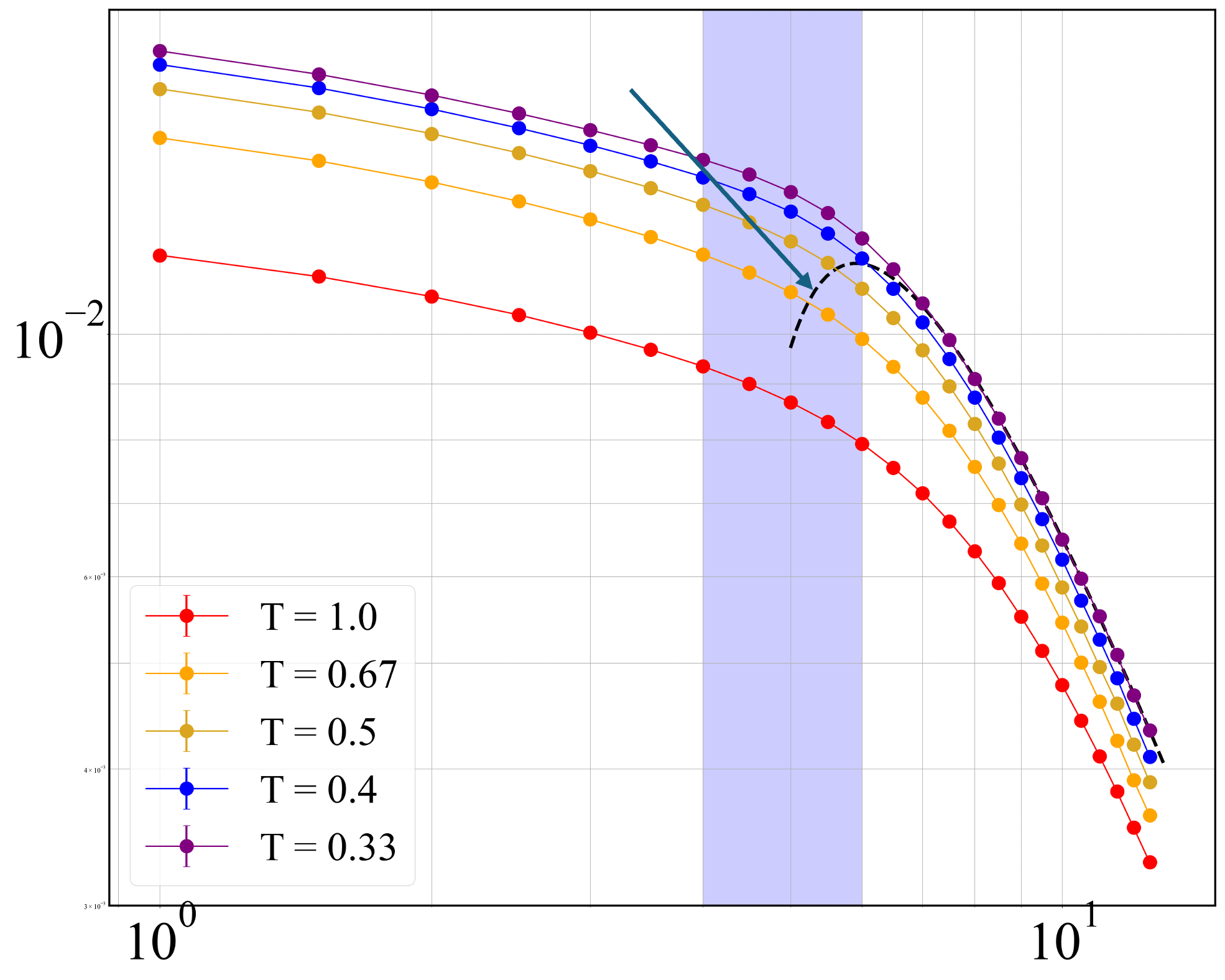}};
%\node[above=of img3,node distance=0cm,yshift=-1.7cm,xshift=0.0cm]{\large{(c) $|i-j|=1$, $n = 1$}};
\node[above=of img4,node distance=0cm,yshift=-1.2
cm,xshift=0.0cm]{\small{(d) $C_{hd}(1)$}};
%\node[left=of img3,node distance=0cm,rotate=90,anchor=center,yshift=-0.9cm,xshift=0.0cm]{\scriptsize{ $C_{hd}(1)$}};
\node[below=of img4,node distance=0cm,yshift=1.2cm,xshift=0.0cm]{\small{$U$}};
\node[left=of img4,node distance=0cm,yshift=1.4cm,xshift=4.8cm]{\tiny{$y \sim U^{-2}-U^{-4}$}};
%%%%%%%%%%%%%%%%%%%%%%%%%%%%%%%%
\end{tikzpicture}
\caption{Correlation functions at half-filling. \textbf{(a)} Number fluctuations $\beta \sum_{r}C_{nn}(r)$. At weak coupling, number fluctuations are suppressed with temperature, as expected in a metal. In the strong coupling regime, number fluctuations are thermally activated as in an insulator. The zero crossings of $\frac{\partial \tilde{\kappa}}{\partial T}$ in Fig \ref{gap_scales}(a) define the crossover window (red shaded region) where $\sum_{r}C_{nn}(r)$ at different temperatures cross, signifying a charge gap opening in TDOS. \textbf{(b)} Normalized moment correlator between neighboring sites, $\tilde{C}_{mm}(1) = C_{mm}(1)/(m^2)^2$. The blue-shaded region denotes the interaction strength window in which a gap opens in the single-particle density of states in the temperature range we consider, Fig \ref{gap_scales}(d). The nearest neighbor moment-moment correlations peak in this window, indicating that short-range moment correlations are responsible for forming a gap in LDOS. \textbf{(c)} Average value of the holon-doublon correlation function between neighboring sites, $p_{hd} = \sum_{\langle ij \rangle}\langle h_id_j\rangle$.  Beyond the LDOS gap opening window, $p_{hd} \sim t^2/U^2$ as predicted by first-order perturbation theory (refer to text), signaling that the system has no double occupancy. \textbf{(d)}  Nearest neighbor holon-doublon connected correlations $C_{hd}(1)$.  Similar to $p_{hd}$, $C_{hd}$ shows a scaling $U^2-U^4$ due to singly occupied sites in the ground state.} 
\label{Correlators_half_filling}
\end{figure*}

(ii) Metal-to-insulator transitions can also be characterized by the behavior of single particle green's function \cite{park2008cluster,schafer2015fate}, 
$G(r,r',\tau) = -\langle \mathcal{T}_{\tau}c_{r}(\tau)c^{\dagger}_{r'}\rangle $. The quasiparticle density of states, $N(\omega) = \frac{-1}{\pi}\text{Im}G(r=0,\omega)$ directly probes the metallicity; $N(\omega) = 0$ has a peak at $\omega = 0$ for a metal. On crossing over to an insulating state beyond $U_{cr}^{\rm LDOS}(T)$, $N(\omega)$ develops a gap at $\omega=0$. 
In Fig \ref{gap_scales}(c), we track the evolution of the LDOS at $T=0.67$. Compared to $U^{\rm TDOS}_{cr} \sim 3.8$, a ``pseudogap" like feature develops in LDOS at a higher interaction strength, $U > 5.5$. 

In a Fermi liquid, the electron compressibility is related to the density of states at the Fermi surface, $\frac{\partial n}{\partial \mu} = \frac{N(0)}{(1+F^{s}_{0})}$, with $F^{s}_{0}$ being the Landau parameter. However, in case of deviations from Fermi liquid theory, the two predictors of metal-insulator transition can show marked differences. This marked difference allows us to construct an approximate phase diagram in Fig \ref{gap_scales}(d). In the intermediate temperature regime, three distinct phases emerge at half-filling showing an extended metal-to-insulator crossover, based on gap formation in TDOS vs LDOS: In region A, the system is metallic and gapless in both TDOS and LDOS. In region B, the system has thermally activated density fluctuations, as in an insulator, but exhibits no gap in $N(\omega)$ at low energies, as in a metal. In region C, the system has thermally activated density fluctuations and the presence of a gap in $N(\omega)$ at low energies, consistent with an insulating behavior.

In Ref \cite{kim2020spin}, an extended crossover was reported between a Fermi liquid to a non-Fermi liquid to an insulator. The crossover was observed at $T \leq T_{FL}$, where quasiparticle properties are meaningful, and $T\leq T_{spin}$, where quasi-long-range AFM correlations develop. In contrast, the phase diagram in Fig \ref{gap_scales}(d) shows the presence of an extended crossover at intermediate to high temperatures that are well above $T_{FL}$ and $T_{spin}$\cite{paiva2010fermions}.

To study the role of microscopic degrees of freedom across the extended crossover in Fig \ref{gap_scales}(d), we look at equal-time correlation functions. Locally, a site can either have a holon $h_i = (1-n_{i\uparrow})(1-n_{i\downarrow})$,  doublon $d_i = n_{i\uparrow}n_{i\downarrow}$ or singlon/local moment, $m^2_i = (n_{i\uparrow}-n_{i\downarrow})^2$. Their mutual correlations can be described by the respective connected correlation functions between a pair of sites $\langle i,i+r \rangle$ as,

\begin{align}
    C_{nn}(r) &= \frac{1}{N_s}\sum_{i}\langle n_{i} n_{i+r}\rangle - \langle n_{i}\rangle \langle n_{i+r}\rangle, \nonumber \\
    C_{hd}(r) &= \frac{1}{N_s}\sum_{i}\langle h_{i} d_{i+r}\rangle - \langle h_{i}\rangle \langle d_{i+r}\rangle \nonumber \\
    C_{mm}(r) &= \frac{1}{N_s}\sum_{i}\langle m^2_{i} m^2_{i+r}\rangle - \langle m^2_{i}\rangle \langle m^2_{i+r}\rangle
\end{align}

The $U,T$ dependence of the above correlation functions at half-filling reveals the following:

(i) The density-density correlations satisfy the fluctuation-dissipation theorem, $\tilde{\kappa} = \frac{\partial n}{\partial \mu} = \beta \sum_{r}C_{nn}(r)$. As expected, the total number fluctuations capture the gap opening in TDOS, shown in Fig \ref{Correlators_half_filling}(a). The number fluctuations curve for temperatures,  $T \in [0.33,1]$, cross in the red shaded region predicted by $U^{\rm TDOS}_{cr}(T)$ from Fig \ref{gap_scales}(b). 

(ii) As short-range spin correlations are thought to be responsible for generating pseudogap \cite{macridin2006pseudogap,kyung2006pseudogap,boschini2020emergence,krien2022explaining}, we test this directly by $C_{mm}(1)$, which in cold atom experiments\cite{cheuk2016observation} measures the tendency of the system to bunch up local moments between neighboring sites. Since the magnitude of $C_{mm}(1)$ is influenced by the magnitude of local moments, that grows with $U$ \cite{paiva2001signatures}, we extract the degree of correlation by normalizing it to $\tilde{C}_{mm}(1) = C_{mm}(1)/\langle m^2 \rangle^2$. As shown in Fig \ref{Correlators_half_filling}(b), normalized moment correlations between nearest neighbor sites is maximal in the blue shaded region, which is determined by $U_{cr}^{LDOS}(T)$ in the temperature range we study(Fig \ref{gap_scales}(d)).

\begin{figure*}[t]
\begin{tikzpicture}
%%%%%%%%%%%%%%%%%%%%%%%%%%%%%
\node(img1){\includegraphics[width=4.0cm,height=3.0cm]{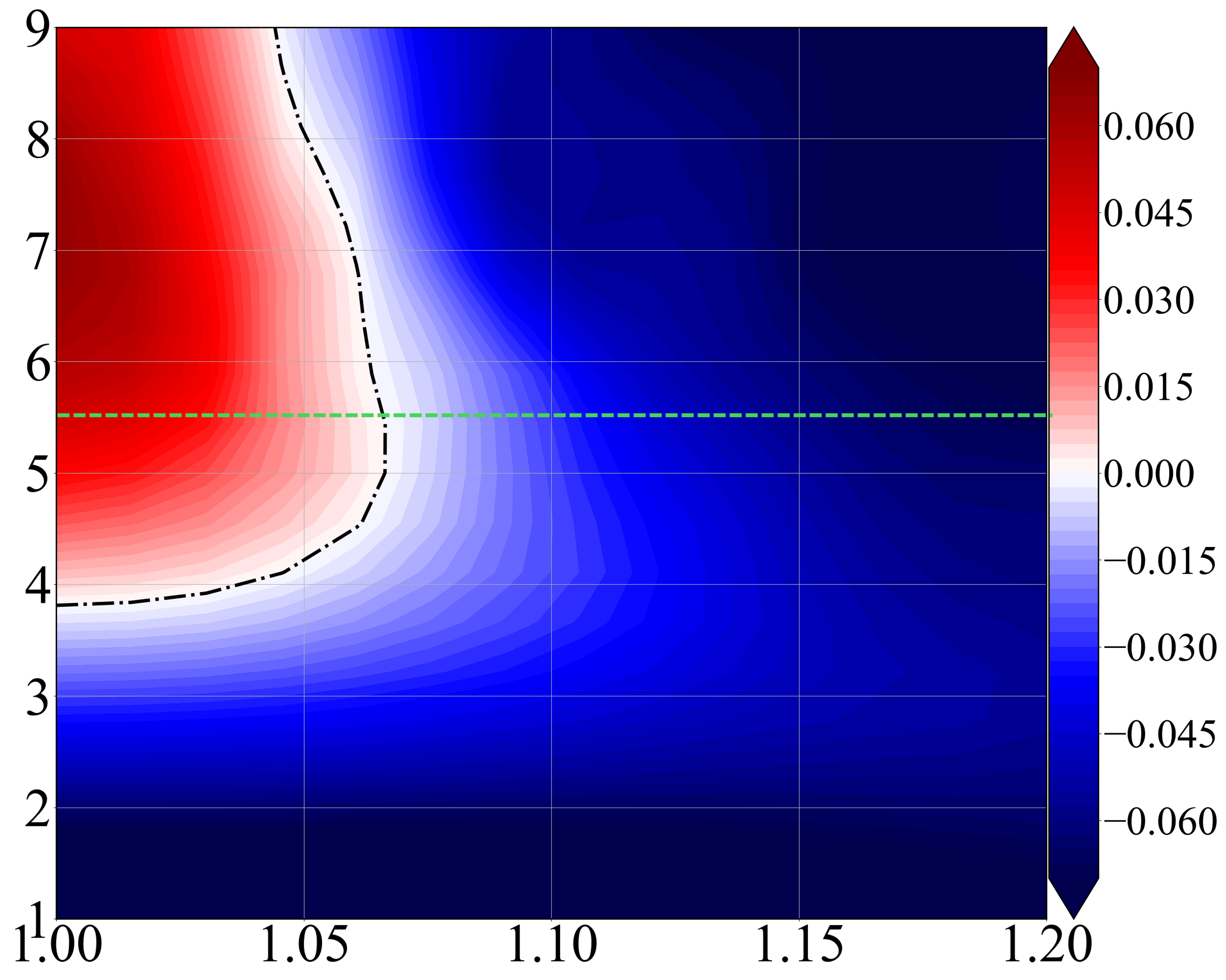}};
%\node[left=of img1,node distance=0cm,rotate=0,anchor=center,yshift=0.0cm,xshift=0.9cm]{\large{ $\frac{\partial \tilde{\kappa}}{\partial T}$}};
\node[left=of img1,node distance=0cm,rotate=0,anchor=center,yshift=0.0cm,xshift=0.9cm]{\normalsize{ $U$}};
\node[below=of img1,node distance=0cm,yshift=1.2cm,xshift=0.0cm]{\normalsize{$n$}};
%\node[left=of img1,node distance=0cm,rotate=0,anchor=center,yshift=-1.0cm,xshift=1.9cm]{\small{ $(a)$}};
\node[above=of img1,node distance=0cm,rotate=0,anchor=center,yshift=-0.9cm,xshift=0.0cm]{\small{(a) $\frac{\partial \tilde{\kappa}}{\partial T}$, $T=0.67$}};
\node[left=of img1,node distance=0cm,rotate=0,anchor=center,yshift=0.1cm,xshift=1.8cm]{\tiny{\textcolor{black}{Insulator}}};
\node[left=of img1,node distance=0cm,rotate=0,anchor=center,yshift=-0.3cm,xshift=3.4cm]{\tiny{\textcolor{white}{Metal}}};
%%%%%%%%%%%%%%%%%%%%%%%%%%%%%%%%%%%%%%%%%%%%
\node(img2)[right=of img1,xshift=-0.75cm]{\includegraphics[width=4.0cm,height=3.0cm]{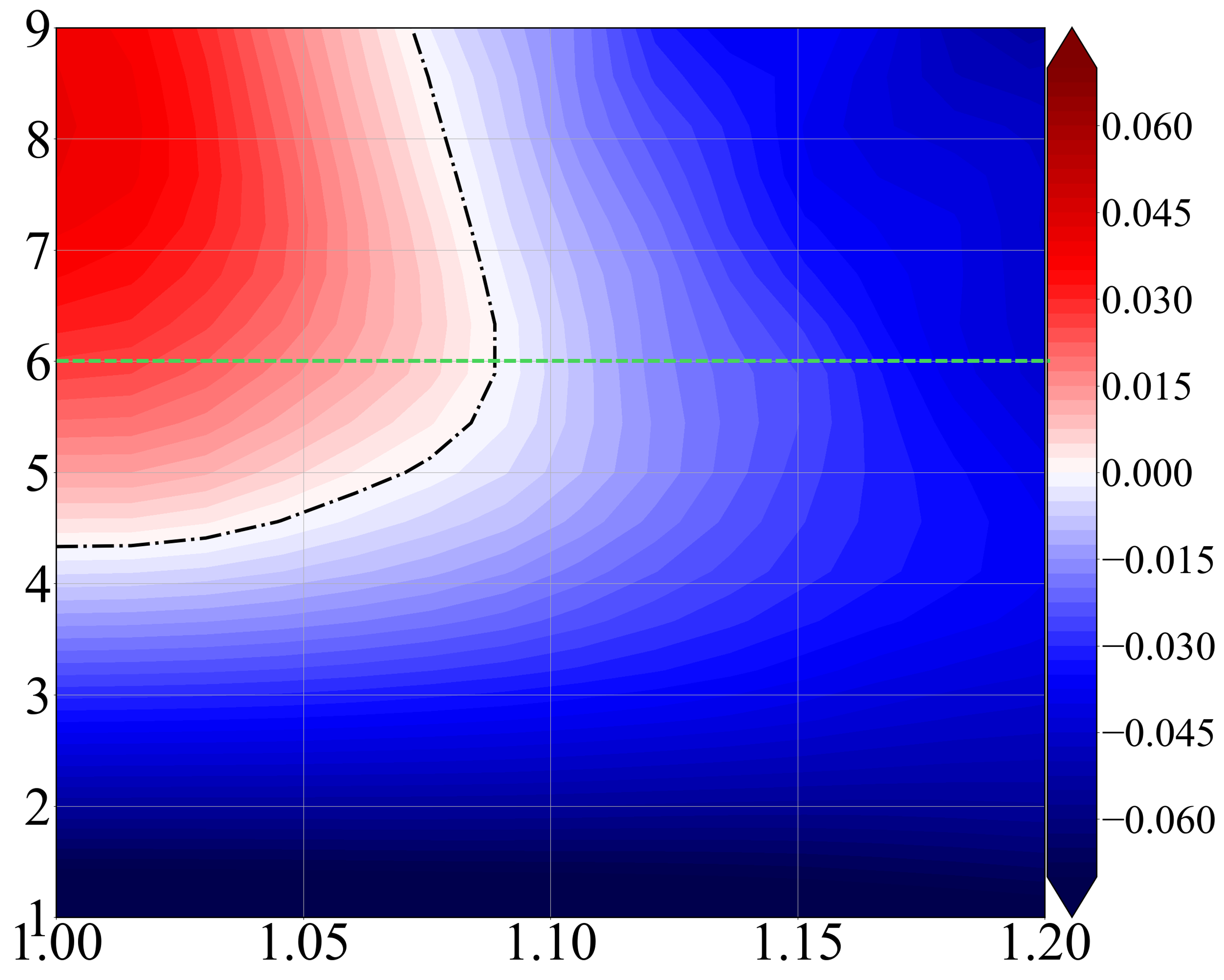}};
%\node[left=of img2,node distance=0cm,rotate=90,anchor=center,xshift=0.0cm,yshift=-0.9cm]{\small{ $C_{nn}(0)$}};
%\node[left=of img2,node distance=0cm,rotate=0,anchor=center,yshift=1.0cm,xshift=4.2cm]{\small{ $(b)$}};
\node[left=of img2,node distance=0cm,rotate=0,anchor=center,yshift=0.0cm,xshift=0.9cm]{\normalsize{ $U$}};
\node[above=of img2,node distance=0cm,rotate=0,anchor=center,yshift=-0.9cm,xshift=0.0cm]{\small{(b) $\frac{\partial \tilde{\kappa}}{\partial T}$, $T = 1.0$}};
\node[below=of img2,node distance=0cm,yshift=1.2cm,xshift=0.0cm]{\normalsize{$n$}};
\node[left=of img2,node distance=0cm,rotate=0,anchor=center,yshift=0.3cm,xshift=2.0cm]{\tiny{\textcolor{black}{Insulator}}};
\node[left=of img2,node distance=0cm,rotate=0,anchor=center,yshift=-0.3cm,xshift=3.4cm]{\tiny{\textcolor{white}{Metal}}};
%%%%%%%%%%%%%%%%%%%%%%%%%%%%%%%%%%%%%%%%%%%%

\node(img3)[right=of img2, xshift = -0.75cm,yshift=0.0cm]{\includegraphics[width=3.8cm,height=3.0cm]{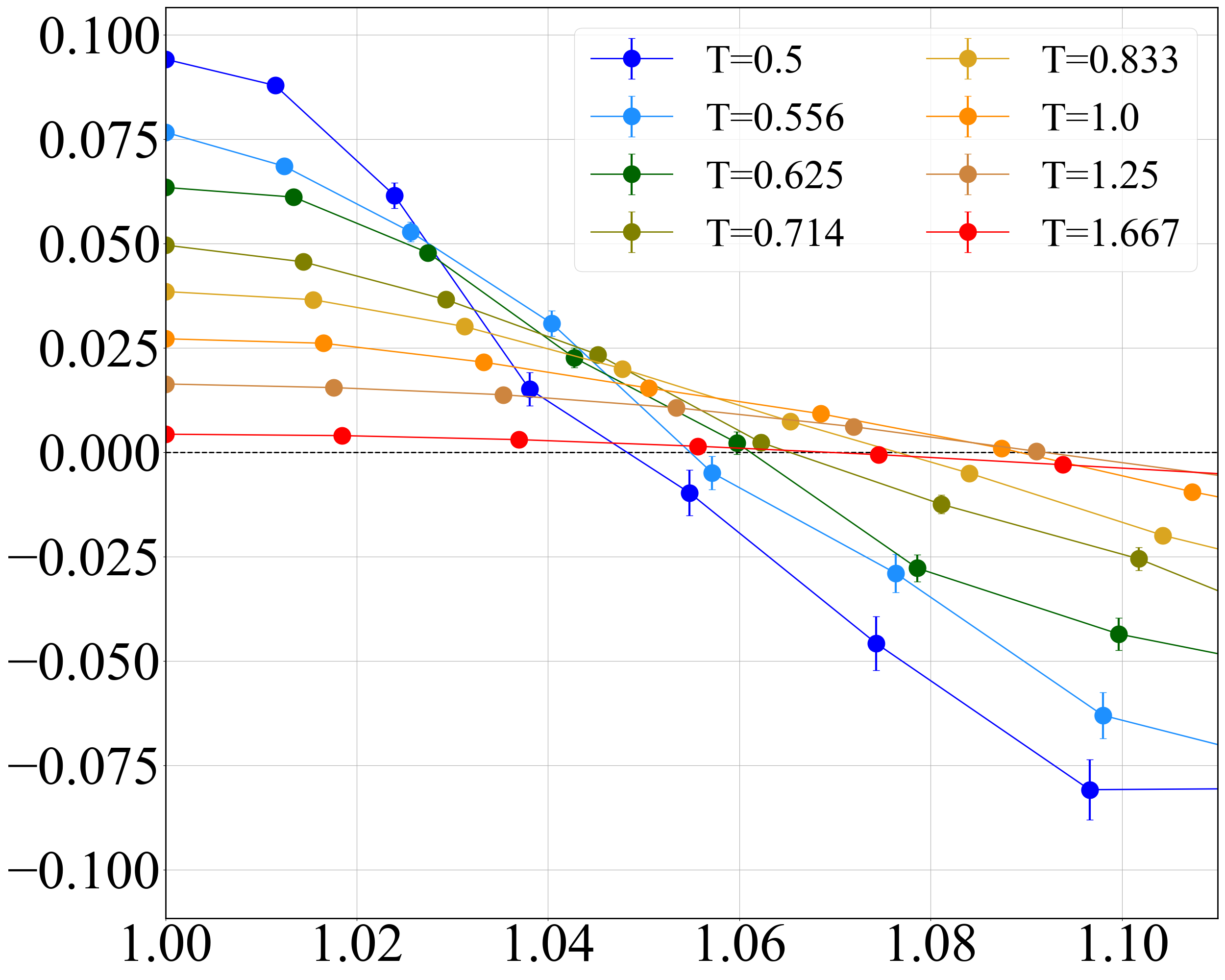}};
\node[left=of img3,node distance=0cm,rotate=0,anchor=center,yshift=0.0cm,xshift=0.9cm]{\large{ $\frac{\partial \tilde{\kappa}}{\partial T}$}};
\node[below=of img3,node distance=0cm,yshift=1.2cm,xshift=0.0cm]{\normalsize{$n$}};
%\node[left=of img5,node distance=0cm,rotate=0,anchor=center,yshift=-1.0cm,xshift=1.9cm]{\small{ $(d)$}};
\node[above=of img3,node distance=0cm,rotate=0,anchor=center,yshift=-3.7cm,xshift=-0.5cm]{\small{(c)} \scriptsize{$U=6.0$}};
%%%%%%%%%%%%%%%%%%%%%%%%%%%%%%%%
\node(img4)[right=of img3,xshift=-0.75cm]{\includegraphics[width=3.6cm,height=3.0cm]{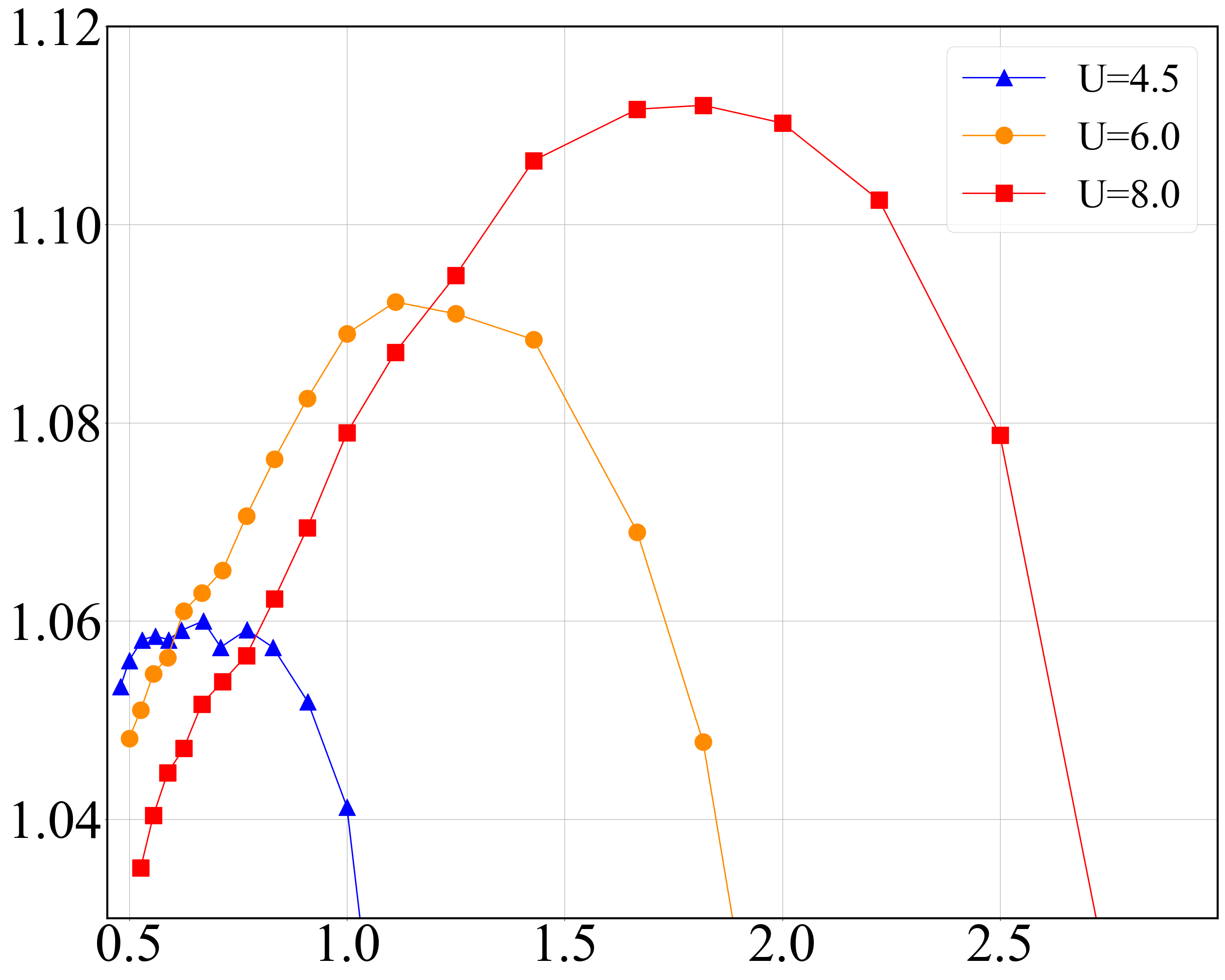}};
\node[left=of img4,node distance=0cm,rotate=0,anchor=center,yshift=0.0cm,xshift=0.9cm]{\normalsize{ $n_{cr}$}};
%\node[left=of img6,node distance=0cm,rotate=0,anchor=center,yshift=1.0cm,xshift=4.2cm]{\small{ $(e)$}};
\node[above=of img4,node distance=0cm,rotate=0,anchor=center,yshift=-3.7cm,xshift=1.5cm]{\small{(d)}};
\node[below=of img4,node distance=0cm,yshift=1.2cm,xshift=0.0cm]{\normalsize{$T$}};
%%%%%%%%%%%%%%%%%%%%%%%%%%%%%
\end{tikzpicture}
\captionof{figure} {Insulator-to-metal crossover at finite doping driven by a charge gap in TDOS. \textbf{(a)} $\frac{\partial \tilde{\kappa}}{\partial T}$ identifying an insulator-to-metal crossover at $T=0.67$. \textbf{(b)} Insulator-to-metal crossover at $T=1.0$. For both (a) and (b), the insulating phase around half-filling grows with $U$, when the system is doped from the half-filling in region A of Fig \ref{gap_scales}(d). In contrast, the insulating phase shrinks with $U$ when it is doped from the half-filling point in region (C). The bright green line in both cases marks $U_{cr}^{TDOS}(T)$ for the respective temperatures. \textbf{(c)} $\frac{\partial \tilde{\kappa}}{\partial T}$ at $U=6.0$, showing the evolution of the insulating phase near half filling, with temperature. The insulating phase initially grows, then decreases, and finally vanishes as the Mott gap in TDOS closes. \textbf{(d)} Crossover density $n_{cr}$ vs temperature for multiple interaction strengths. In the strong coupling regime, the Mott Insulating phase near half-filling grows with temperature, until a scale set by interaction, before turning around and vanishing as the Mott gap closes.}
\label{gap_closing_MIT_correlators}
\end{figure*}
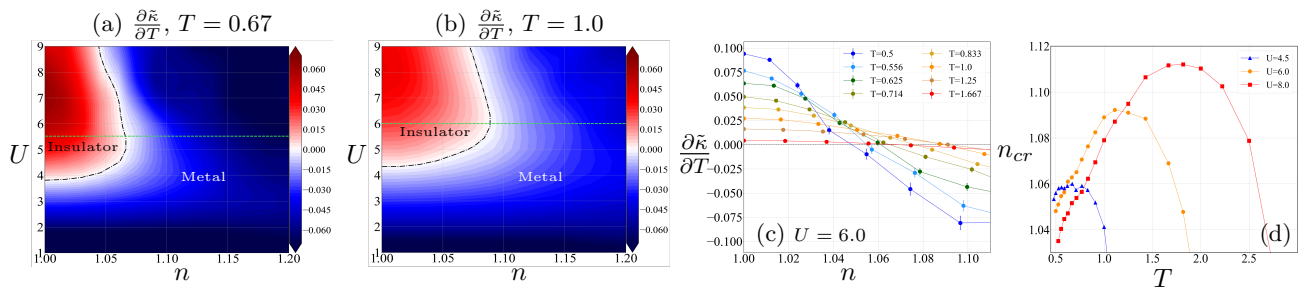

(iii) The change of nature of the system across the crossover from region B to C in Fig \ref{gap_scales}(d) can be probed in cold atom experiments by holon-doublon correlation functions, shown in Fig \ref{Correlators_half_filling}(c) and (d). In the $U\rightarrow \infty$ limit, the ground state $|\psi^{(0)}\rangle $ at half filling is a Mott Insulator with no double occupancy. The first order correction to the ground state $|\psi^{(1)}\rangle $, containing a doublon and a holon on nearest neighbor sites, is $ |\psi\rangle = |\psi^{(0)}\rangle-\frac{t}{U}|\psi^{(1)}\rangle $. This implies that the doublon occupation $\langle d \rangle $ and holon occupation $\langle h\rangle$ scale as $t^2/U^2$, and the probability to find a holon-doublon pair between neighboring sites\cite{endres2013single} $p_{hd}$, and the connected holon doublon correlator $C_{hd}(1)$ should scale as:

\begin{align}
    C_{hd}(1) &= \sum_{\langle ij \rangle}\langle \psi|h_{i}d_{j}|\psi\rangle-\langle \psi|h_{i}|\psi\rangle\langle \psi|d_{j}|\psi\rangle \sim \frac{t^2}{U^2}-\frac{t^4}{U^4} \nonumber \\
        p_{hd} &= \sum_{ij}\langle \psi |h_{i} d_{j}|\psi \rangle \sim \frac{t^2}{U^2}
\end{align}

The $U$ dependence of $p_{hd}$ in Fig \ref{gap_scales}(c) and $C_{hd}(1)$ in Fig \ref{gap_scales}(d) strongly suggest that once a pseudogap forms in LDOS, the system eliminates double occupancy in the ground state, with a finite doublon number coming from thermal fluctuations. Thus the crossover across $U_{cr}^{\rm LDOS}(T)$ is characterized by \textit{short range} correlations that capture the effects of strong onsite coulomb repulsion; whereas the crossover across $U_{cr}^{\rm TDOS}(T)$ is characterized by density fluctuations at \textit{large length scales}, as we shall see in the following section. As we show below, this distinction also has implications for correlation functions away from half-filling.

\section{Correlations away from half filling}
\label{subsec_density_density_correlator}

In this section, we study the evolution of long-range (density-density) and short-range (holon-doublon, moment-moment) correlation functions with interaction and temperature, as the system is doped away from half-filling in various regions of the phase diagram shown in Fig \ref{gap_scales}(d).

\subsection{Density correlations away from half-filling}

The insulator-to-metal crossover at finite doping, as identified by the temperature dependence of TDOS, is shown in Fig. \ref{gap_closing_MIT_correlators} for $T=0.67$(a) and $T=1.0$(b). The black dashed line marks the separatrix $n_{cr}(U,T)$ where the crossover takes place due to $\frac{\partial \tilde{\kappa}}{\partial T} \big |_{n_{cr}}=0$. The insulating phase boundary has the following features:

(i) At a fixed temperature, the interaction strength dependence of $n_{cr}(U,T)$ is sensitive to the nature of the insulating state at half-filling. When the system has a gap in TDOS but no gap in LDOS at half-filling (region B in Fig. \ref{gap_scales}(d)), $n_{cr}(U,T)$ grows with increasing $U$. In contrast, when the system is gapped in both TDOS and LDOS at half-filling (Region C), $n_{cr}(U,T)$ decreases with increasing $U$. This is seen for both $T=0.67$ and $T=1.0$ in Fig \ref{gap_closing_MIT_correlators}.

(ii) The insulator-to-metal crossover at finite doping also shows a non-monotonic dependence on temperature, as can be seen for $U=6.0$ in Fig. \ref{gap_closing_MIT_correlators}(c). The crossover density $n_{cr}(U,T)$ initially increases with $T$, until $T \sim O(t)$; it then turns down and decreases monotonically with $T$, reaching half-filling when $\frac{\partial \tilde{\kappa}}{\partial T}=0$ and the gap in TDOS closes. A comparison between different $U$ in Fig. \ref{gap_closing_MIT_correlators}(d) shows that for both intermediate and strong coupling, $n_{cr}(U,T)$ initially increases up to a temperature scale set by $U$. Beyond it, $n_{cr}(U,T)$  monotonically decreases until the gap in TDOS closes.

\begin{figure}[t]
\begin{tikzpicture}
%%%%%%%%%%%%%%%%%%%%%%%%%%%%%
\node(img1){\includegraphics[width=4.0cm,height=3.0cm]{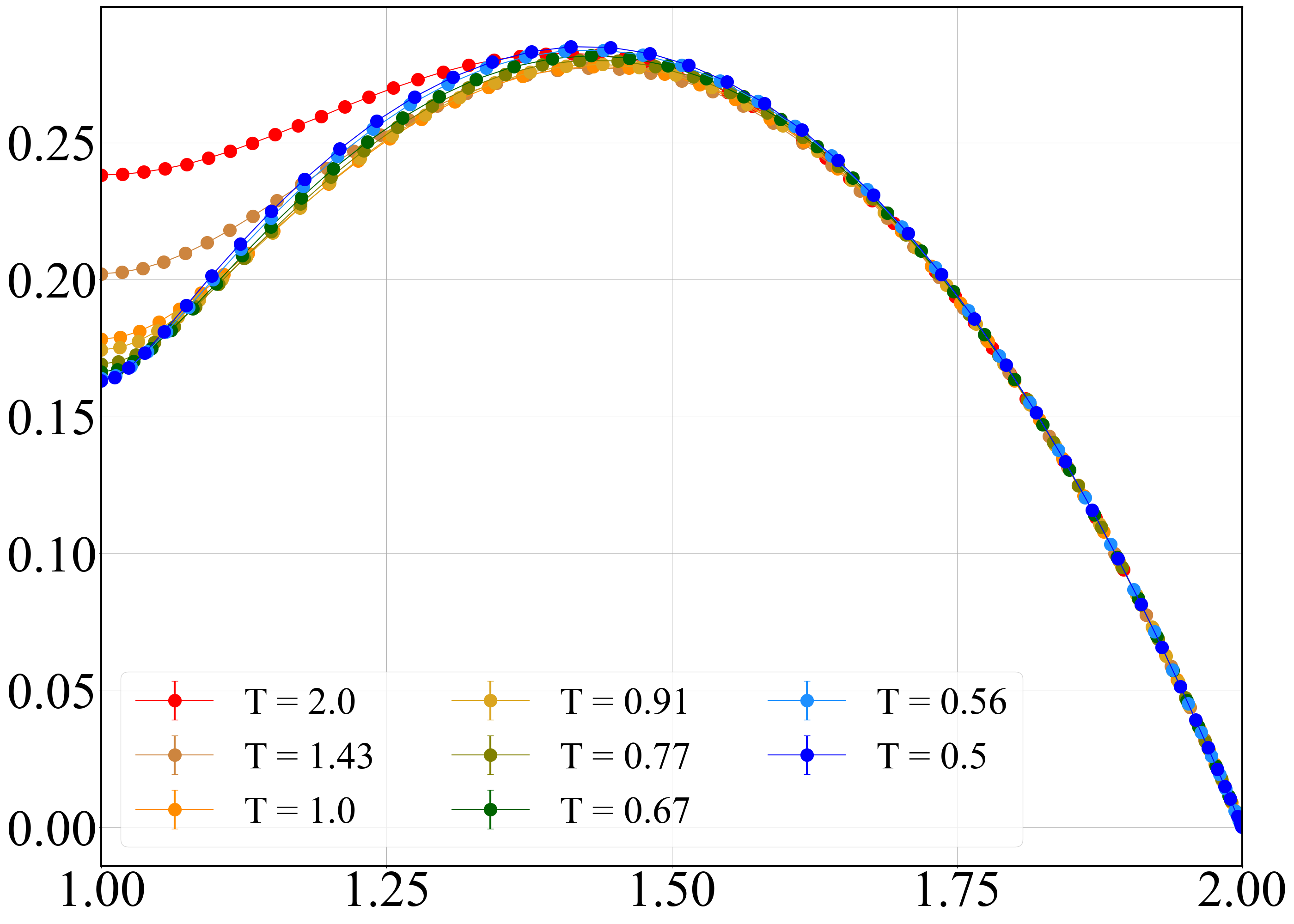}};
%\node[left=of img1,node distance=0cm,rotate=0,anchor=center,yshift=0.0cm,xshift=0.9cm]{\large{ $\frac{\partial \tilde{\kappa}}{\partial T}$}};
\node[below=of img1,node distance=0cm,yshift=1.2cm,xshift=0.0cm]{\normalsize{$n$}};
%\node[left=of img1,node distance=0cm,rotate=0,anchor=center,yshift=-1.0cm,xshift=1.9cm]{\small{ $(a)$}};
\node[above=of img1,node distance=0cm,rotate=0,anchor=center,yshift=-0.9cm,xshift=0.0cm]{\small{(a) $C_{nn}(0)$, $U = 6.0$}};
%%%%%%%%%%%%%%%%%%%%%%%%%%%%%%%%%%%%%%%%%%%%
\node(img2)[right=of img1,xshift=-1.05cm]{\includegraphics[width=4.0cm,height=3.0cm]{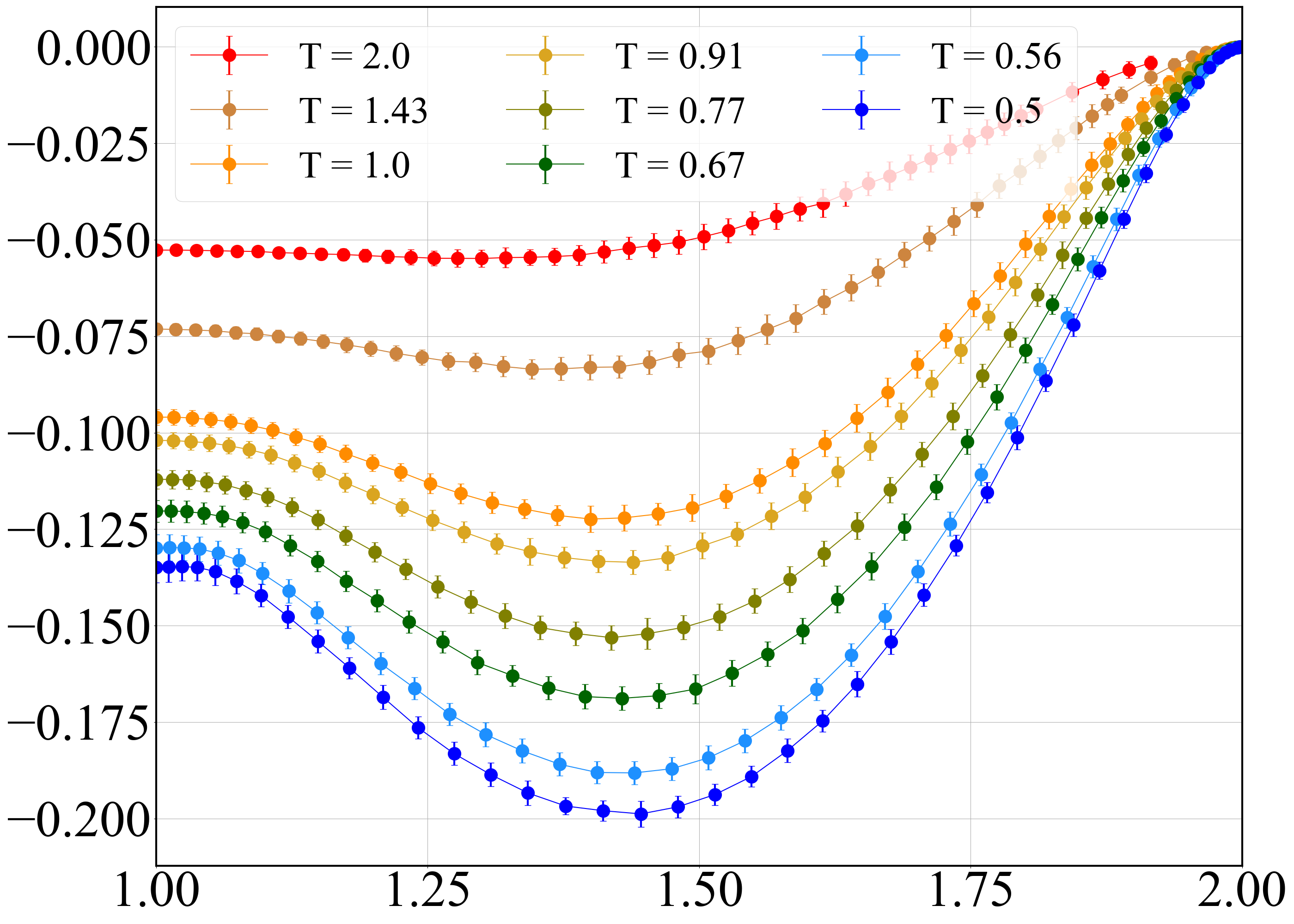}};
%\node[left=of img2,node distance=0cm,rotate=90,anchor=center,xshift=0.0cm,yshift=-0.9cm]{\small{ $C_{nn}(0)$}};
%\node[left=of img2,node distance=0cm,rotate=0,anchor=center,yshift=1.0cm,xshift=4.2cm]{\small{ $(b)$}};
\node[above=of img2,node distance=0cm,rotate=0,anchor=center,yshift=-0.9cm,xshift=0.0cm]{\small{(b) $C_{nn}^{\rm nl}$, $U = 6.0$}};
\node[below=of img2,node distance=0cm,yshift=1.2cm,xshift=0.0cm]{\normalsize{$n$}};
%%%%%%%%%%%%%%%%%%%%%%%%%%%%%%%%%%%%%%%%%%%%
\node(img3)[below=of img1, yshift = 0.4cm,xshift  = 0.0cm]{\includegraphics[width=3.8cm,height=3.0cm]{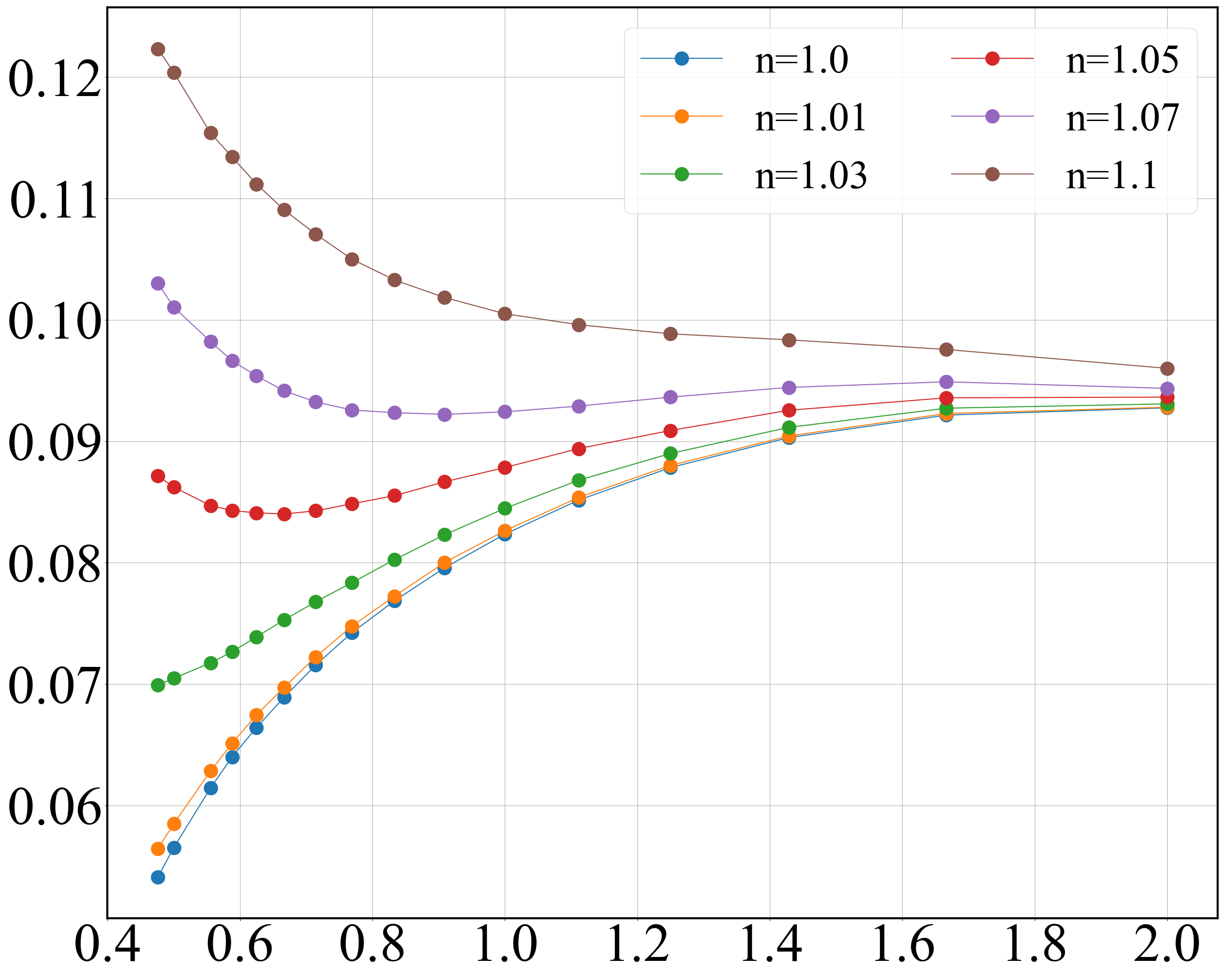}};
%\node[left=of img5,node distance=0cm,rotate=0,anchor=center,yshift=0.0cm,xshift=0.9cm]{\large{ $\frac{\partial \tilde{\kappa}}{\partial T}$}};
\node[below=of img3,node distance=0cm,yshift=1.2cm,xshift=0.0cm]{\normalsize{$T$}};
%\node[left=of img5,node distance=0cm,rotate=0,anchor=center,yshift=-1.0cm,xshift=1.9cm]{\small{ $(d)$}};
\node[above=of img3,node distance=0cm,rotate=0,anchor=center,yshift=-0.9cm,xshift=0.0cm]{\small{(c) $\tilde{\kappa}=\beta C^{\rm nl}_{nn}+\beta C_{nn}(0)$}};
%%%%%%%%%%%%%%%%%%%%%%%%%%%%%%%%
\node(img4)[right=of img3,xshift=-0.85cm]{\includegraphics[width=3.8cm,height=3.0cm]{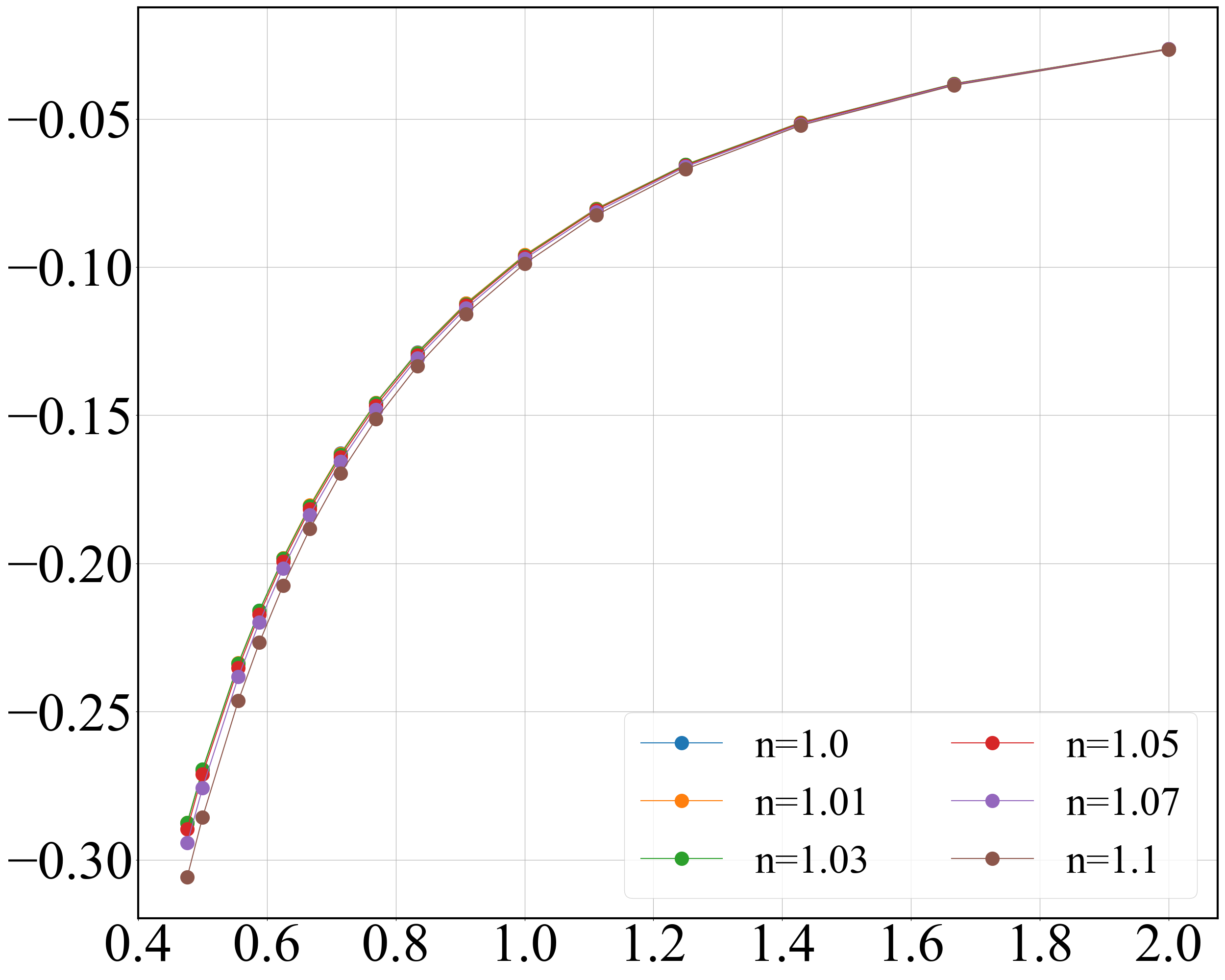}};
%\node[left=of img6,node distance=0cm,rotate=90,anchor=center,xshift=0.0cm,yshift=-0.9cm]{\small{ $C_{nn}(0)$}};
%\node[left=of img6,node distance=0cm,rotate=0,anchor=center,yshift=1.0cm,xshift=4.2cm]{\small{ $(e)$}};
\node[above=of img4,node distance=0cm,rotate=0,anchor=center,yshift=-0.9cm,xshift=0.0cm]{\small{(d) $\beta C_{nn}^{\rm nl}$}};
\node[below=of img4,node distance=0cm,yshift=1.2cm,xshift=0.0cm]{\normalsize{$T$}};
%%%%%%%%%%%%%%%%%%%%%%%%%%%%%%%%%%%%%%%%%%%%
\node(img5)[below=of img3,yshift=0.4cm]{\includegraphics[width=3.8cm,height=3.0cm]{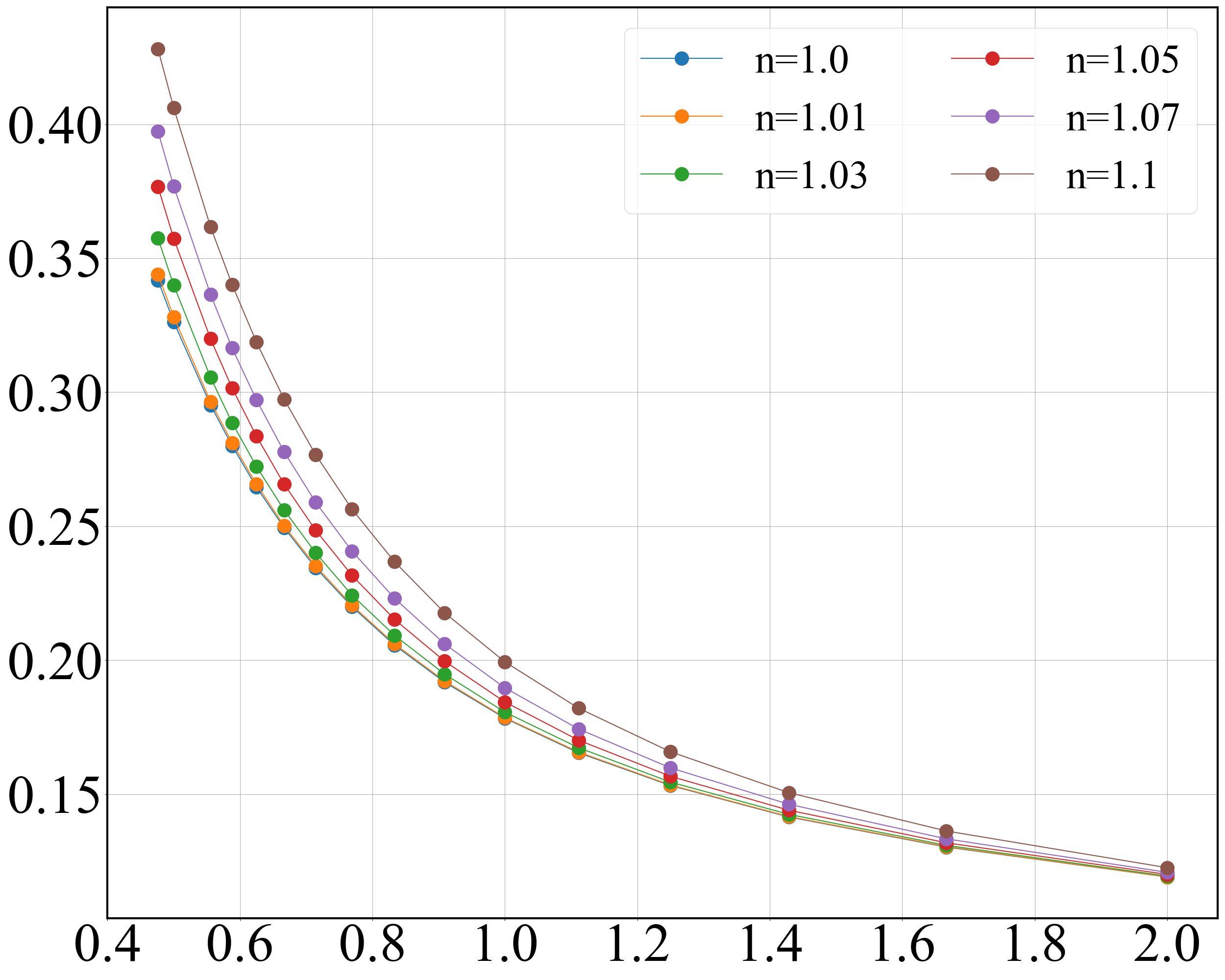}};
%\node[left=of img6,node distance=0cm,rotate=90,anchor=center,xshift=0.0cm,yshift=-0.9cm]{\small{ $C_{nn}(0)$}};
%\node[left=of img6,node distance=0cm,rotate=0,anchor=center,yshift=1.0cm,xshift=4.2cm]{\small{ $(e)$}};
\node[above=of img5,node distance=0cm,rotate=0,anchor=center,yshift=-0.9cm,xshift=0.0cm]{\small{(e) $\beta C_{nn}(0)$}};
\node[below=of img5,node distance=0cm,yshift=1.2cm,xshift=0.0cm]{\normalsize{$T$}};
%%%%%%%%%%%%%%%%%%%%%%%%
\node(img6)[right=of img5,xshift=-0.85cm]{\includegraphics[width=3.8cm,height=3.0cm]{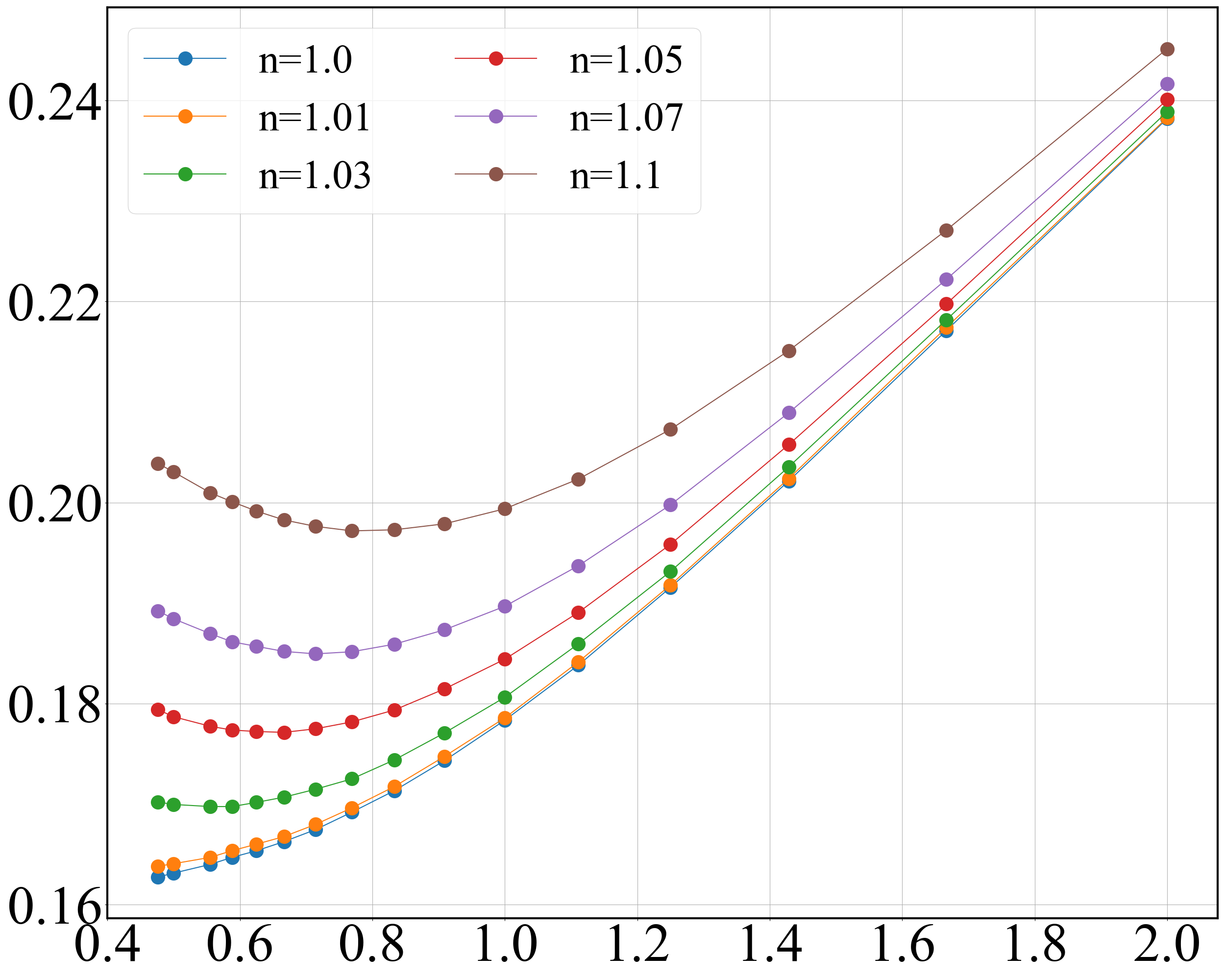}};
%\node[left=of img6,node distance=0cm,rotate=90,anchor=center,xshift=0.0cm,yshift=-0.9cm]{\small{ $C_{nn}(0)$}};
%\node[left=of img6,node distance=0cm,rotate=0,anchor=center,yshift=1.0cm,xshift=4.2cm]{\small{ $(e)$}};
\node[above=of img6,node distance=0cm,rotate=0,anchor=center,yshift=-0.9cm,xshift=0.0cm]{\small{(f) $C_{nn}(0)$}};
\node[below=of img6,node distance=0cm,yshift=1.2cm,xshift=0.0cm]{\normalsize{$T$}};
%%%%%%%%%%%%%%%%%%%%%%%
\end{tikzpicture}
\captionof{figure} {Density-density correlations at $U = 6.0$. \textbf{(a)} Local correlations $C_{nn}(0)$ are temperature independent at large densities. Near half-filling, $C_{nn}(0)$ shows very little temperature variation at intermediate temperatures due to strong onsite coulomb repulsion; it increases beyond $T \sim t$. \textbf{(b)} Non-local correlations $C_{nn}^{\rm nl}$, on the other hand, show considerable temperature variation at all densities, due to a decrease of the thermal De Broglie wavelength that mediate these fluctuations\cite{walsh2019critical}. \textbf{(c)} $\tilde{\kappa} = \beta C^{\rm tot}_{nn}$ at fixed densities. Close to half-filling, ($1.03 < n < 1.1$) a metallic state can turn insulating with increasing temperature, leading to the anomalous rise of $n_{cr}(U,T)$. \textbf{(d)} Non-local contributions to $\tilde{\kappa}$, $\beta C_{nn}^{\rm nl}$. \textbf{(e)} Local contribution to $\tilde{\kappa}$, $\beta C_{nn}(0)$ at fixed densities. At low temperatures, the local contributions increase with doping, whereas the non-local contributions show little doping dependence. \textbf{(f)} Local density fluctuation $C_{nn}(0)$ at fixed doping. The minima at low $T$ correspond to $d(T)$ minima and shift to higher temperatures with increasing density. }
\label{temp_Dependence_density_correlations}
\end{figure}
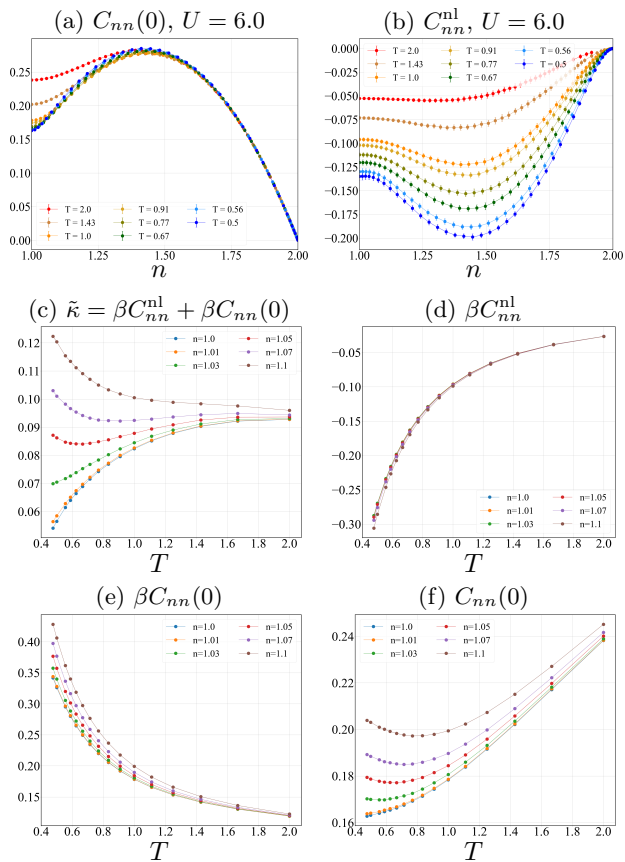

The anomalous increase of $n_{cr}(U,T)$ can be understood by looking at the onsite and non-local density fluctuation contributions to $\tilde{\kappa}$ separately. The fluctuation-dissipation theorem enables us to separate the components of  $\tilde{\kappa}$ as $\tilde{\kappa} = \beta [C_{nn}(0)+C_{nn}^{\rm nl}]$, with the local and non-local pieces of the density correlations defined as:

\begin{align}  
    &C_{nn}(0) = \frac{1}{N_s}\sum_{i}\langle n^2_{i}\rangle - \langle n_i\rangle^2, \nonumber \\
    &C_{nn}^{\rm nl} = \frac{1}{N_s}\sum_{i,r>0}\langle n_{i}n_{i+r}\rangle-\langle n_i\rangle^2 
\end{align}
%As inferred in Ref \cite{cocchi2016equation}, the local density fluctuations does not account for all the number fluctuations; the non local fluctuations are of the same order of magnitude as the local ones. 
While the local correlations are always positive, the nonlocal correlations are negative, suggesting repulsion between density pairs. At $U=0$, this is due to Pauli exclusion, giving rise to the ``Pauli exclusion hole" \cite{hartke2020doublon}. The repulsion gets stronger with $U$, giving rise to the ``correlation hole"\cite{mahan2013many}. The total thermodynamic fluctuation $\delta n^2 = \delta n^2_{\rm onsite}+\delta n^2_{\rm non local}$ arises due to a subtle interplay between these two effects\cite{walsh2019critical}. Fig. \ref{temp_Dependence_density_correlations}(a) and (b) show the onsite and non-local density correlations for $U=6.0$ from intermediate to high temperature, until the gap in TDOS closes at $T \sim 1.9$. Two comments are in order before we look at the interplay of $C_{nn}(0)$ and $C_{nn}^{\rm nl}$ in determining $n_{cr}(U,T)$.

First,  the onsite density-density correlation can be expanded as $C_{nn}(0) = n+2d-n^2$. For large densities, $C_{nn}(0)$ is temperature independent, as seen in Fig \ref{temp_Dependence_density_correlations}(a). This is due to a similar temperature dependence of the doublon number and the density. However, in the vicinity of half-filling, $n \approx n^2$ and the doublon number displays little variation at low temperatures due to strong onsite repulsion. It starts to increase only when $T \sim t$, leading to the temperature variation of $C_{nn}(0)$. 

Second, the non-local density correlations show significant temperature dependence at all densities, due to the decrease of the thermal De Broglie wavelength ($\lambda_T \propto \frac{1}{\sqrt{T}}$) that mediates these correlations. The minima at $n = 0.5$ (and hence $n = 1.5$, since $C_{nn}(r)$ is particle hole symmetric) is a consequence of strong onsite Coulomb repulsion \cite{hartke2020doublon}. Hubbard repulsion between unlike spins reduces available lattice sites for each spin species by half; the negative correlations are thus maximized when there are as many empty sites as occupied sites, at $n=0.5$. Increasing $T$ weakens interaction effects and suppresses the minima. For all temperatures considered, the non-local correlations show minimal variation with respect to doping near half-filling, due to interparticle distance $\lambda_p \propto \frac{1}{\sqrt{n}}$ being larger than $\lambda_T$. It monotonically shrinks as particle density $n$ increases; with $C_{nn}^{\rm nl}$ starting to show doping dependence when $\lambda_p$ becomes comparable to $\lambda _T$.

The temperature dependence of $\tilde{\kappa}$ is shown in Fig \ref{temp_Dependence_density_correlations}(c), up to temperatures where the gap in TDOS closes. The system is an insulator very close to half-filling, and metallic at higher densities ($n > 1.1)$. The behavior at intermediate doping is more complex; the system starts as a metal at a lower temperature, but can become insulating at a higher temperature; this leads to the anomalous rise of $n_{cr}(T;U)$ seen in Fig \ref{gap_closing_MIT_correlators}(d). A comparison between the onsite contribution to $\tilde{\kappa}$, $\beta C_{nn}(0)$, and the non-local contribution, $\beta C_{nn}^{nl}$, (Fig \ref{temp_Dependence_density_correlations}(d-e)) shows that $\beta C_{nn}^{\rm nl}$ has little density dependence, due to competition between $l_p$ and $\lambda_T$ as noted before. In contrast, $\beta C_{nn}(0)$ shows a more pronounced density dependence at low temperatures leading to the non-monotonic temperature dependence of $\tilde{\kappa}$ at intermediate density (Fig \ref{temp_Dependence_density_correlations}(c)).

In order to investigate the low-temperature doping dependence of $\beta C_{nn}(0)$, we plot the onsite density correlation $C_{nn}(0)$ as a function of $T$ in Fig. \ref{temp_Dependence_density_correlations}(f), which reveals a striking feature. %Slightly away from half filling,
For some densities, a minimum develops in  $C_{nn}(0)$ at low temperatures. %; as doping increases the minimum moves to higher temperatures. 
Since $C_{nn}(0) = n+2d-n^2$, at a fixed density, the temperature dependence of $C_{nn}(0)$ is entirely due to average doublon number $d(T)$. At half-filling, a shallow minima occurs in $d(T)$ \cite{paiva2010fermions}; the minima shifts to higher temperatures with increasing doping. As thermally activated doublon number ($\frac{\partial d}{\partial T}>0$) is a signature of localization \cite{paiva2010fermions,dare2007interaction}, comparison between $\tilde{\kappa}$ and $C_{nn}(0)$ in Fig. \ref{temp_Dependence_density_correlations}(c) and (f) suggests that at intermediate doping, an anomalous state can be accessed by increasing $T$, such that it is metallic ($\frac{\partial \tilde{\kappa}}{\partial T}<0$) but has interaction induced localization of electrons ($\frac{\partial d}{\partial T} >0$) as in an insulator (Appendix \ref{app_c}).

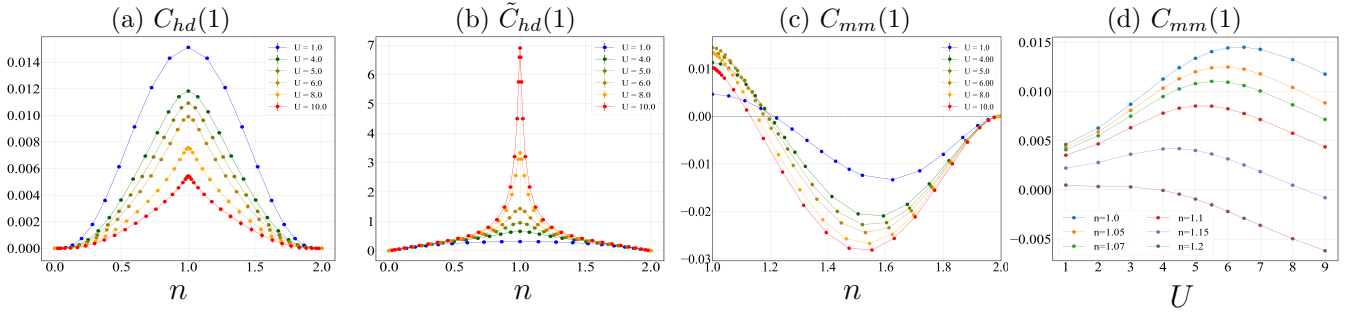
\begin{figure*}[t]
\begin{tikzpicture}

\node (img1) {\includegraphics[width=4.4cm,height=3.2cm]{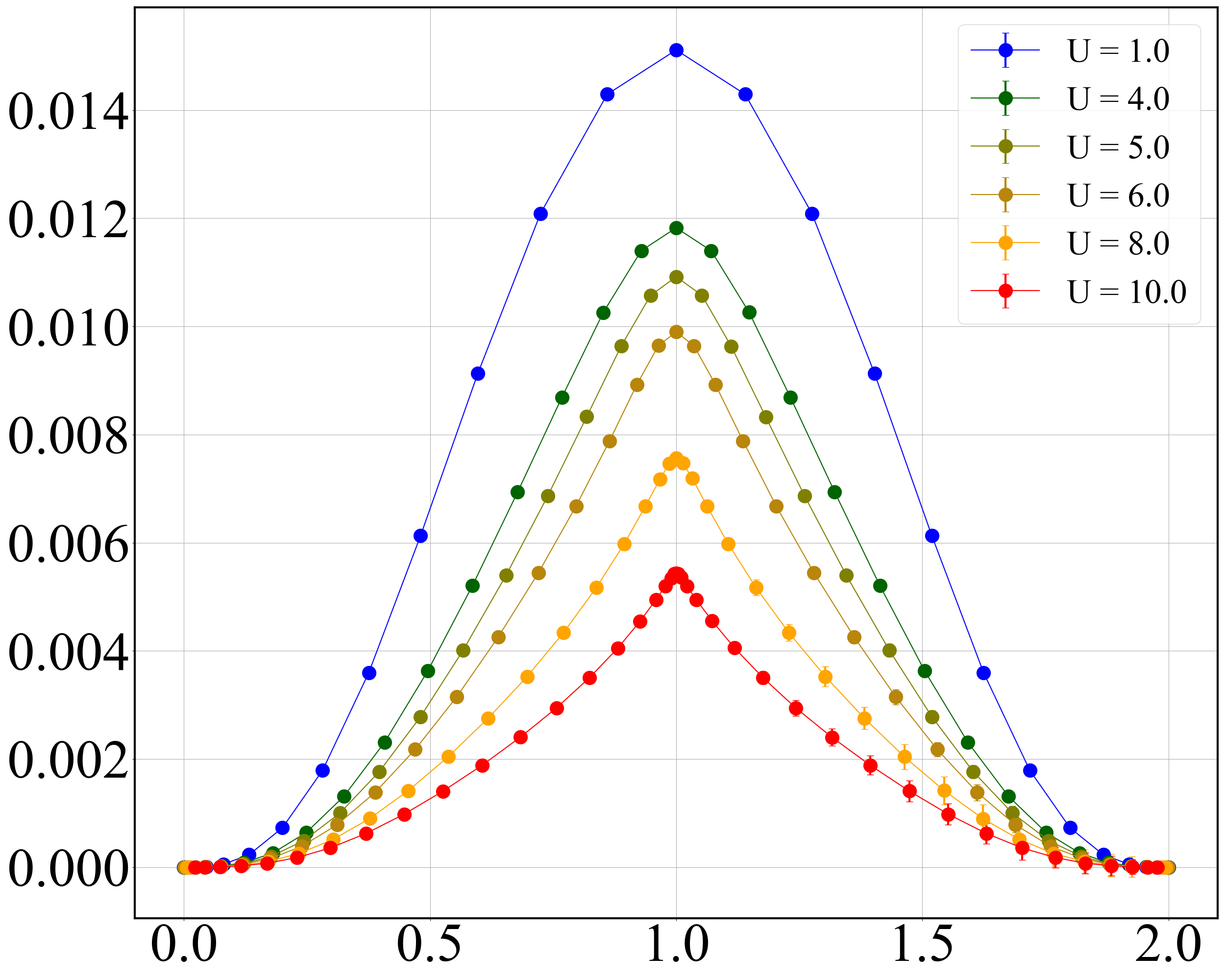}};
%\node[below=of img1,node distance=0cm,yshift=-1.7cm,xshift=0.0cm]{\small{(a)}};
%\node[left=of img1,node distance=0cm,rotate=90,anchor=center,yshift=-0.8cm,xshift=0.0cm]{\small{ $\sum_{r}C_{nn}(r)$}};
\node[above=of img1,node distance=0cm,yshift=-1.2
cm,xshift=0.0cm]{\normalsize{(a) $C_{hd}(1)$}};
\node[below=of img1,node distance=0cm,yshift=1.1cm,xshift=0.1cm]{\large{$n$}};
%%%%%%%%%%%%%%%%%%%%%%%%%%%%%%%%%%%%%%%%%
\node (img2) [right=of img1,xshift=-0.85cm]{\includegraphics[width=4.0cm,height=3.2cm]{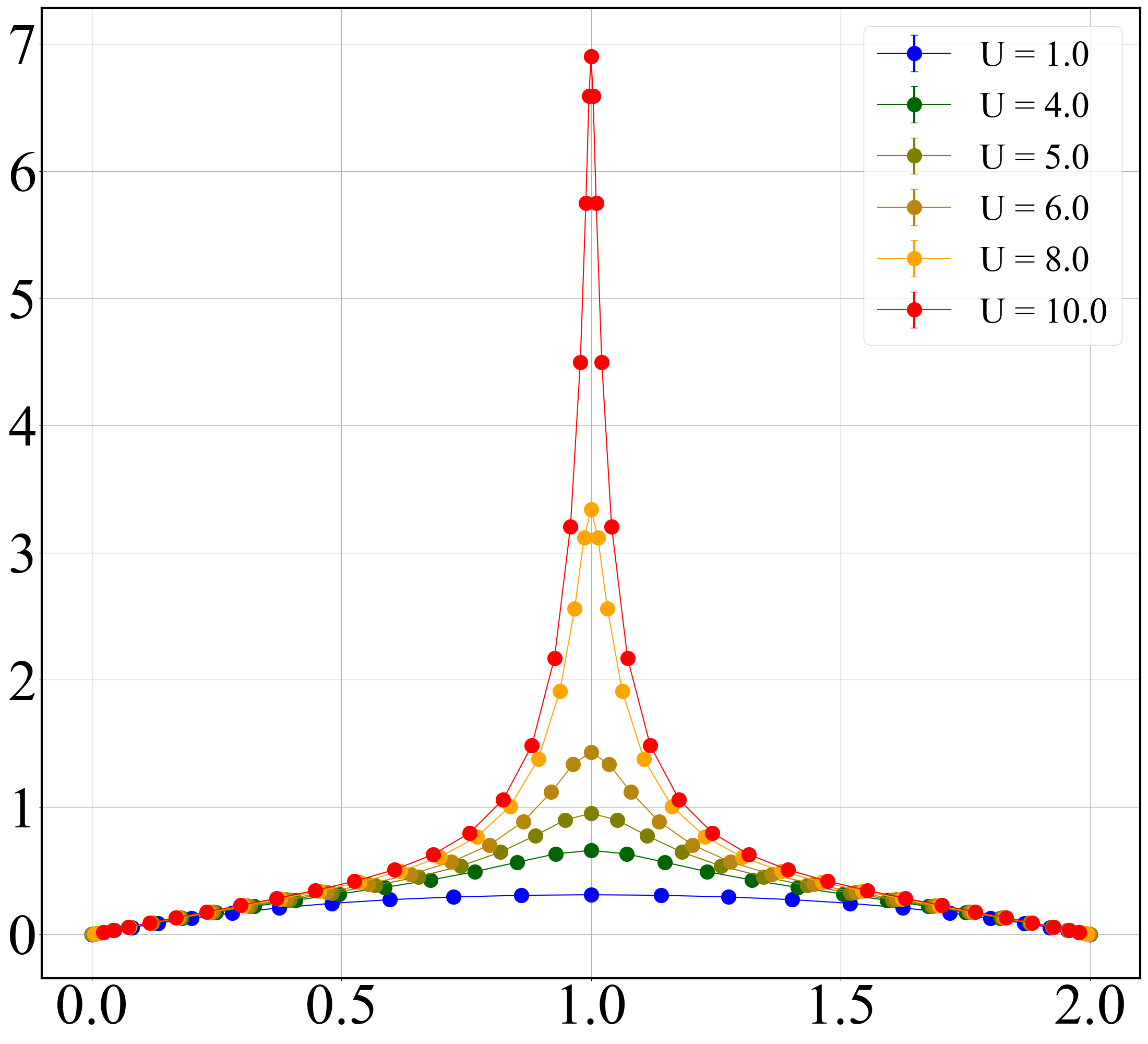}};
%\node[above=of img3,node distance=0cm,yshift=-1.7cm,xshift=0.0cm]{\large{(c) $|i-j|=1$, $n = 1$}};
\node[above=of img2,node distance=0cm,yshift=-1.2
cm,xshift=0.0cm]{\normalsize{(b) $\tilde{C}_{hd}(1)$}};
\node[below=of img2,node distance=0cm,yshift=1.1cm,xshift=0.1cm]{\large{$n$}};
%%%%%%%%%%%%%%%%%%%%%%%%%%%%%%%%%%%%%%%%
\node (img3) [right=of img2,xshift=-1.05cm]{\includegraphics[width=4.4cm,height=3.2cm]{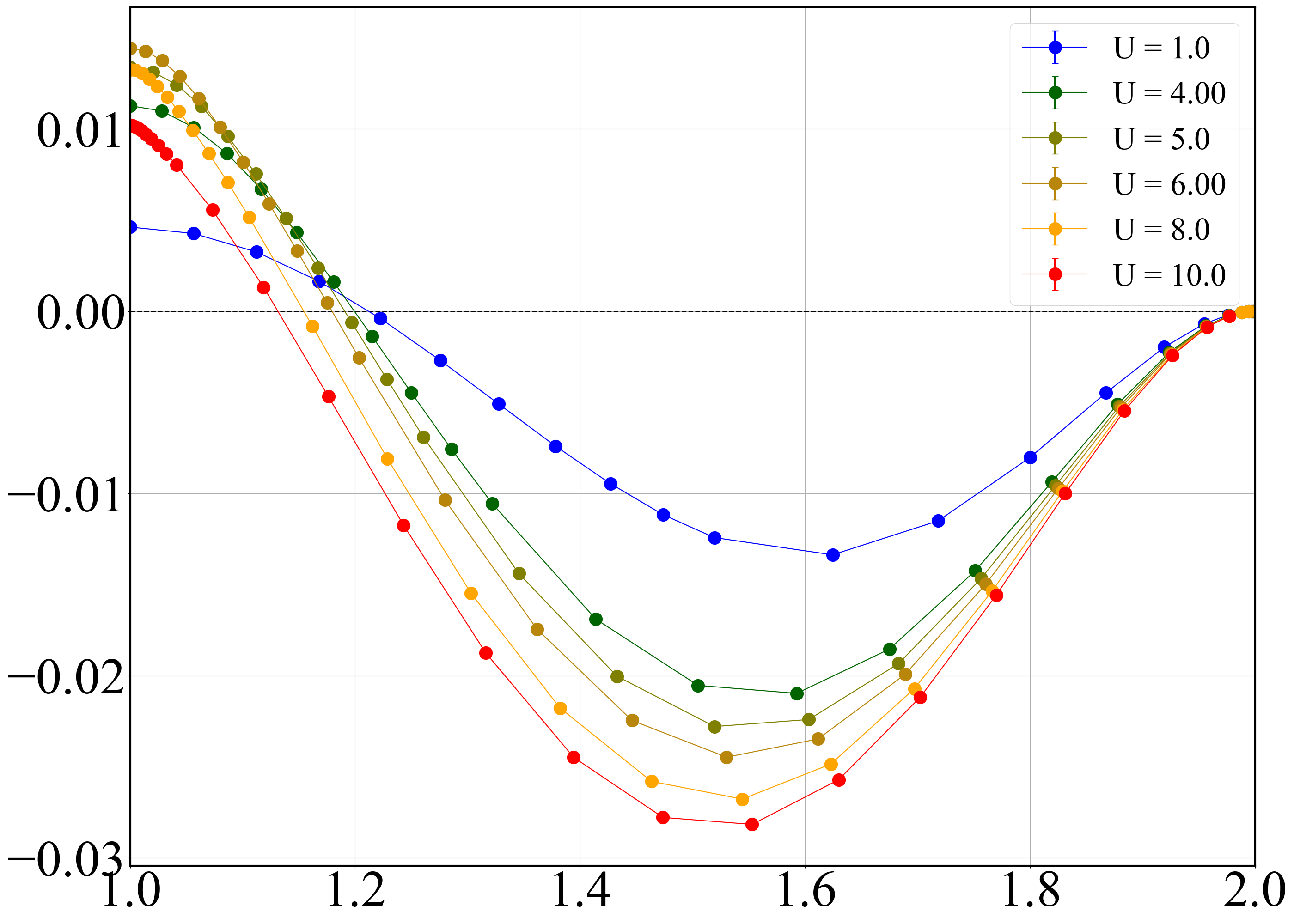}};
%\node[left=of img2,node distance=0cm,rotate=90,anchor=center,yshift=-0.9cm,xshift=0.0cm]{\scriptsize{ $\sum_{\langle ij \rangle}\langle h_id_j\rangle$}};
\node[above=of img3,node distance=0cm,yshift=-1.2
cm,xshift=0.0cm]{\normalsize{(c) $C_{mm}(1)$}};
%\node[left=of img2,node distance=0cm,yshift=1.0cm,xshift=4.4cm]{\scriptsize{$y \sim U^{-2}$}};
\node[below=of img3,node distance=0cm,yshift=1.1cm,xshift=0.1cm]{\large{$n$}};
%%%%%%%%%%%%%%%%%%%%%%%%%%%%%%%%%%%%%%%%%%%
\node (img4) [right=of img3,xshift=-1.25cm]{\includegraphics[width=4.4cm,height=3.2cm]{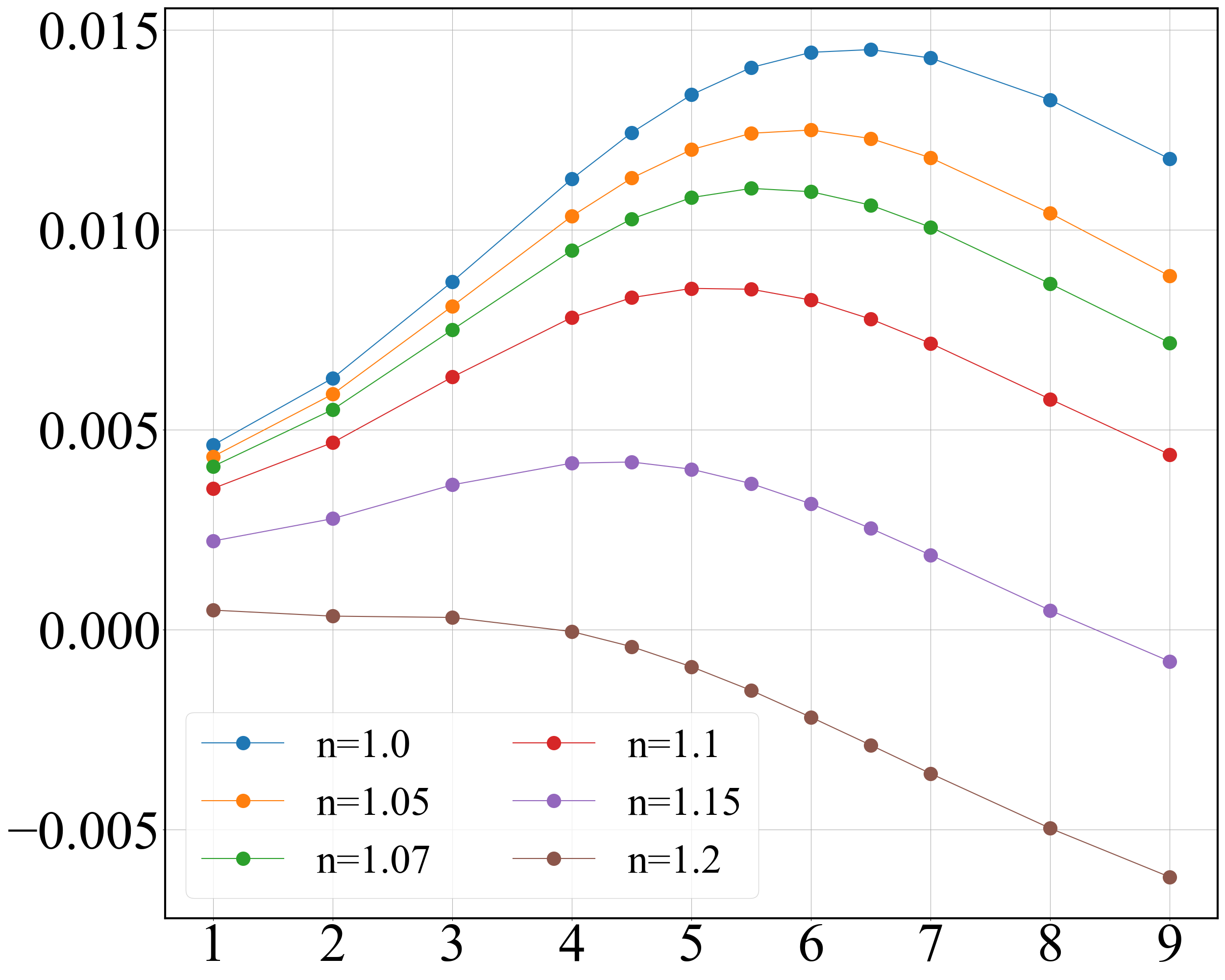}};
%\node[above=of img3,node distance=0cm,yshift=-1.7cm,xshift=0.0cm]{\large{(c) $|i-j|=1$, $n = 1$}};
%\node[left=of img3,node distance=0cm,rotate=90,anchor=center,yshift=-0.9cm,xshift=0.0cm]{\scriptsize{ $C_{hd}(1)$}};
\node[above=of img4,node distance=0cm,yshift=-1.2
cm,xshift=0.0cm]{\normalsize{(d) $C_{mm}(1)$}};
\node[below=of img4,node distance=0cm,yshift=1.1cm,xshift=0.1cm]{\large{$U$}};
%%%%%%%%%%%%%%%%%%%%%%%%%%%%%%%%

%%%%%%%%%%%%%%%%%%%%%%%%
%\node (img5) [right=of img3,xshift=-0.35cm,yshift = 0.8cm]{\includegraphics[width=1.4cm,height=1.1cm]{Graphs_new/U_comp_Normalized_Moment_moment_correlation_connected_1st_nearest_neighbor_averaged_normal_averaged_half_filling_N_10_dtau_0.05.png}};
%\node[left=of img5,node distance=0cm,rotate=90,anchor=center,yshift=-1.0cm,xshift=0.0cm]{\tiny{ $\tilde{C}_{mm}(1)$}};
%\node[below=of img5,node distance=0cm,yshift=1.2cm,xshift=0.0cm]{\tiny{$U$}};
%%%%%%%%%%%%%%%%%%%%%%%%%%%%%%%%
\end{tikzpicture}
\caption{Holon-doublon and moment-moment correlation function at $T=0.67$. \textbf{(a)} Holon-doublon correlations $C_{hd}(1)$ are always maximal at half-filling due to a maximum number of doublon and holon pairs being present. $C_{hd}(1)$ decreases uniformly with $U$ due to onsite repulsion cost to form a doublon. \textbf{(b)} Normalized doublon-holon correlation $\tilde{C}_{hd}(1) = C_{hd}(1)/\langle h_i\rangle \langle d_j\rangle$, which captures the ``bunching" strength of doublon and holon pairs. The binding strength in the weak coupling regime is small; crossing over to the strongly interacting regime, doublon-holon near-neighbor correlations increase half-filling, as expected in a Mott Insulator. \textbf{(c)} $C_{nn}(1)$ and $C_{hd}(1)$ compete to generate moment correlations between neighboring sites $C_{mm}(1)$. Singlet formation is preferred near half-filling; away from half-filling, density repulsion dominates leading to ``anti-bunching" of singlet pairs.
\textbf{(d)} Moment correlations at fixed doping. Close to half-filling, $C_{mm}(1)$ is positive and increases with $U$ until a gap in LDOS forms at half-filling. Beyond this, quantum fluctuations due to tunneling %$t$ of the ground state with singly occupied sites 
lead to a reduction of $C_{mm}(1)$, although $m_i^2$ grows with $U$.}  
\label{moment_correlations}
\end{figure*}

%\textcolor{blue}{Stoped here}

\subsection{Holon-doublon and moment-moment correlations away from half filling}

We now turn to short-range correlation functions that are associated with the crossover across $U_{cr}^{\rm LDOS}(T)$. From the local constraint $m^2_{i}+h_{i}+d_{i}=1$, we can write the contributions to the moment-moment correlations as

\begin{align}
    C_{mm}(r) = 4C_{hd}(r)+C_{nn}(r)
    \label{moment_correlator}
\end{align}

A positive $C_{mm}(1)$ implies bunching of local moments, while a negative $C_{mm}(1)$ means repulsion between neighboring ones. In Ref \cite{cheuk2016observation}, where the holon density is defined as $h_{i}=\sum_{\sigma}(1-n_{i,\sigma})$, $C_{mm}(1)$ was shown to originate from a competition between holon-doublon attraction and holon-holon/doublon-doublon repulsion.   However, with $h_i = \prod_{\sigma}(1-n_{i\sigma})$,  the above equation shows that singlet pairs between neighboring sites form when the fluctuation of doublon-holon pairs into singlet pairs overcomes the density repulsion between the singlet pairs. This can be seen by noting that $C_{hd}(1)$ is always positive, and $C_{nn}(1)$ is always negative, so singlet pair formation is favored when $4|C_{hd}(1)|>|C_{nn}(1)|$.

Holon-doublon correlations are maximal at half-filling (Fig \ref{moment_correlations}(a)), since bunching of doublon and holons into pairs minimizes repulsion between isolated doublon-doublon or holon-holon pairs (refer to Appendix \ref{app_d}). The doublon-holon correlations monotonically decrease with $U$ at all densities, due to energy cost from onsite repulsion $U$ to form a double occupancy. It is interesting to note, however,  that the normalized holon-doublon correlation function $\tilde{C}_{hd}(1) = C_{hd}(1)/\langle h_i\rangle \langle d_j\rangle$ sharply increases with $U$ as shown in Fig \ref{moment_correlations}(d).

The binding strength between holon-doublon pairs also indicates the nature of the system. On crossover to the strong coupling limit ($U>U_{cr}^{\rm LDOS}$), the system is a Mott insulator with singly occupied sites; a finite $t$ can generate strongly bound holon-doublon pairs by virtual hopping over short distances \cite{hartke2020doublon}. With increasing $U$, delocalization of the doublon becomes more energy expensive; hence the holon-doublon pairs become more strongly bound, causing $\tilde{C}_{hd}(1)$ to grow with $U$ near half filling. 

The interplay between attractive holon-doublon pairs and repulsive density-density pairs generates the nearest neighbor moment-moment correlations $C_{mm}(1)$(Fig \ref{moment_correlations}(b)), which has the following features:

(i) In a finite doping window around half filling, $C_{mm}(1)$ is positive, and the system tends to ``bunch up" local moments \cite{cheuk2016observation}; we dub this the ``optimal window" for local moment near-neighbor pair formation. In this regime, holon-doublon attraction dominates over density-density repulsion. Away from half filling, density-density repulsion dominates leading to ``anti-bunching" of moments; $C_{mm}(1)$ closely follows $C_{nn}(1)$. The optimal window shrinks monotonically with $U$ at a fixed temperature; this is a consequence of holon-doublon correlations decreasing with $U$ due to strong onsite repulsion. Shrinking of the optimal doping window with $U$ also implies that at fixed doping, the system may bunch up moments at small $U$, but repel them at large $U$ (shown for $n=1.15$ in Fig \ref{moment_correlations}(d)).  The short range nature of the doublon holon correlations also imply that the system bunches up local moments only on short length scales
comparable to lattice spacing near half filling (refer to Appendix \ref{app_e}).

(ii) Inside the ``optimal" doping window, $C_{mm}(1)$ exhibits a non monotonic behavior with respect to $U$, shown in Fig \ref{moment_correlations}(c). Close to half-filling, moment correlations increase with $U$, but decrease beyond the metal to insulator crossover across $U_{cr}^{\rm LDOS}(T)$. This is not intuitive, as the local moment $m_i^2$ increases with $U$. The decrease of $C_{mm}(1)$ at large $U$ can be best understood from %the nature of the ground state at half filling beyond $U_{cr}^{\rm LDOS}(T)$. In a Mott insulator, the ground state is singly occupied. 
the virtual hopping of electrons that generate the super-exchange scale $J \sim t^2/U$ in the strong coupling limit and reduces the degree of localization and hence the moment-moment correlations. This is similar to the mechanism generating the low-temperature maximum of $m^2$ seen in Ref \cite{paiva2001signatures}. Doping the system away from half-filling further reduces moment-moment correlations, due to the formation of isolated doublons that interfere with moment pair formation (Appendix \ref{app_d})

\subsection{Comparison with experiment}
\label{sec_experimental_compare}

We conclude this section by comparing our numerical results with experimental findings from a cold atomic Fermi gas, subjected to Hubbard like interactions in an optical trap. Correlation functions and thermodynamics are observed by a bilayer quantum gas microscope capable of site resolved density readout of the gas in 2D in a single fluorescence image \cite{hartke2020doublon}. Imaging a large number of sites gives access to equation of state, and local and non-local correlation functions such as density, doublon, holon and singlon pairs. The temperature of the gas can be determined from the fluctuation-dissipation theorem, by simultaneously measuring the compressibility and total density fluctuations. Although the experiment measurements are performed at $U/t \approx 7.1, 11.8,18.3,25.5$, we  only show the correlators at $U/t = 11.8$. Experimental fluctuation thermometry of the gas yields $T/t = 0.69$ for $U/t = 11.8$ \cite{hartke2020doublon}. % 

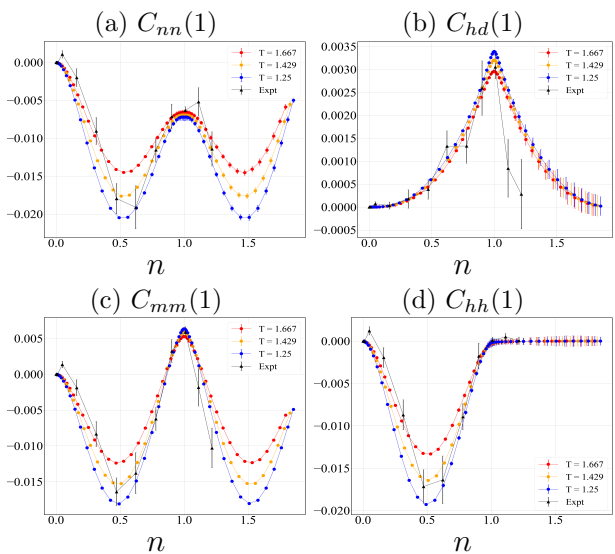
\begin{figure}[ht]
\begin{tikzpicture}

\node (img1) {\includegraphics[width=4cm,height=2.8cm]{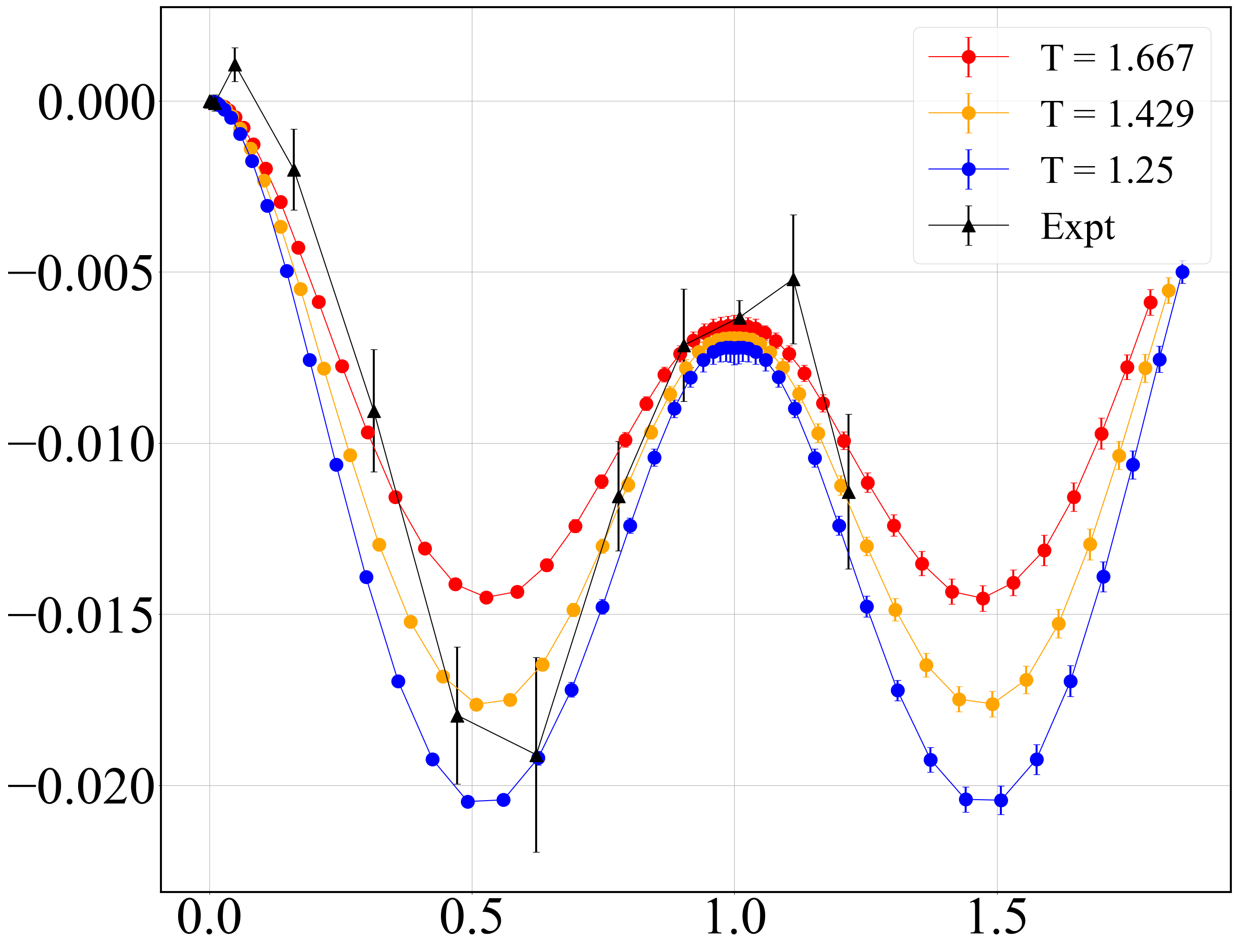}};
\node[above=of img1,node distance=0cm,yshift=-1.20cm,xshift=0.0cm]{\normalsize{(a) $C_{nn}(1)$}};
\node[below=of img1,node distance=0cm,yshift=1.1cm,xshift=0.0cm]{\large{$n$}};
%%%%%%%%%%%%%%%%%%%%%%%%%%%%%%%%
\node (img2) [right=of img1,xshift=-1.15cm]{\includegraphics[width=4cm,height=2.8cm]{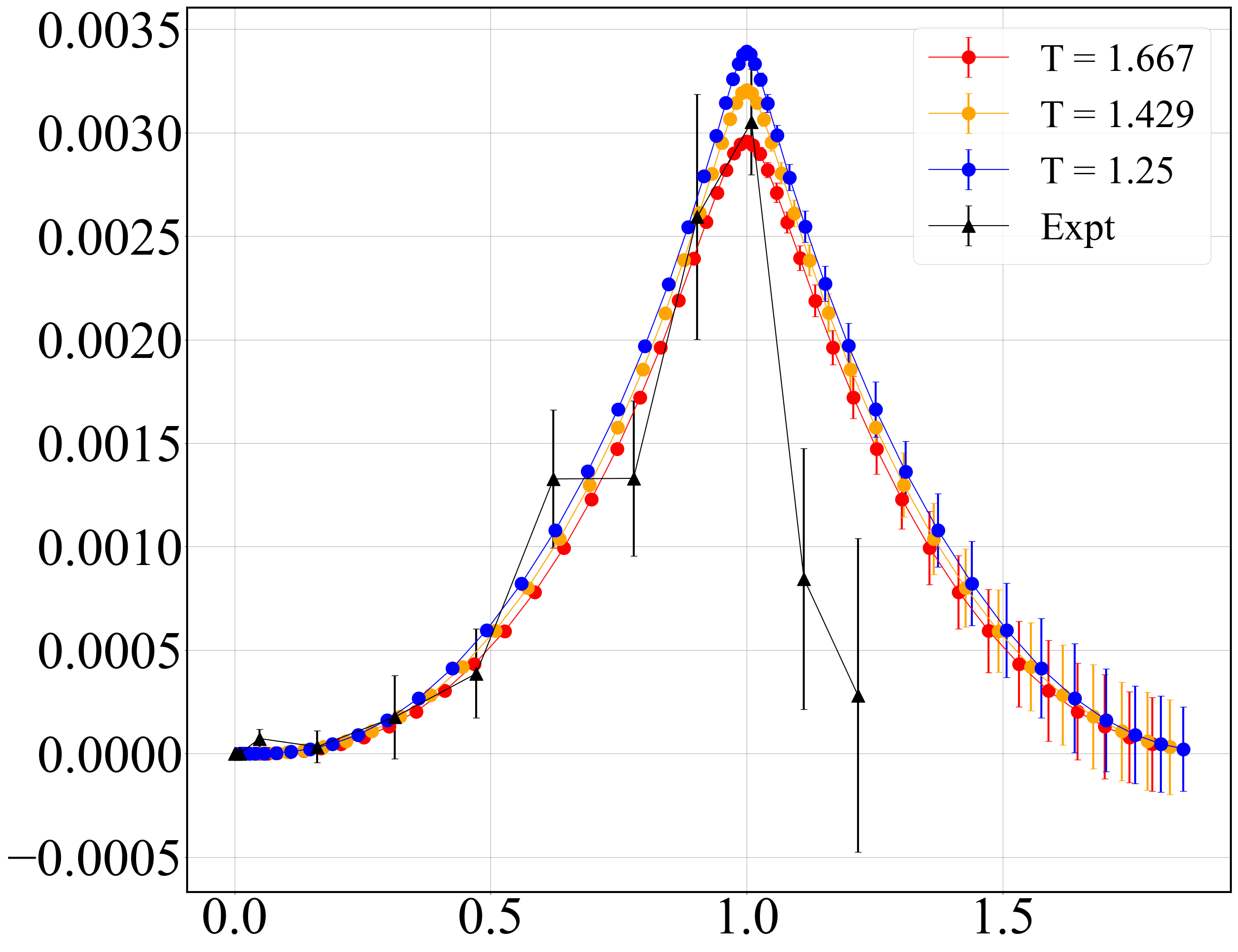}};
\node[above=of img2,node distance=0cm,yshift=-1.20cm,xshift=0.0cm]{\normalsize{(b) $C_{hd}(1)$}};
\node[below=of img2,node distance=0cm,yshift=1.1cm,xshift=0.0cm]{\large{$n$}};
%%%%%%%%%%%%%%%%%%%%%%%%%%%%%%%%%%%%%%%%%%%%%
\node (img3) [below=of img1,yshift=0.35cm]{\includegraphics[width=4cm,height=2.8cm]{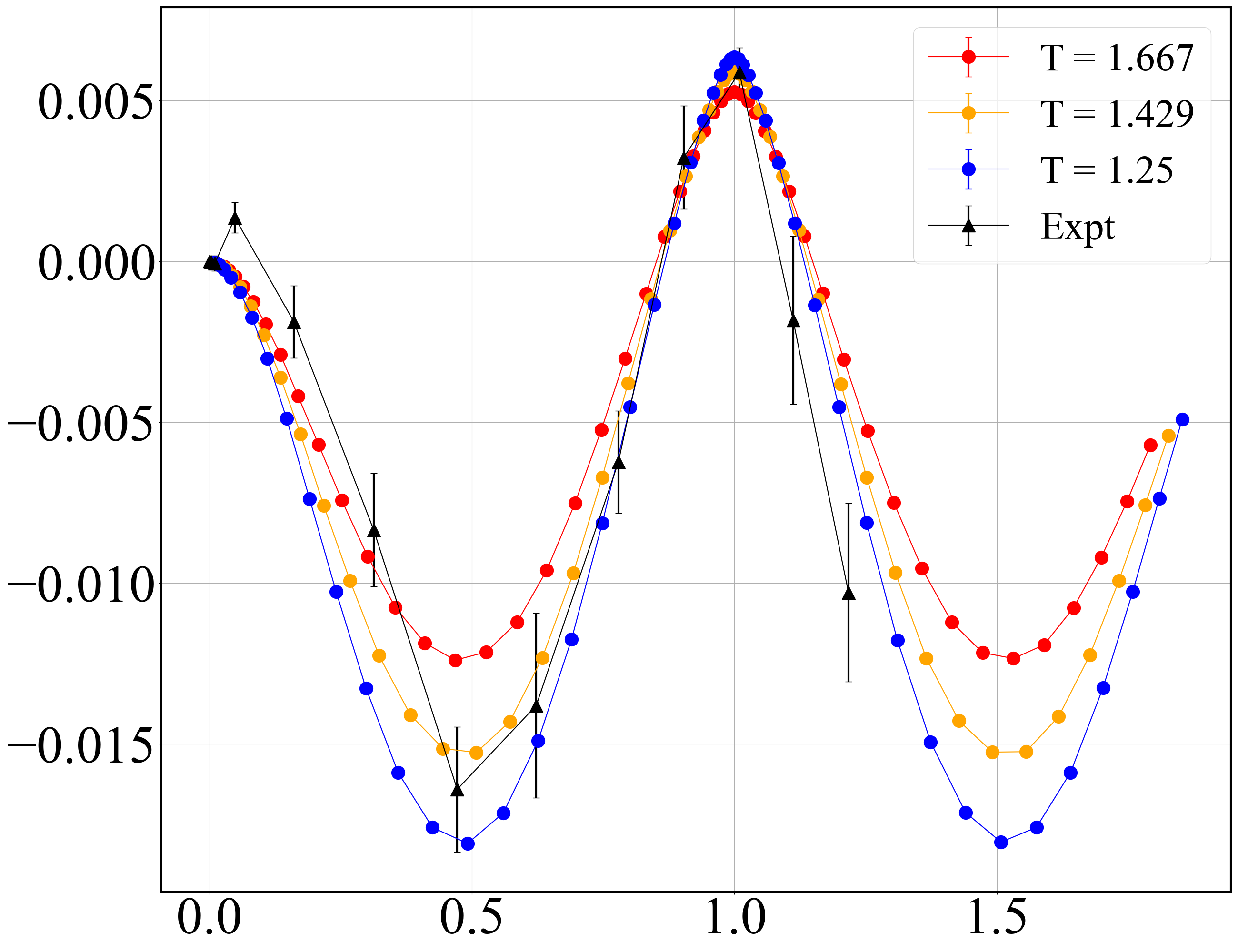}};
\node[above=of img3,node distance=0cm,yshift=-1.20cm,xshift=0.0cm]{\normalsize{(c) $C_{mm}(1)$}};
\node[below=of img3,node distance=0cm,yshift=1.1cm,xshift=0.0cm]{\large{$n$}};
%%%%%%%%%%%%%%%%%%%%%%%%%%%%%%%%%%%%%%%%%%%%%
\node (img4) [right=of img3,xshift = -1.15cm] {\includegraphics[width=4cm,height=2.8cm]{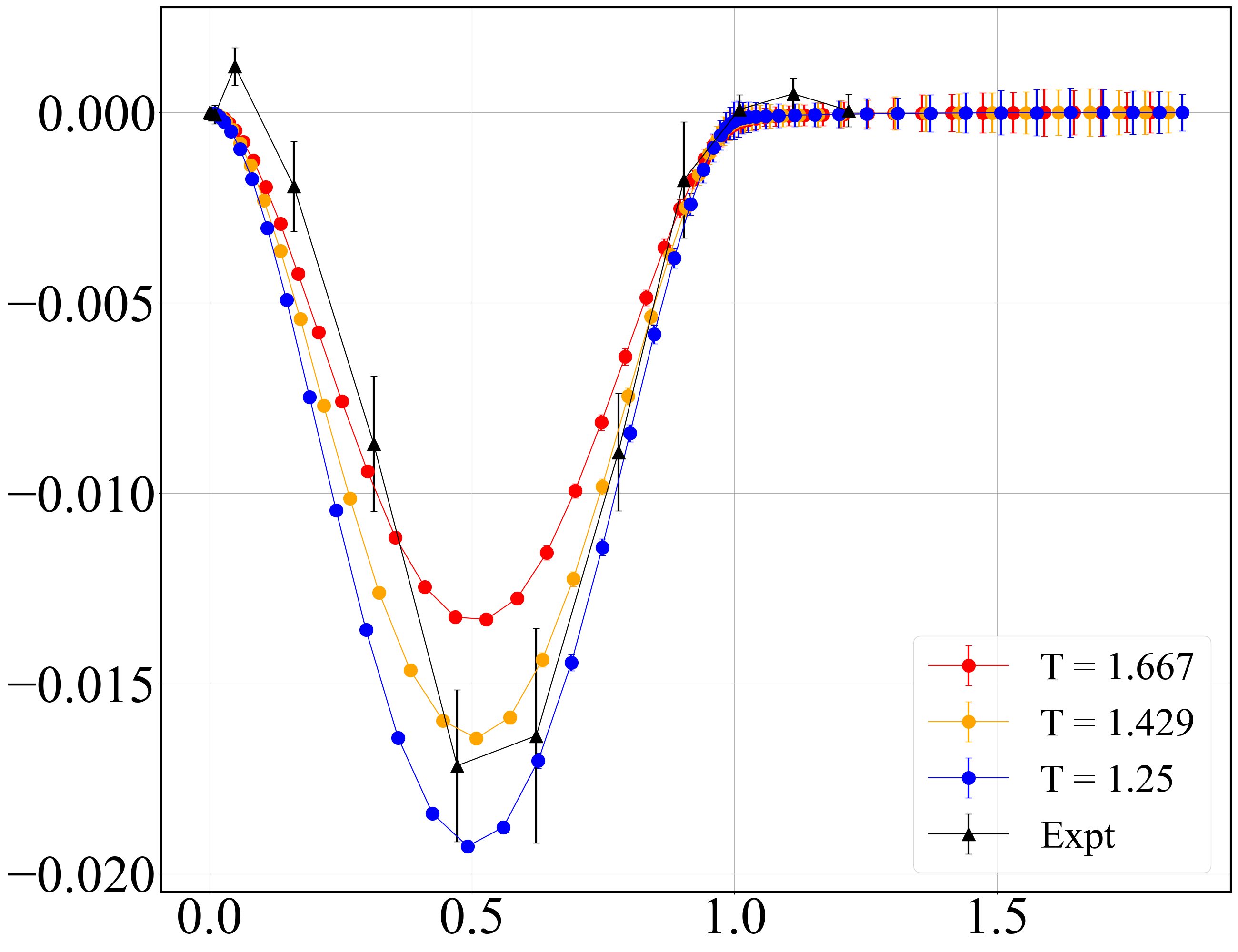}};
\node[above=of img4,node distance=0cm,yshift=-1.20cm,xshift=0.0cm]{\normalsize{(d) $C_{hh}(1)$}};
\node[below=of img4,node distance=0cm,yshift=1.1cm,xshift=0.0cm]{\large{$n$}};
%%%%%%%%%%%%%%%%%%%%%%%%%%%%
\end{tikzpicture}
\captionof{figure}{Comparison of experimental nearest neighbor correlation functions \cite{hartke2020doublon} with QMC simulations at  $U = 11.8$ (a) Density density correlation function. (b) Holon doublon correlation function. (c) Moment moment correlation function. (d) Holon-holon correlation function. The temperature estimate of the experimental data is $T \approx 1.43$, by comparing with DQMC data, %which is lower than the temperature considered ($T = 1.6$) in Ref \cite{hartke2020doublon} for comparing equation of state data with Ref \cite{varney2009quantum}, and 
higher than the ones obtained by fluctuation thermometry, $T \sim 0.69$.} 
\label{Experimental_compare}
\end{figure}

For $U/t = 11.8$, we compare the experimental data with numerical correlation functions in the temperature range $T/t = 1.25,1.43,1.67$, as  shown in Fig \ref{Experimental_compare} (a)-(d). Out of the three temperatures, the experimental data seem to agree best with the numerical results at $T/t = 1.43$.

\section{Conclusion}
\label{sec_conclusion}

In this work, we have calculated equal time correlation functions for the repulsive Fermi Hubbard model and studied their signatures across weak to strong coupling. Our analysis identifies the following key observation:
\\

(a) There is an extended crossover from a metallic to insulating state, at intermediate to high temperatures above the spin ordering temperature $T_{spin}$. The crossover occurs by successive gap openings in the thermodynamic density of states (TDOS) and local density of states (LDOS), that occur at different interaction strengths. 
\\

(b) Equal time density-density, holon-doublon and moment-moment correlation functions track the metal-to-insulator crossover, both at and away from half-filling. While density-density correlations track TDOS gap opening, holon-doublon and moment-moment correlations track onset of a pseudogap in LDOS. 
\\

(c) Moment moment correlation functions arise due to competition between doublon holon attraction and density density repulsion. In the Mott Insulating phase, moment correlations are weakened by virtual hopping of electrons generating the superexchange scale, although individual local moments grow in size.
\\

Our calculations have focused exclusively on the single band Hubbard model. The fate of the metal to insulator crossover in presence of multiple bands (in particular, flatbands) remains an interesting area to be explored \cite{roylieb}. With the experimental realization of the Lieb lattice \cite{slot2017experimental,mukherjee2015observation} (which hosts a flat band and two dispersive bands), experimental study of a metal to insulator crossover in conjunction with static correlations, as outlined in this work, can be performed. It will be interesting to test the universality of the results shown here for such models, and see how they get modified with nontrivial band topology, as in Kagome based systems.

Comparison of our results with recent experimental findings \cite{hartke2020doublon} establishes the delicate interplay of microscopic degrees of freedom in generating nontrivial correlations. Our calculations provide a systematic way of isolating various phases in the extended metal to insulator crossover by measurement of suitable correlation functions in cold atom experiments. Finally, we also provide testable predictions that can be used to benchmark quantum simulations of the repulsive Hubbard model in experimental setups.

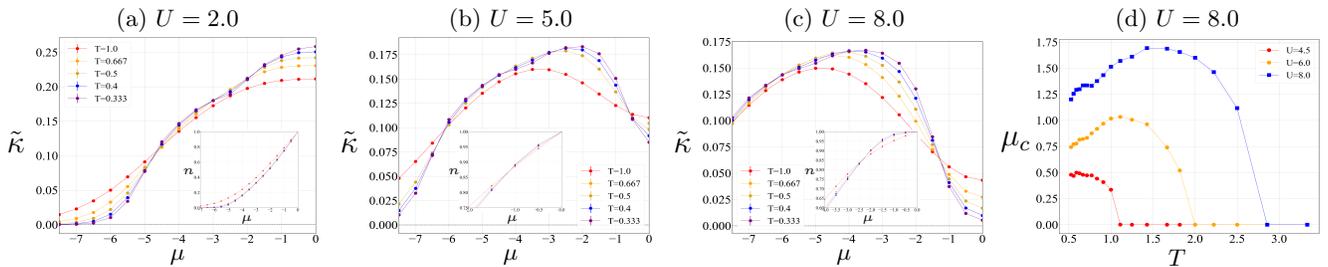
\begin{figure*}[t]
\begin{tikzpicture}
%%%%%%%%%%%%%%%%%%%%%%%%%%%%%%%%%%%%%%%
%%%%%%%%%%%%%%%%%%%%%%%%%%%%%%%%%%%%%%%
\node (img1)
{\includegraphics[width=3.8cm,height=2.8cm]{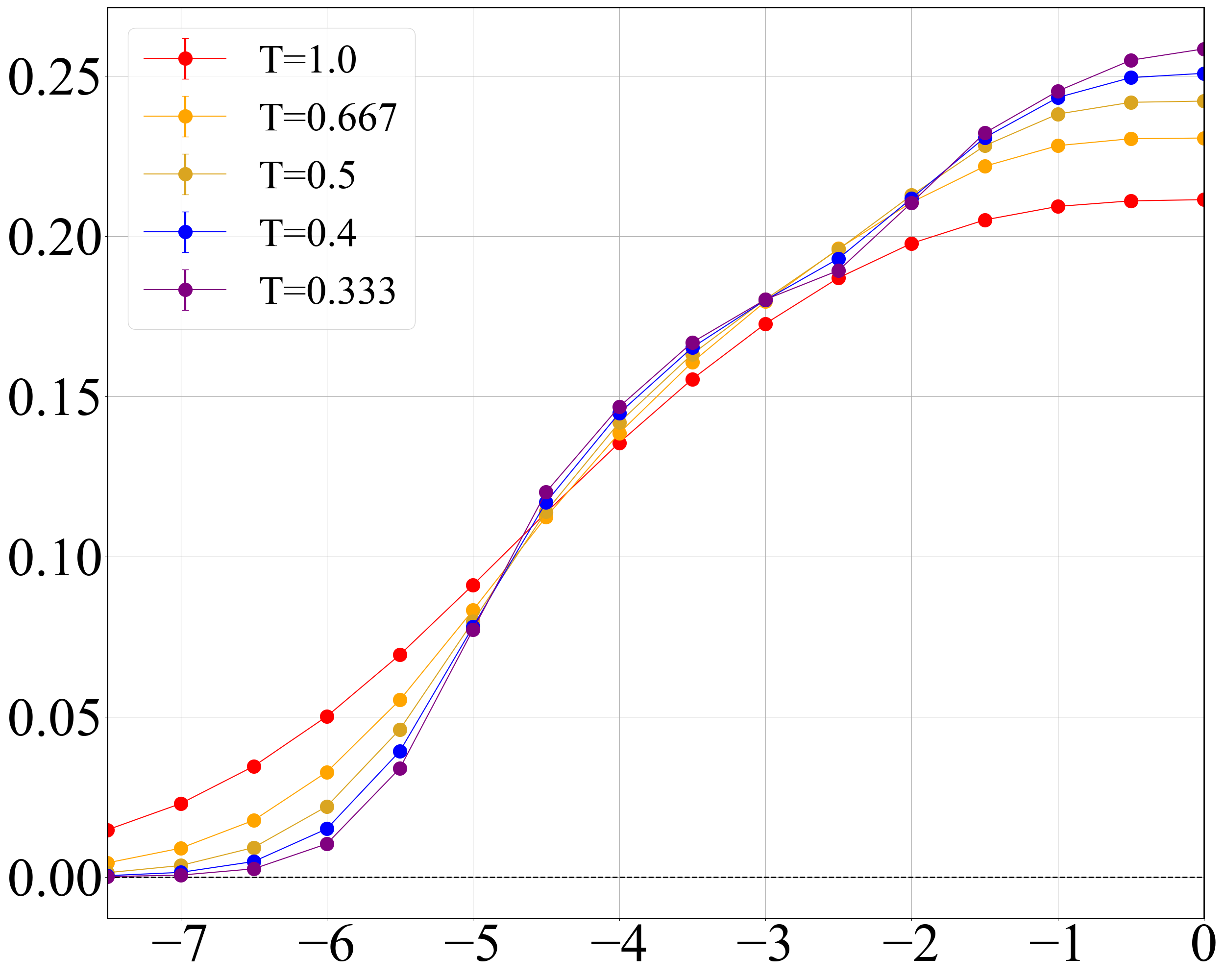}};
%\node[left=of img3,node distance=0cm,yshift=2.0cm,xshift=1.6cm]{\large{(b)}};
\node[left=of img1,node distance=0cm,rotate=0,anchor=center,yshift=0.0cm,xshift=0.8cm]{\large{ $\tilde{\kappa}$}};
\node[below=of img1,node distance=0cm,yshift=1.2cm,xshift=0.0cm]{\small{$\mu$}};
%\node[left=of img2,node distance=0cm,rotate=0,anchor=center,yshift=0.0cm,xshift=3.8cm]{\small{ $U=2.0$}};
\node[above=of img1,node distance=0cm,rotate=0,anchor=center,yshift=-0.9cm,xshift=0.0cm]{\small{(a) $U=2.0$}};
%\node[above=of img3,node distance=0cm,yshift=-1.3cm,xshift=0.0cm]{\large{(a)}\Large{ $\frac{\partial \tilde{\kappa}}{\partial T}$}\large{, $T = 0.67$}};
%%%%%%%%%%%%%%%%%%%%%%%%%%%%%%%%%%%%%%%
\node (img2)[right=of img1, xshift = -0.6cm]
{\includegraphics[width=3.8cm,height=2.8cm]{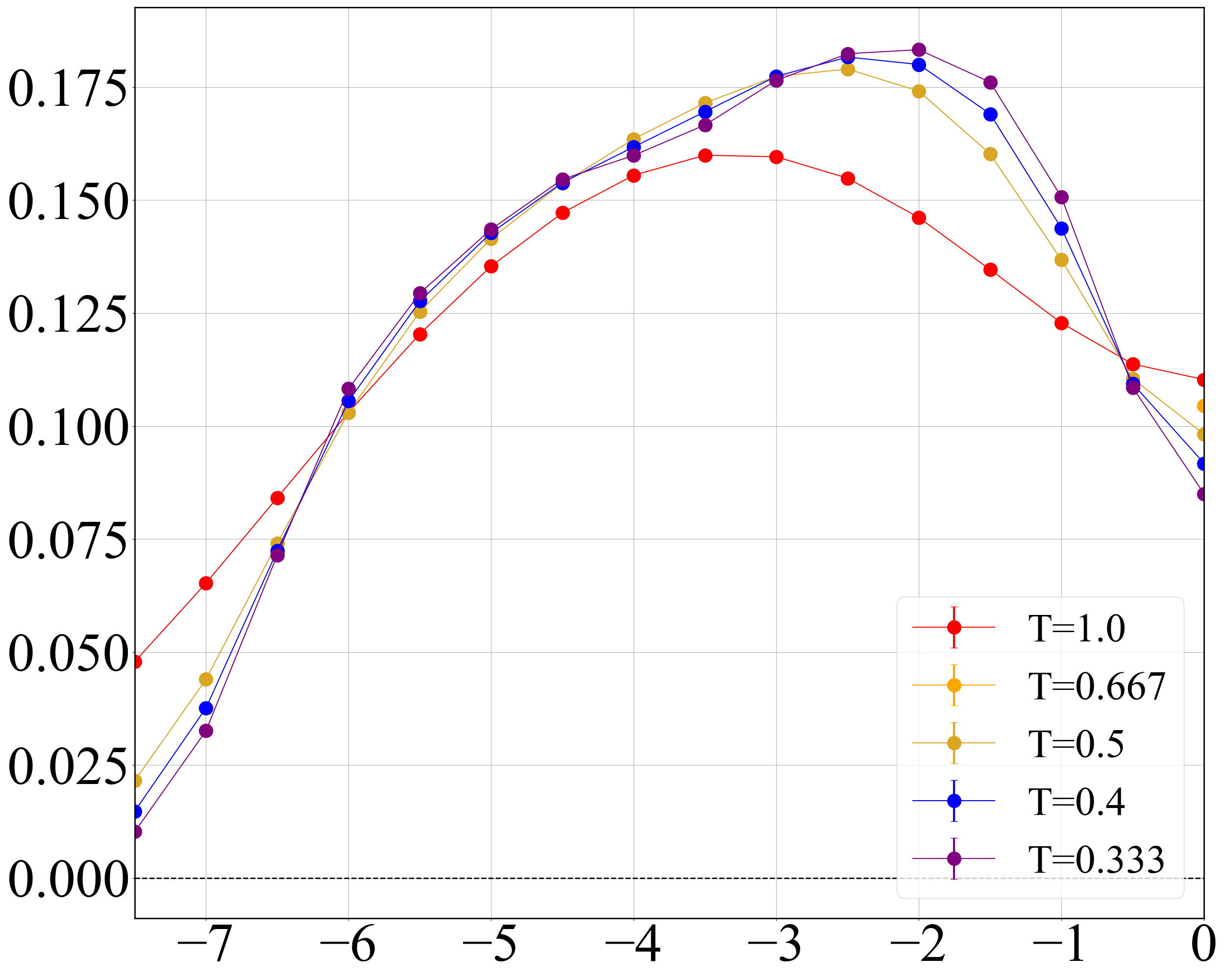}};
%\node[left=of img3,node distance=0cm,yshift=2.0cm,xshift=1.6cm]{\large{(b)}};
\node[left=of img2,node distance=0cm,rotate=0,anchor=center,yshift=0.0cm,xshift=0.8cm]{\large{ $\tilde{\kappa}$}};
\node[below=of img2,node distance=0cm,yshift=1.2cm,xshift=0.0cm]{\small{$\mu$}};
%\node[left=of img3,node distance=0cm,rotate=0,anchor=center,yshift=0.0cm,xshift=3.8cm]{\small{ $U=5.0$}};
\node[above=of img2,node distance=0cm,rotate=0,anchor=center,yshift=-0.9cm,xshift=0.0cm]{\small{(b) $U=5.0$}};
%\node[above=of img3,node distance=0cm,yshift=-1.3cm,xshift=0.0cm]{\large{(a)}\Large{ $\frac{\partial \tilde{\kappa}}{\partial T}$}\large{, $T = 0.67$}};
%%%%%%%%%%%%%%%%%%%%%%%%%%%%%%%%%%%%%%%
\node (img3)[right=of img2, xshift = -0.6cm]
{\includegraphics[width=3.8cm,height=2.8cm]{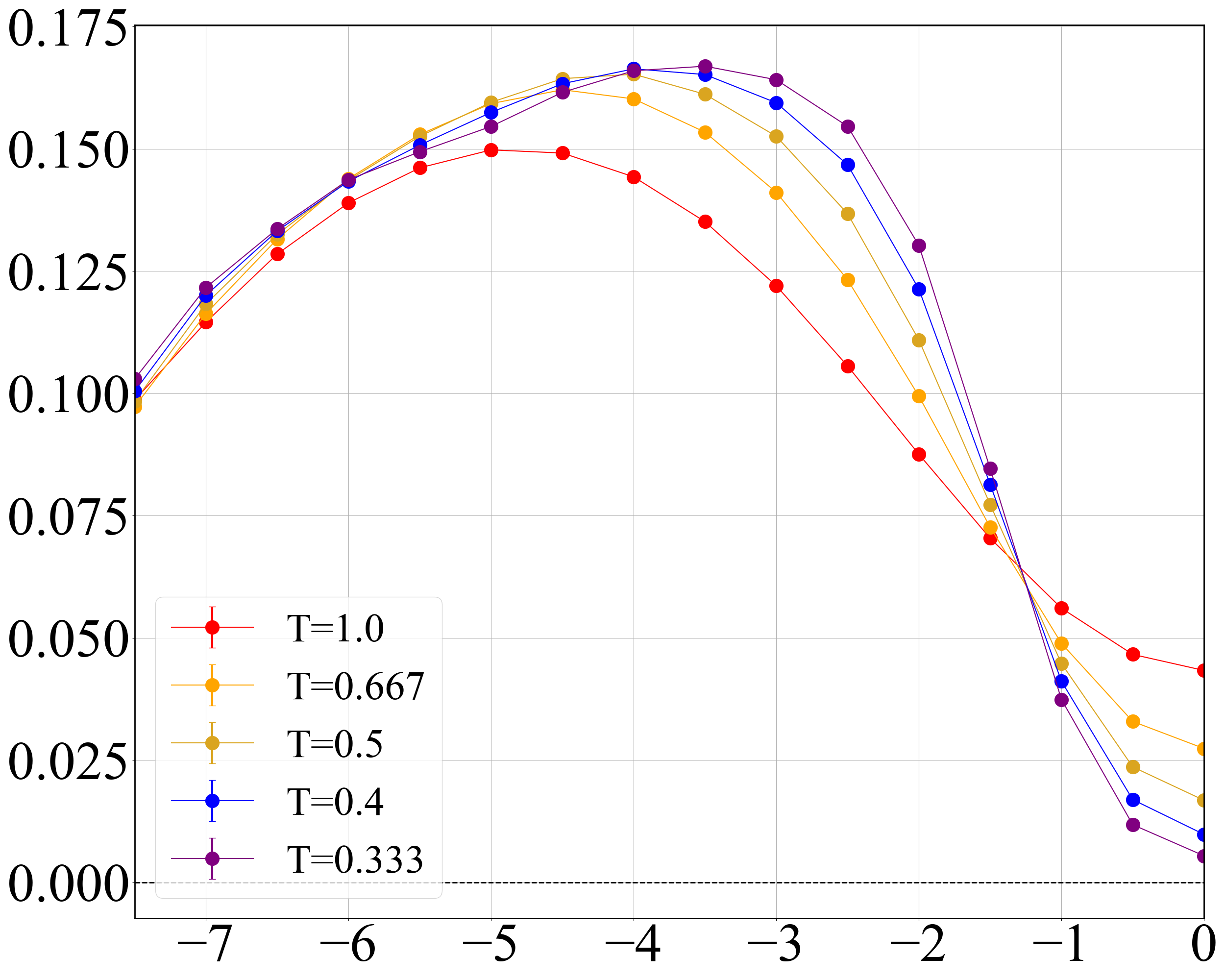}};
%\node[left=of img3,node distance=0cm,yshift=2.0cm,xshift=1.6cm]{\large{(b)}};
\node[left=of img3,node distance=0cm,rotate=0,anchor=center,yshift=0.0cm,xshift=0.8cm]{\large{ $\tilde{\kappa}$}};
\node[below=of img3,node distance=0cm,yshift=1.2cm,xshift=0.0cm]{\small{$\mu$}};
%\node[left=of img4,node distance=0cm,rotate=0,anchor=center,yshift=0.0cm,xshift=3.0cm]{\small{ $U=8.0$}};
\node[above=of img3,node distance=0cm,rotate=0,anchor=center,yshift=-0.9cm,xshift=0.0cm]{\small{(c) $U=8.0$}};
%%%%%%%%%%%%%%%%%%%%%%%%%%%%%%%%%%%%%%%
\node (img5) [left=of img1,xshift=4.75cm,yshift = -0.4cm]{\includegraphics[width=1.4cm,height=1.1cm]{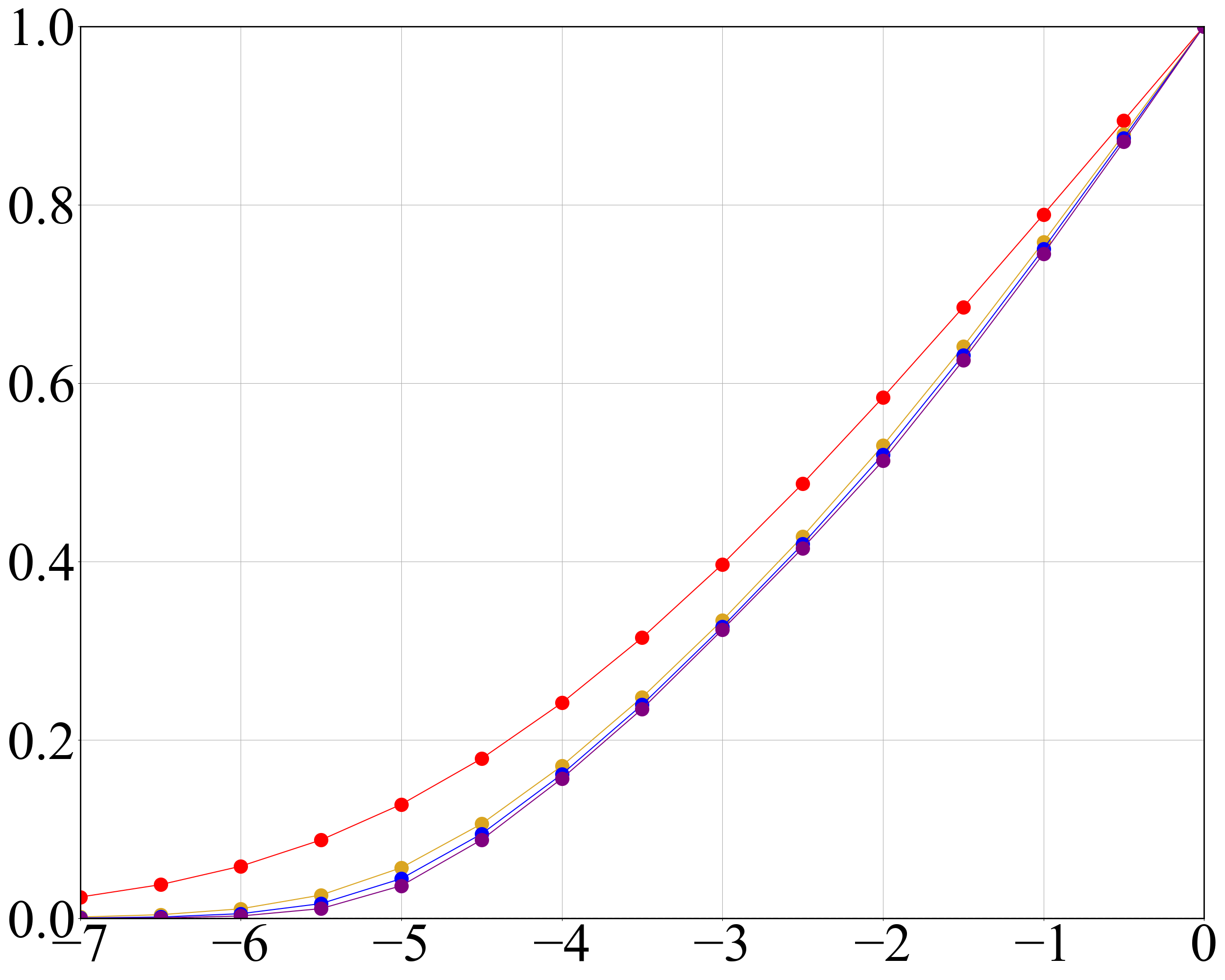}};
\node[left=of img5,node distance=0cm,rotate=0,anchor=center,yshift=0.0cm,xshift=1.0cm]{\tiny{ $n$}};
\node[below=of img5,node distance=0cm,yshift=1.2cm,xshift=0.0cm]{\tiny{$\mu$}};
%%%%%%%%%%%%%%%%%%%%%%%%%%%%%%%%%%%%%%%%%
\node (img6) [left=of img2,xshift=3.85cm,yshift = -0.4cm]{\includegraphics[width=1.4cm,height=1.1cm]{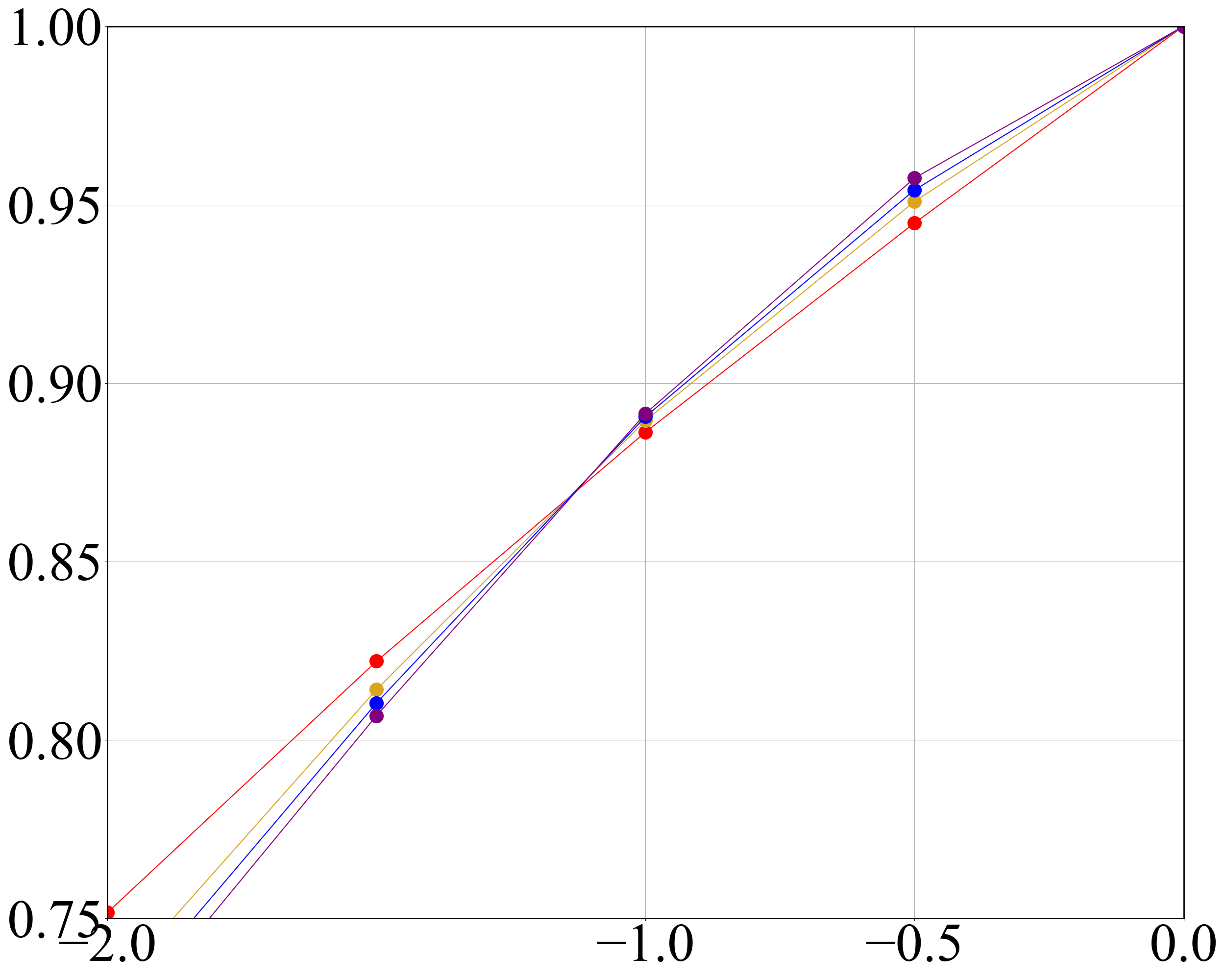}};
\node[left=of img6,node distance=0cm,rotate=0,anchor=center,yshift=0.0cm,xshift=1.0cm]{\tiny{ $n$}};
\node[below=of img6,node distance=0cm,yshift=1.2cm,xshift=0.0cm]{\tiny{$\mu$}};
%%%%%%%%%%%%%%%%%%%%%%%%%%%%%%%%%%%%%%%%%
\node (img7) [left=of img3,xshift=4.15cm,yshift = -0.4cm]{\includegraphics[width=1.4cm,height=1.1cm]{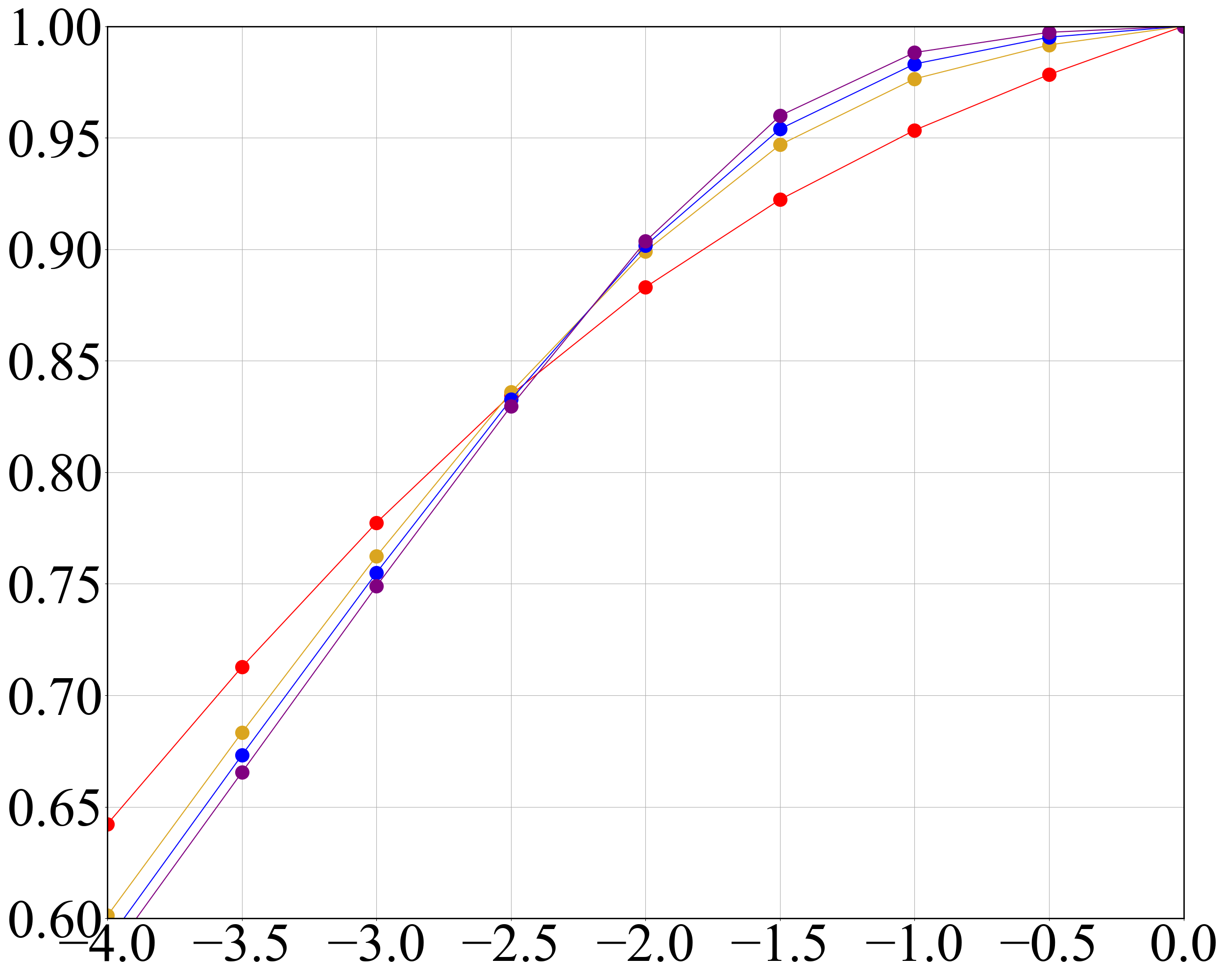}};
\node[left=of img7,node distance=0cm,rotate=0,anchor=center,yshift=0.0cm,xshift=1.0cm]{\tiny{ $n$}};
\node[below=of img7,node distance=0cm,yshift=1.2cm,xshift=0.0cm]{\tiny{$\mu$}};
%%%%%%%%%%%%%%%%%%%%%%%%%%%%%%%%%%%%%%%%%
\node (img9)[right=of img3, xshift = -0.6cm]
{\includegraphics[width=3.8cm,height=2.8cm]{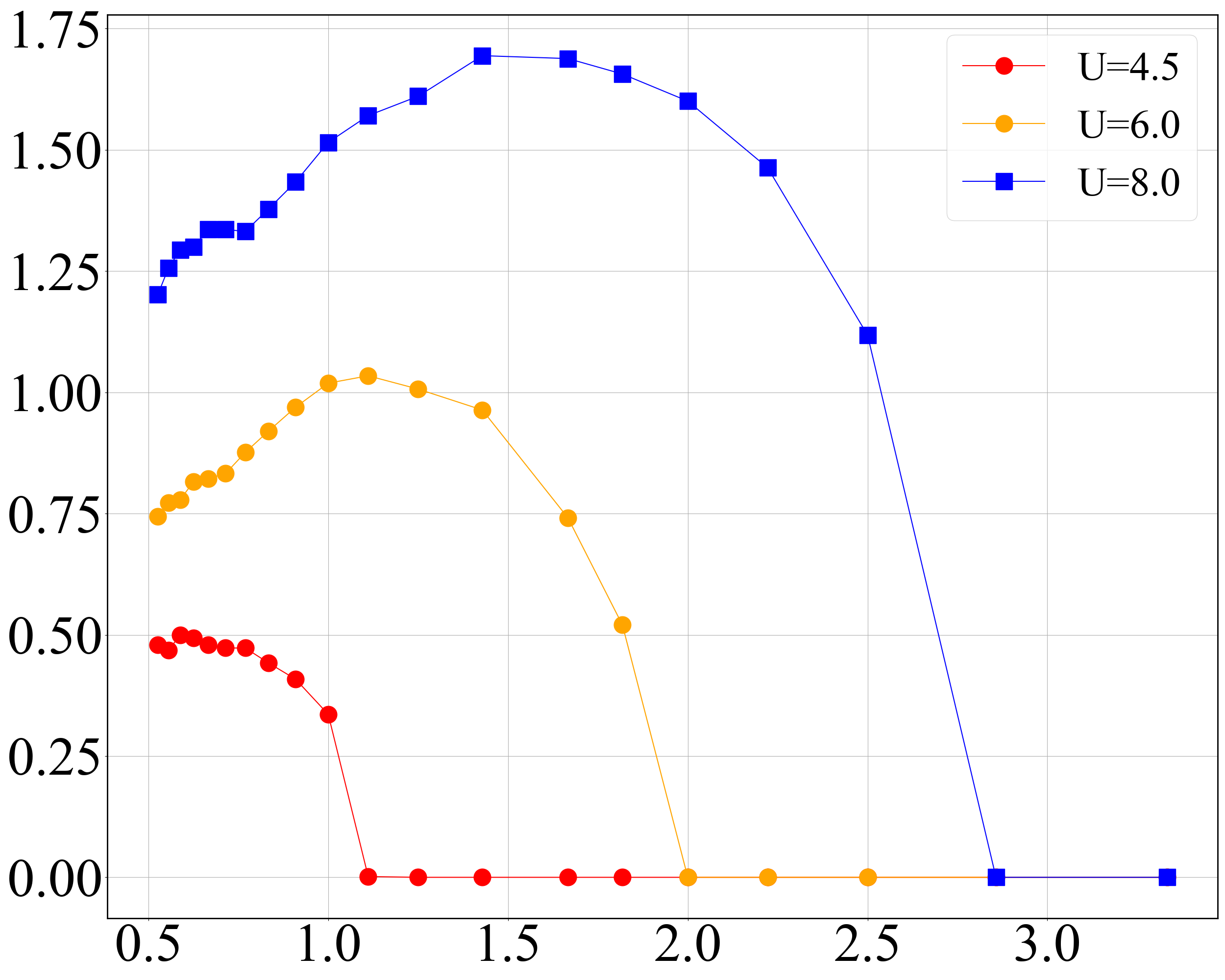}};
%\node[left=of img3,node distance=0cm,yshift=2.0cm,xshift=1.6cm]{\large{(b)}};
\node[left=of img9,node distance=0cm,rotate=0,anchor=center,yshift=0.0cm,xshift=0.8cm]{\large{ $\mu_c$}};
\node[below=of img9,node distance=0cm,yshift=1.2cm,xshift=0.0cm]{\small{$T$}};
%\node[left=of img4,node distance=0cm,rotate=0,anchor=center,yshift=0.0cm,xshift=3.0cm]{\small{ $U=8.0$}};
\node[above=of img9,node distance=0cm,rotate=0,anchor=center,yshift=-0.9cm,xshift=0.0cm]{\small{(d) $U=8.0$}};
%%%%%%%%%%%%%%%%%%%%%%%%%%%
\end{tikzpicture}
\caption{Opening of charge gap in the thermodynamic density of states(TDOS) at finite temperature. \textbf{(a)} TDOS $\tilde{\kappa} = \frac{\partial n}{\partial \mu}$ at $U=2.0$. Near half filling, $\tilde{\kappa}$ decreases with temperature similar to a metal. At very low doping, $\tilde{\kappa}$ at different temperatures cross, showing the formation of a band insulator. \textbf{(b)} TDOS at $U=5.0$. A Mott insulating state develops around half filling, with $\tilde{\kappa}$ increasing with temperature. \textbf{(d)} TDOS in the strong coupling regime, $U=8.0$, showing a pronounced charge gap (region of $\mu$ with thermally activated $\tilde{\kappa}$). The insets show the $n$ vs $\mu$ curves at respective $U$. \textbf{(d)} Charge gap $\mu_c(U,T)$ (defined in main text) for different $U$ as a function of temperature. $\mu_c(U,T)$ increases initially with temperature, up to a scale set by $U$. Increasing $T$ beyond this point melts the Mott insulating phase around half-filling, with $\mu_c$ going to zero when the charge gap in TDOS closes.}
\label{isobestic_points}
\end{figure*}

\section{Acknowledgement}

We would like to thank Thomas Hartke, Ningyuan Jia and Martin Zweirlein 
%for numerous insightful discussions during various stages of the work, and 
for providing data 
on various correlation functions in Fig \ref{Experimental_compare}. We would also like to thank Abhisek Samanta for feedback on the manuscript and numerous insightful discussions. S.R and N.T would like to acknowledge support from National Science Foundation through grant number NSF-DMR grant no 2138905.  T.P. acknowledges financial support from 
Funda\c{c}\~ao Carlos Chagas Filho de Amparo \`a Pesquisa do Estado do Rio de Janeiro grant numbers E-26/200.959/2022  and E-26/210.100/2023; and from CNPq grant numbers 308335/2019-8, 403130/2021-2,  and 442072/2023-6; and also Instituto Nacional de Ci\^encia e Tecnologia de Informa\c c\~ao Qu\^antica (INCT-IQ). Numerical calculations have been carried out using high-performance computing resources of the Unity cluster at Ohio State University.

\appendix
%\newpage
%\newpage
\section{Isobestic points and finite temperature charge gap}
\label{app_a}

As shown in Fig \ref{gap_scales}, the appearance of zeros of $\frac{\partial \tilde{\kappa}}{\partial T}$ at half-filling implies the opening of a charge gap in TDOS. This results in a doping window around half-filling where $\tilde{\kappa}$ increases with temperature, shown in Fig \ref{isobestic_points}. The doping window, which we dub the finite temperature charge gap, increases as $U$ is varied from weak to strong coupling. The value of the charge gap $\mu_c(U,T)$ is defined from the condition $\mu_c(U,T): \frac{\partial \tilde{\kappa}}{\partial T} \big|_{\mu_c} = 0$.
 
An immediate consequence of this is the appearance of ``isosbestic" points, where $n$ vs $\mu$ curves show little temperature variation at specific $\mu$ points, $\frac{\partial n}{\partial T}\big |_{\mu} = 0$, shown in Fig \ref{isobestic_points}(a)-(c) and their insets. In the weak coupling regime ($U=2$), where the system is TDOS gapless in the temperature range we study, such ``isosbestic" points are absent. The existence of isosbestic points can be shown from a Maxwell relation based on the entropy density, $s =\frac{1}{T}( \varepsilon_k+\varepsilon_p-\mu n)$, where $\varepsilon_k$ is the kinetic energy density, $\varepsilon_p$ is the potential energy density, $s$ is the entropy density and $n$ is the number density. The appearance of $\frac{\partial \tilde{\kappa}}{\partial T} = 0$ guarantees appearance of points where $\frac{\partial s}{\partial \mu} = \frac{\partial n}{\partial T} = 0$ \cite{lenihan2021entropy}, which are the ``isosbestic" points,  through $\frac{\partial ^2 s}{\partial \mu^2} = \frac{\partial}{\partial T}\big(\frac{\partial n}{\partial \mu}\big) $. As shown in Fig \ref{isobestic_points}(b-c) and their insets, this is indeed the case; presence of a charge gap at $U = 5.0, 8.0$ is accompanied by appearance of points (marked by arrows) where $n$ vs $\mu$ curves at different temperatures cross. The charge gap $\mu_c$ also shows a non-monotonic dependence with temperature, similar to the crossover density  $n_{cr}$ in Fig \ref{gap_closing_MIT_correlators}(d).

\section{Spectral weight across the anomalous insulator phase at half filling}
\label{app_b}

\begin{figure}[ht]
\begin{tikzpicture}

\node (img1) {\includegraphics[width=4cm,height=3.2cm]{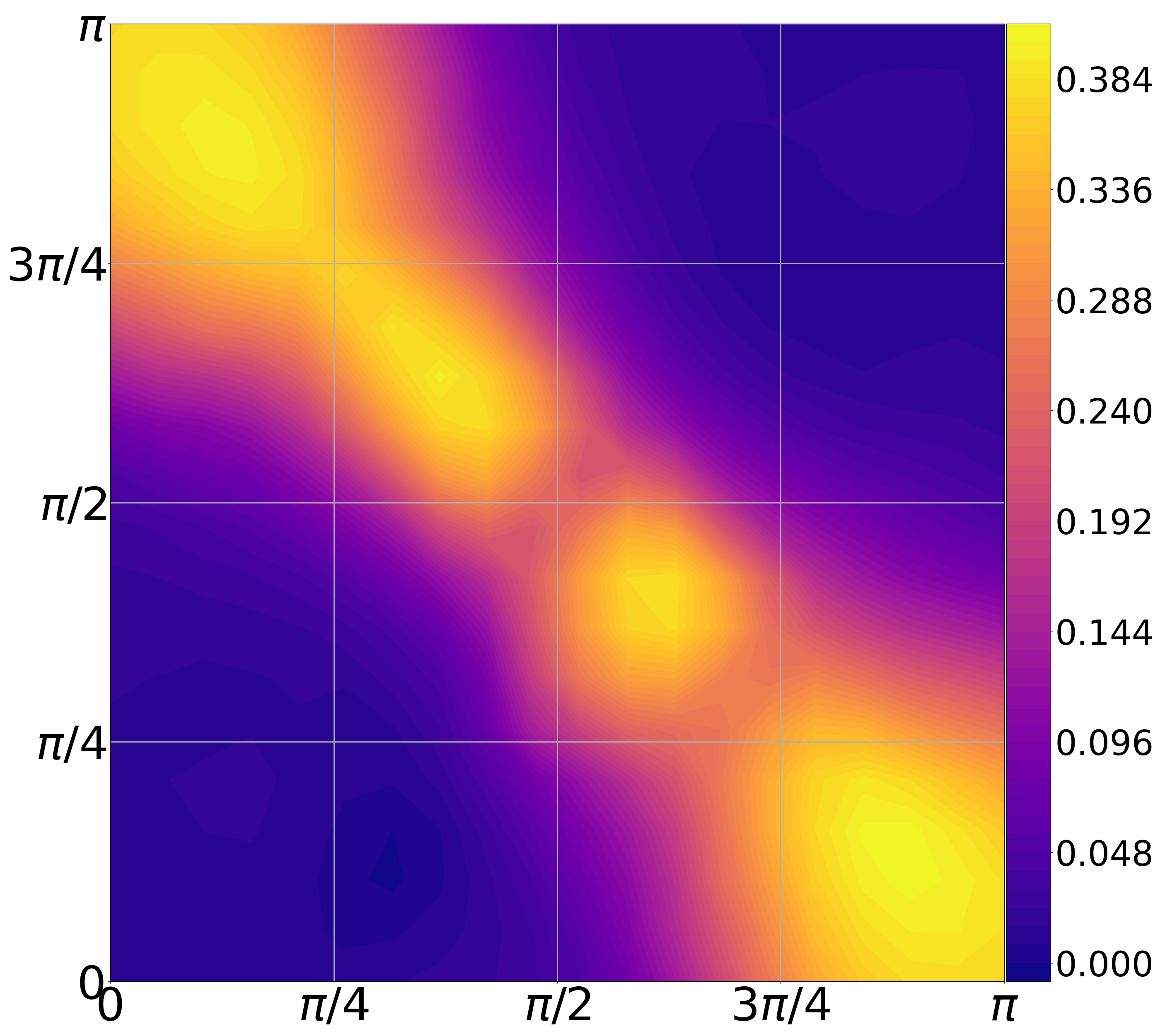}};
\node[above=of img1,node distance=0cm,yshift=-1.80cm,xshift=0.5cm]{\small{\textcolor{white}{(a) $U=4.0$}}};
\node[left=of img1,node distance=0cm,xshift=1.3cm,yshift=0.0cm]{\normalsize{$k_y$}};
%%%%%%%%%%%%%%%%%%%%%%%%%%%%%%%%
\node (img2) [right=of img1,xshift=-1.15cm]{\includegraphics[width=4cm,height=3.2cm]{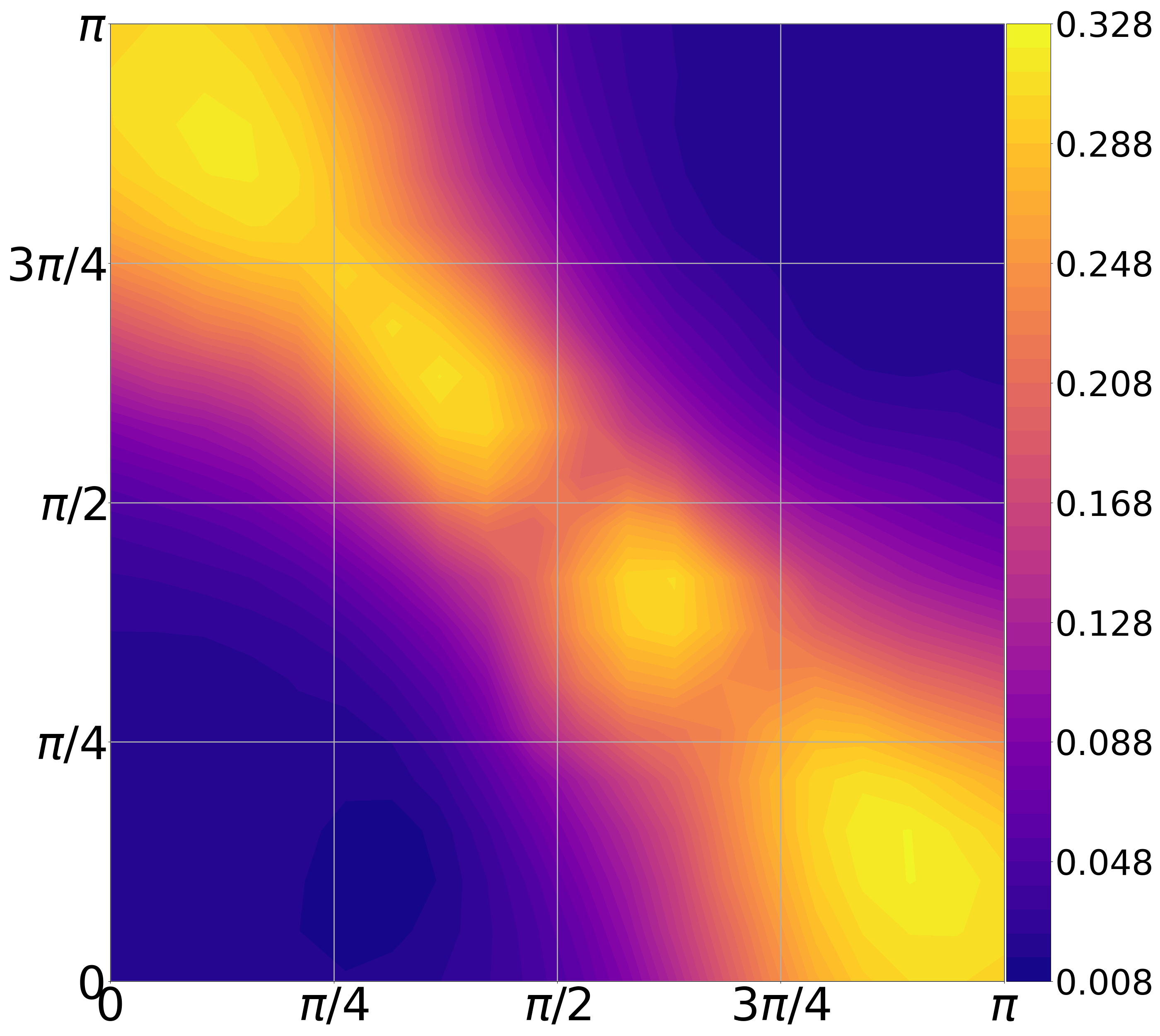}};
\node[above=of img2,node distance=0cm,yshift=-1.80cm,xshift=0.5cm]{\small{\textcolor{white}{(b) $U=4.5$}}};
%\node[above=of img2,node distance=0cm,yshift=-1.20cm,xshift=0.0cm]{\normalsize{(b) $C_{hd}(1)$}};
%\node[below=of img2,node distance=0cm,yshift=1.1cm,xshift=0.0cm]{\large{$n$}};
%%%%%%%%%%%%%%%%%%%%%%%%%%%%%%%%%%%%%%%%%%%%%
\node (img3) [below=of img1,yshift=0.95cm]{\includegraphics[width=4cm,height=3.2cm]{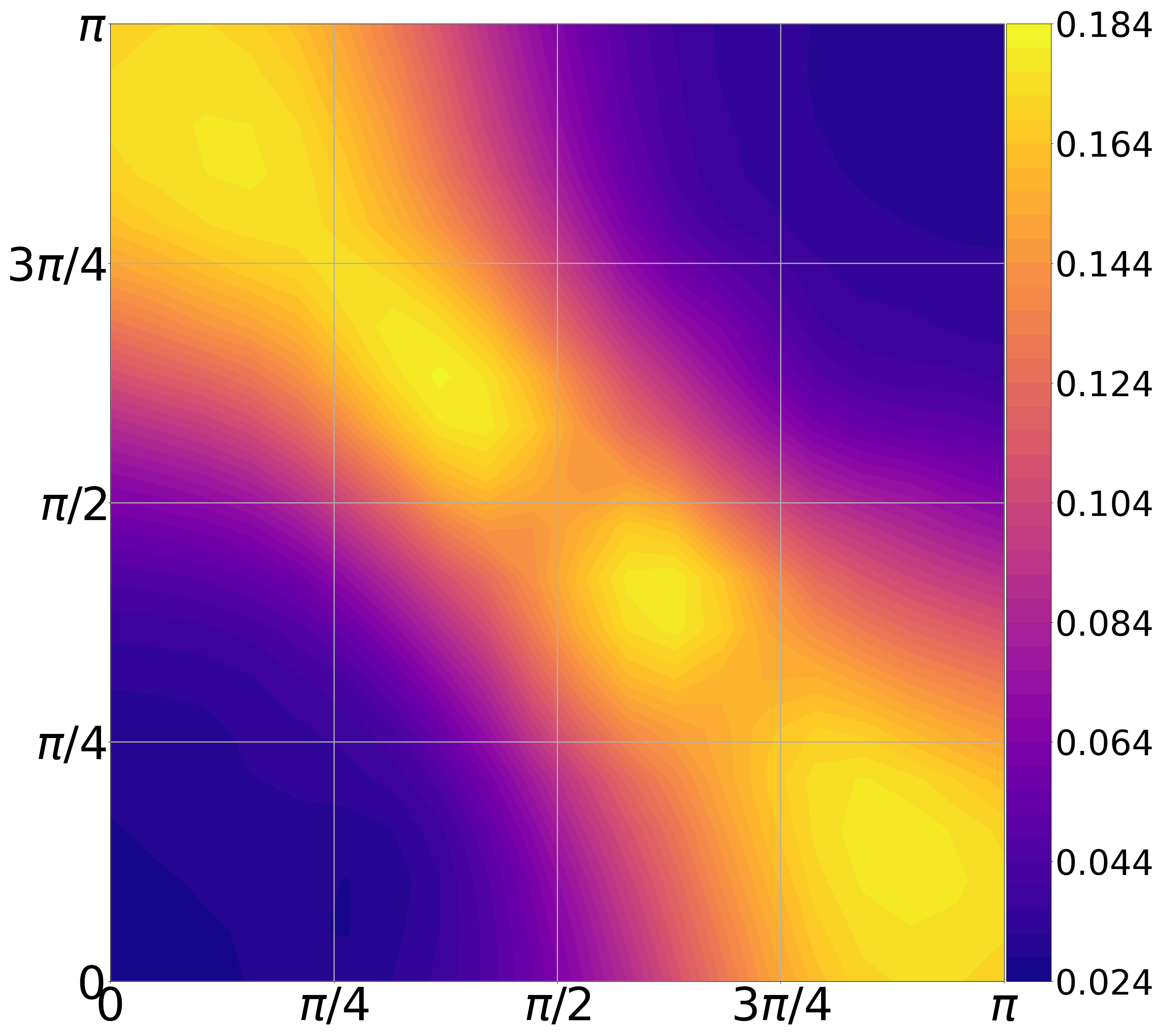}};
%\node[left=of img3,node distance=0cm,xshift=-3.1cm,yshift=0.0cm]{\normalsize{$k_y$}};
\node[above=of img3,node distance=0cm,yshift=-1.80cm,xshift=0.5cm]{\small{\textcolor{white}{(c) $U=5.5$}}};
\node[below=of img3,node distance=0cm,yshift=1.1cm,xshift=0.0cm]{\normalsize{$k_x$}};
\node[left=of img3,node distance=0cm,xshift=1.3cm,yshift=0.0cm]{\normalsize{$k_y$}};
%%%%%%%%%%%%%%%%%%%%%%%%%%%%%%%%%%%%%%%%%%%%%
\node (img4) [right=of img3,xshift = -1.15cm] {\includegraphics[width=4cm,height=3.2cm]{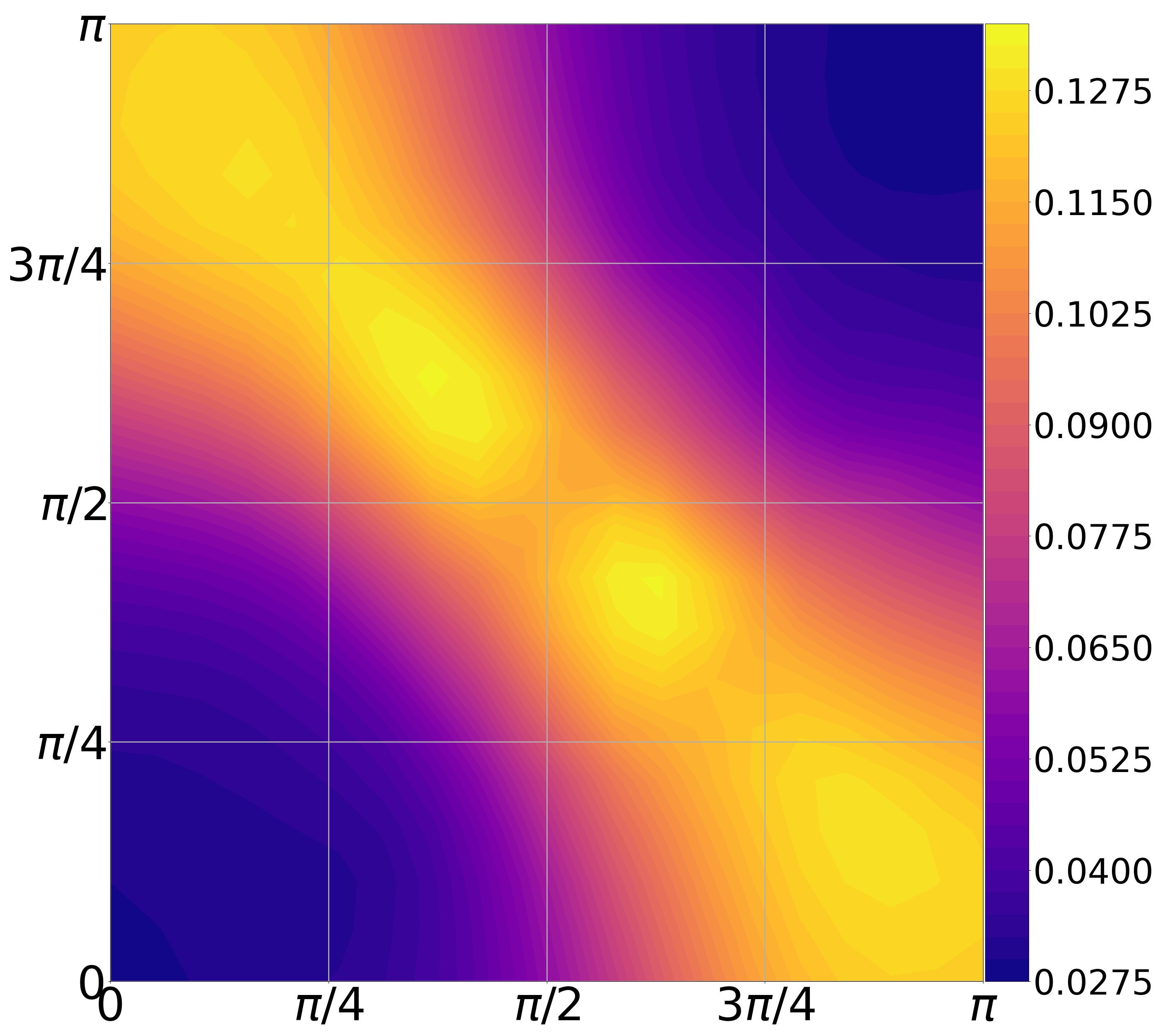}};
\node[above=of img4,node distance=0cm,yshift=-1.80cm,xshift=0.5cm]{\small{\textcolor{white}{(d) $U=6.0$}}};
\node[below=of img4,node distance=0cm,yshift=1.1cm,xshift=0.0cm]{\normalsize{$k_x$}};
%%%%%%%%%%%%%%%%%%%%%%%%%%%%
\end{tikzpicture}
\captionof{figure}{Spectral function $A(k,\omega = 0)$ across phase B (Fig \ref{gap_scales}(d)) at $T=0.67$.\textbf{(a)} $U=4.0$, when a gap has just appeared in TDOS, but LDOS is still gapless. \textbf{(b)} $U=4.5$, inside phase B. \textbf{(c)} $U=5.5$, below the pseduogap opening point in LDOS. \textbf{(d)} $U=6.0$, when a gap has formed in LDOS. As the system is tuned across phase B, there is a uniform suppression of zero energy spectral weight with increasing $U$, \textit{at all points} on the Fermi surface. } 
\label{Spectral_weight_anomalous_phase}
\end{figure}
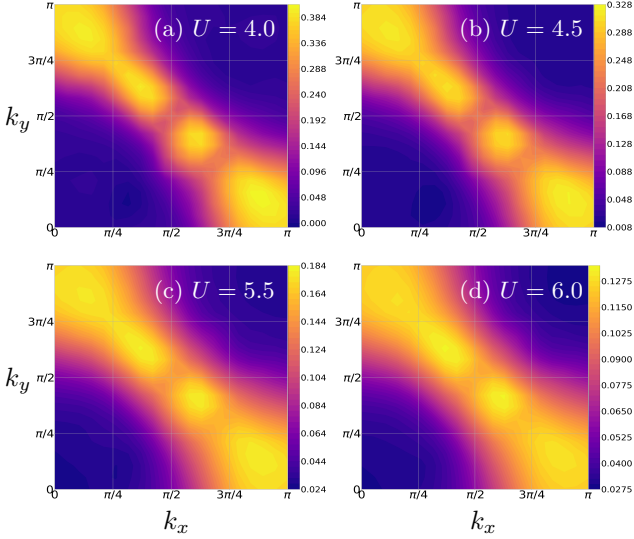

In Ref \cite{vsimkovic2020extended}, an extended crossover from a Fermi Liquid to Non-Fermi liquid (NFL) was identified by the sign of $\Delta_k = Im\Sigma(k,i\omega_0)-Im \Sigma(k,i\omega_1)$; a positive value implies a metal, and a negative value implies an insulator. An NFL regime was identified where $\Delta_k$ changes sign across the Brillouin zone; this would imply that the single particle dispersion is gapped at specific $k$ points, whereas it remains gapless at other $k$ points. However, this can be a feature of low-temperature physics where the system lies below the spin ordering temperature and quasi long range antiferromagnetic correlations form. In the intermediate to high temperature ranges that we study, this is not the case; as shown in Fig \ref{Spectral_weight_anomalous_phase}. Increasing $U$ in the anomalous phase (B) of Fig \ref{gap_scales}(d) leads to a gapped Fermi surface uniformly by suppressing spectral weights at all momenta. This suggests that the anomalous phase we observe, where TDOS is gapped and LDOS is gapless is not a high-temperature effect of the NFL behavior observed below $T_{spin}$ in \cite{vsimkovic2020extended}.

\section{TDOS and Doublon number}
\label{app_c}
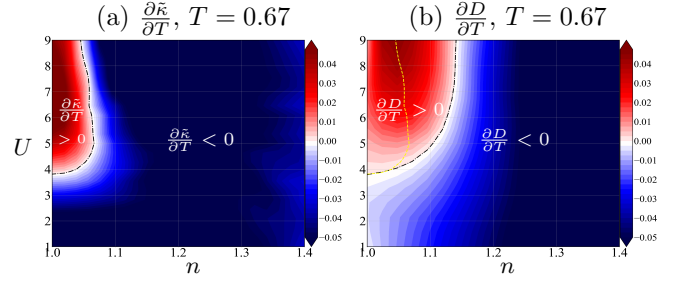
\begin{figure}[t]
\begin{tikzpicture}
%%%%%%%%%%%%%%%%%%%%%%%%%%%%%%%%%%%%%%%
%%%%%%%%%%%%%%%%%%%%%%%%%%%%%%%%%%%%%%%
\node (img1)
{\includegraphics[width=4.0cm,height=3.0cm]{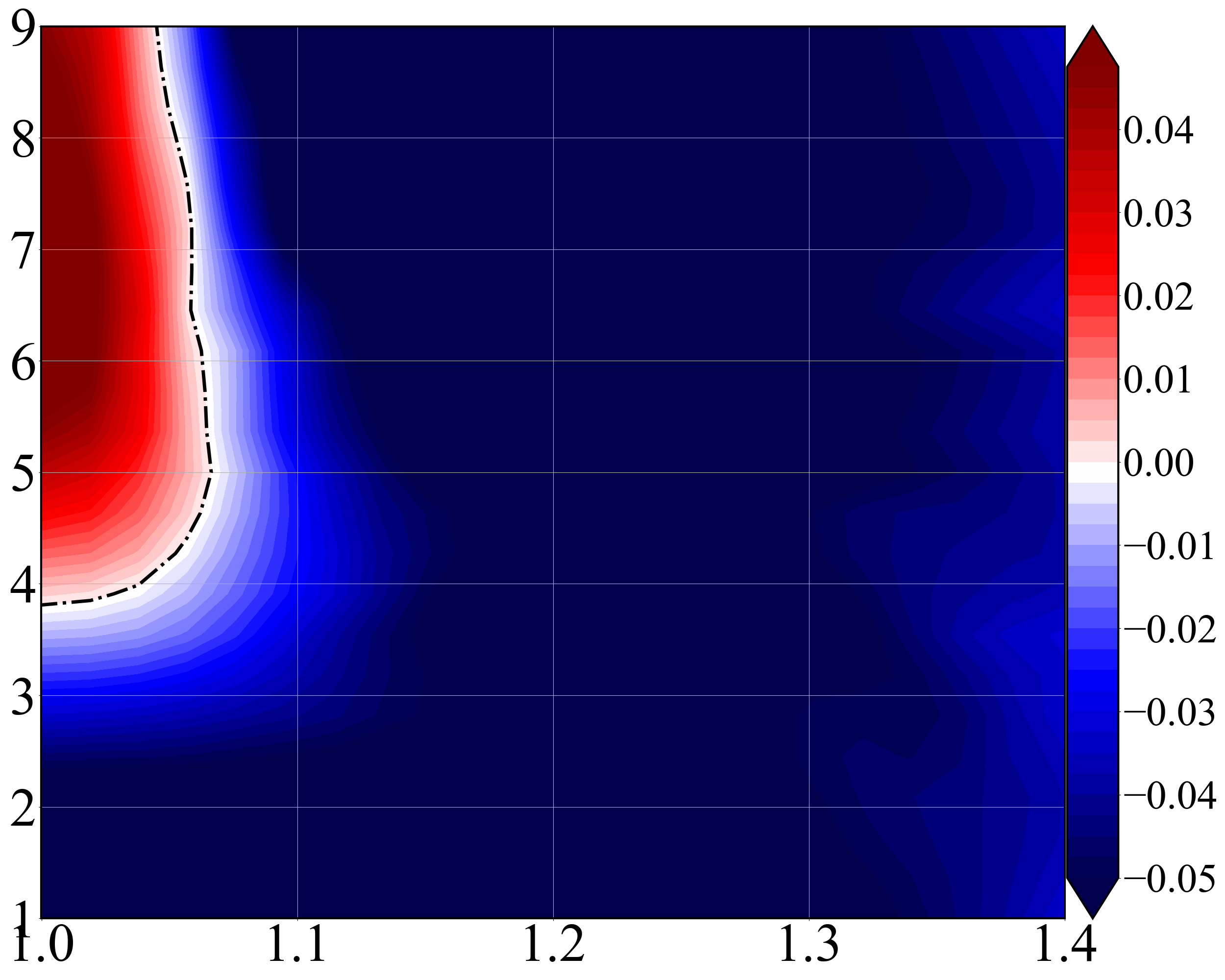}};
%\node[left=of img3,node distance=0cm,yshift=2.0cm,xshift=1.6cm]{\large{(b)}};
\node[left=of img1,node distance=0cm,rotate=0,anchor=center,yshift=0.0cm,xshift=0.8cm]{\small{ $U$}};
\node[below=of img1,node distance=0cm,yshift=1.2cm,xshift=0.0cm]{\small{$n$}};
%\node[left=of img2,node distance=0cm,rotate=0,anchor=center,yshift=0.0cm,xshift=3.8cm]{\small{ $U=2.0$}};
\node[above=of img1,node distance=0cm,rotate=0,anchor=center,yshift=-0.9cm,xshift=0.0cm]{\normalsize{(a) $\frac{\partial \tilde{\kappa}}{\partial T}$, $T = 0.67$}};
%\node[above=of img3,node distance=0cm,yshift=-1.3cm,xshift=0.0cm]{\large{(a)}\Large{ $\frac{\partial \tilde{\kappa}}{\partial T}$}\large{, $T = 0.67$}};
\node[left=of img1,node distance=0cm,rotate=0,anchor=center,yshift=0.5cm,xshift=1.5cm]{\scriptsize{\textcolor{white}{$\frac{\partial \tilde{\kappa}}{\partial T}$}}};
\node[left=of img1,node distance=0cm,rotate=0,anchor=center,yshift=0.1cm,xshift=1.5cm]{\tiny{\textcolor{white}{$>0$}}};
\node[left=of img1,node distance=0cm,rotate=0,anchor=center,yshift=0.1cm,xshift=3.2cm]{\scriptsize{\textcolor{white}{$\frac{\partial \tilde{\kappa}}{\partial T}<0$}}};
%%%%%%%%%%%%%%%%%%%%%%%%%%%%%%%%%%%%%%%
\node (img2)[right=of img1, xshift = -1.05cm]
{\includegraphics[width=4.0cm,height=3.0cm]{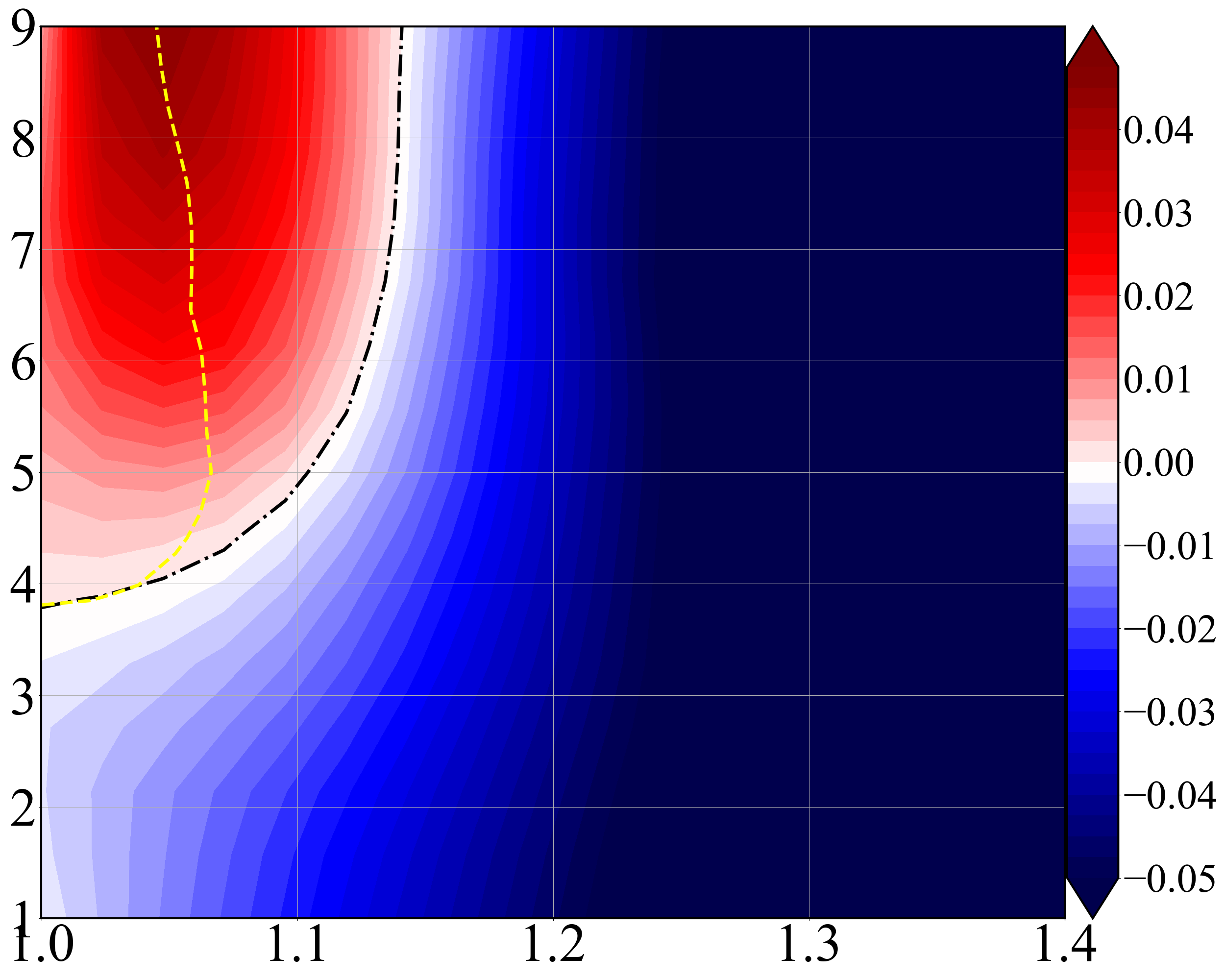}};
%\node[left=of img3,node distance=0cm,yshift=2.0cm,xshift=1.6cm]{\large{(b)}};
%\node[left=of img2,node distance=0cm,rotate=0,anchor=center,yshift=0.0cm,xshift=0.8cm]{\large{ $U$}};
\node[below=of img2,node distance=0cm,yshift=1.2cm,xshift=0.0cm]{\small{$n$}};
%\node[left=of img3,node distance=0cm,rotate=0,anchor=center,yshift=0.0cm,xshift=3.8cm]{\small{ $U=5.0$}};
\node[above=of img2,node distance=0cm,rotate=0,anchor=center,yshift=-0.9cm,xshift=0.0cm]{\normalsize{(b) $\frac{\partial D}{\partial T}$, $T=0.67$}};
\node[left=of img2,node distance=0cm,rotate=0,anchor=center,yshift=0.5cm,xshift=1.8cm]{\scriptsize{\textcolor{white}{$\frac{\partial D}{\partial T}>0$}}};
\node[left=of img2,node distance=0cm,rotate=0,anchor=center,yshift=0.1cm,xshift=3.2cm]{\scriptsize{\textcolor{white}{$\frac{\partial D}{\partial T}<0$}}};
%%%%%%%%%%%%%%%%%%%%%%%%%%%%%%%%%%%%%%
%%%%%%%%%%%%%%%%%%%%%%%%%%%
\end{tikzpicture}
\caption{Phase diagram through TDOS and doublon number in the $U-n$ plane. \textbf{(a)} Temperature variation of TDOS shows an insulating phase ($\frac{\partial \tilde{\kappa}}{\partial T}>0$ around half filling, once the charge gap has opened up. \textbf{(b)} Temperature variation of doublon number shows a localized regime $\frac{\partial D}{\partial T}>0$ around half filling, and a metallic regime with itinerant electrons ($\frac{\partial D}{\partial T}<0.$). The yellow dashed line in (b) marks the locus of $(U,n)$ where $\frac{\partial \tilde{\kappa}}{\partial T} = 0$. The boundaries of the two phase diagrams in (a) and (b) are not in agreement with each other. }
\label{metallic_localized_comp}
\end{figure}

An alternate way to study effects of correlation is to look at the average doublon occupancy, $d_i = \langle n_{i\uparrow}n_{i\downarrow} \rangle$. From the identity $m^2_i = n_i-2d_i$, doublon density and local moment density are anticorrelated at a fixed doping. In a Mott insulator, electrons get pinned down to form local moments, thereby reducing double occupancy; hence doublon number can increase only by thermal fluctuations with increasing temperature \cite{werner2005interaction,dare2007interaction,kim2020spin}. Since $\frac{\partial D}{\partial T}>0$ implies formation of local moments on cooling, this is a signature of Mott localization (with the average doublon number $D = 1/N_s \sum_{i}d_{i}$. In contrast, $\frac{\partial D}{\partial T}<0$ implies that on cooling, electrons are itinerant and can move around to form double occupancy; the states are hence delocalized. In Fig \ref{metallic_localized_comp}, we compare the temperature dependence of TDOS, which probes metallicity, with the temperature dependence of average doublon number, which probes localization of electrons. In the $U-n$ plane, comparison of the temperature dependence of $\tilde{\kappa}$ and $D$ shows a cascade of phases as a function of doping. Very close to half filling, the system is an insulator, with localized electrons. Increasing doping at fixed $U$ leads to an anomalous metallic state, with $\frac{\partial \tilde{\kappa}}{\partial T}<0$, but with $\frac{\partial D}{\partial T}>0$. Further increasing doping results in a metallic state, where electrons are delocalized and can form double occupancy.

\section{Moment-holon and moment-Doublon correlation functions}
\label{app_d}

\begin{figure}[t]
\begin{tikzpicture}
\node (img1) {\includegraphics[width=0.45\linewidth]{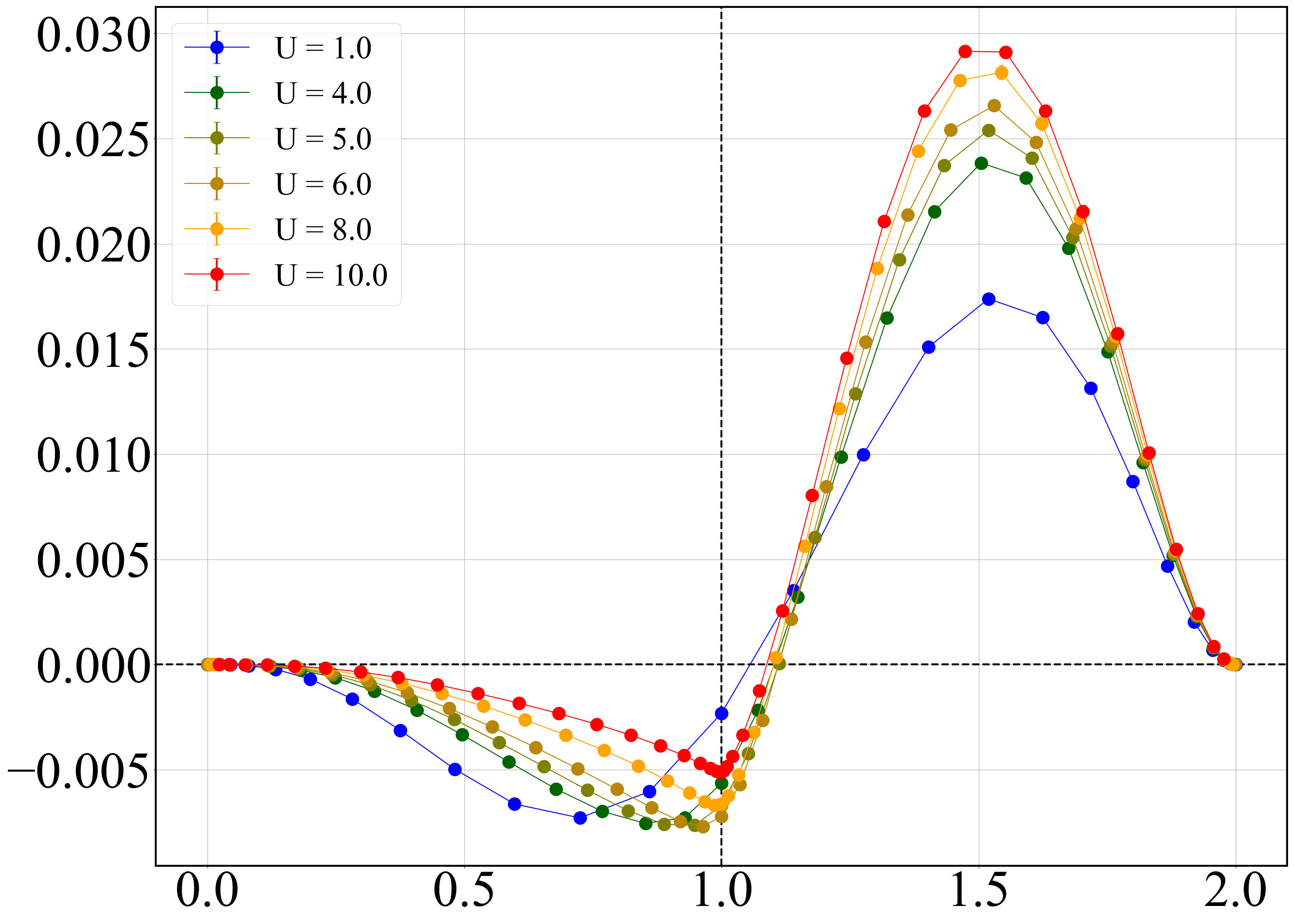}};
\node[above=of img1,node distance=0cm,yshift=-1.2cm,xshift=0.0cm]{\small{(a) $C_{md}(1)$}};
%\node[left=of img1,node distance=0cm,rotate=90,anchor=center,yshift=-1.5cm,xshift=0.0cm]{\small{ }};
\node[below=of img1,node distance=0cm,yshift=1.2cm,xshift=0.1cm]{\small{$n$}};
%%%%%%%%%%%%%%%%%%%%%%%%%%%%%%%%%%%%%%%%%%%%
\node (img2) [right=of img1,xshift=-0.85cm]{\includegraphics[width=0.45\linewidth]{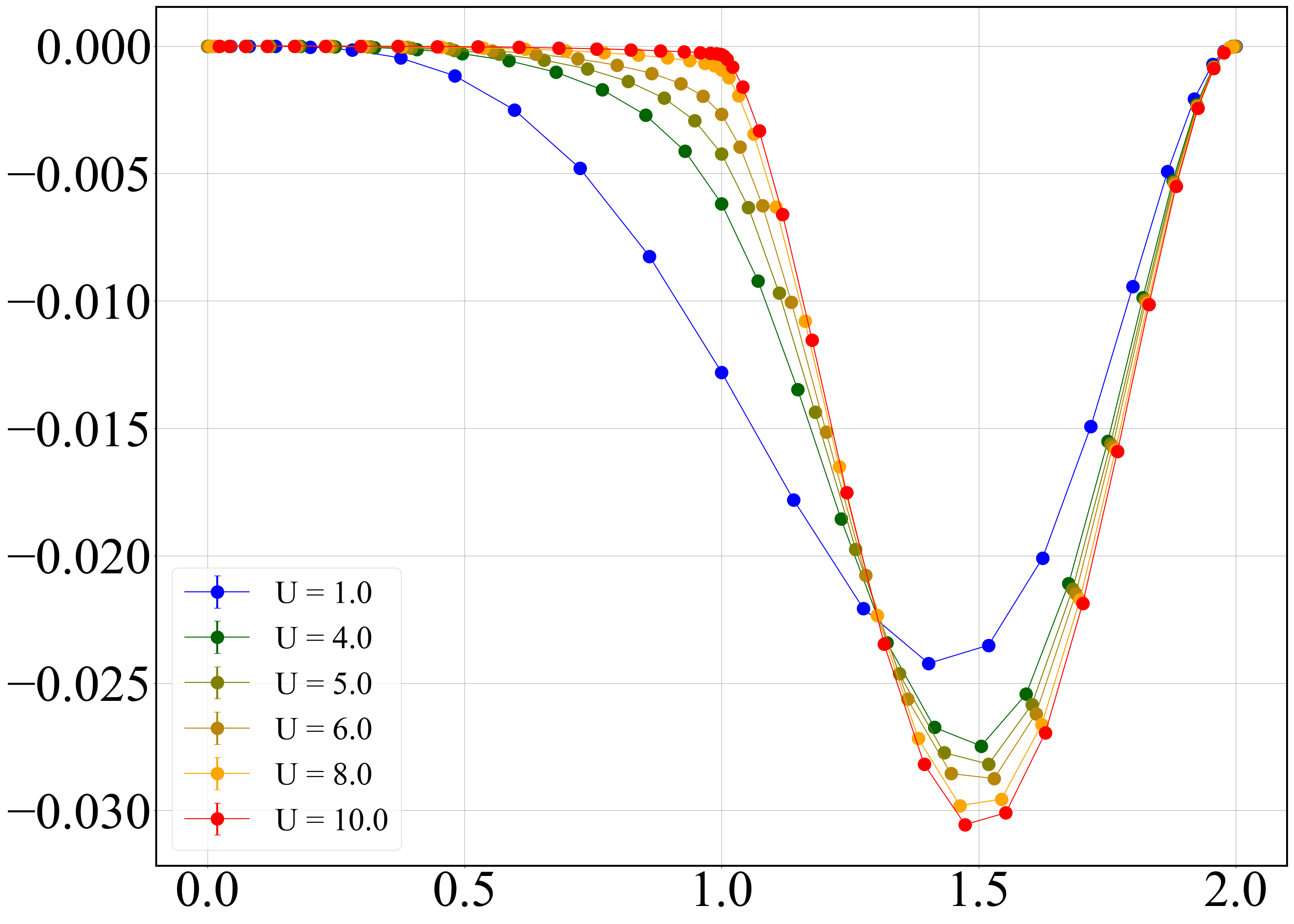}};
\node[above=of img2,node distance=0cm,yshift=-1.2cm,xshift=0.0cm]{\small{(b) $C_{dd}(1)$}};
%\node[left=of img2,node distance=0cm,rotate=90,anchor=center,yshift=-1.0cm,xshift=0.0cm]{\small{ $C_{dd}(r)$}};
%\node[left=of img2,node distance=0cm,yshift=-0.84cm,xshift=4.6cm]{\scriptsize{(b)}};
\node[below=of img2,node distance=0cm,yshift=1.2cm,xshift=0.1cm]{\small{$n$}};
%%%%%%%%%%%%%%%%%%%%%%%%%%%%%%%%%%%%%%%%%%%%
\node (img3)[below=of img1,yshift = 0.2cm,xshift=0cm ]{\includegraphics[width=0.45\linewidth]{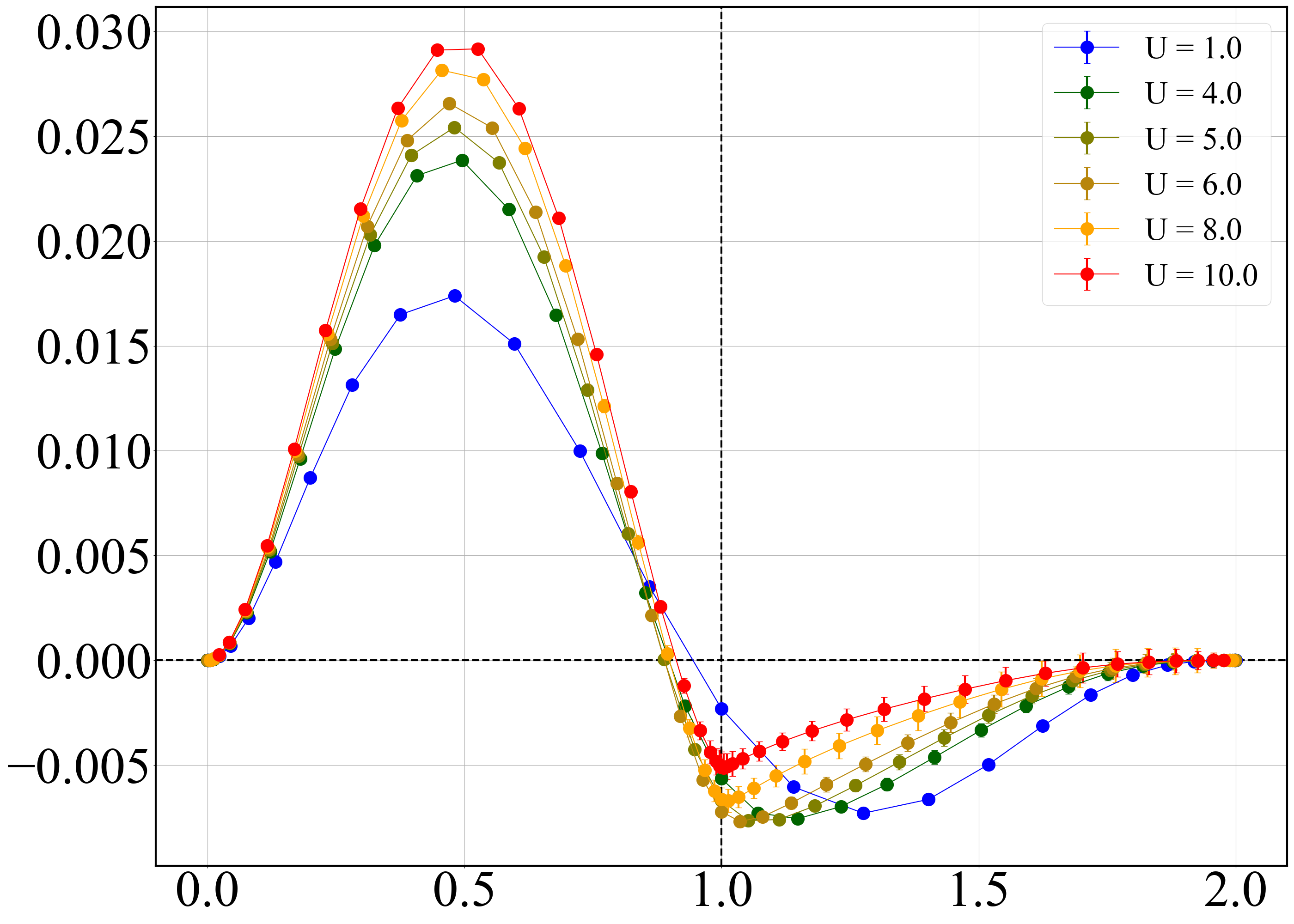}};
\node[above=of img3,node distance=0cm,yshift=-1.2cm,xshift=0.0cm]{\small{(c) $C_{mh}(1)$}};
%\node[left=of img4,node distance=0cm,rotate=90,anchor=center,yshift=-1.0cm,xshift=0.0cm]{\small{ $C_{mh}(r)$}};
\node[below=of img3,node distance=0cm,yshift=1.2cm,xshift=0.1cm]{\small{$n$}};
%%%%%%%%%%%%%%%%%%%%%%%%%%%%%%%%%%%%%%%%%%%%%
\node (img4) [right=of img3,xshift=-0.85cm]{\includegraphics[width=0.45\linewidth]{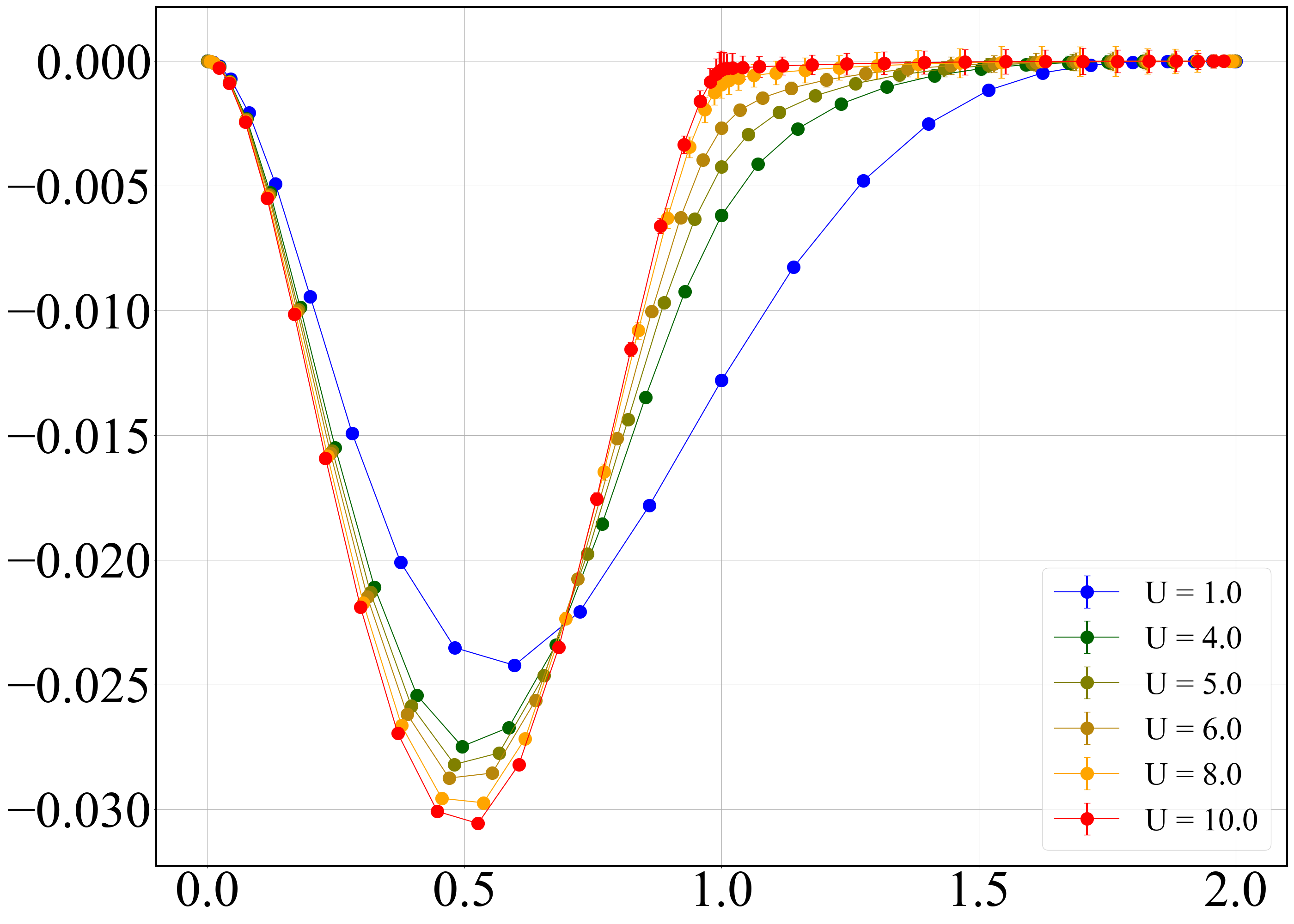}};
%\node[left=of img5,node distance=0cm,rotate=90,anchor=center,yshift=-1.0cm,xshift=0.0cm]{\small{ $C_{hh}(r)$}};
\node[above=of img4,node distance=0cm,yshift=-1.2cm,xshift=0.0cm]{\small{(d) $C_{hh}(1)$}};
\node[below=of img4,node distance=0cm,yshift=1.2cm,xshift=0.1cm]{\small{$n$}};
%%%%%%%%%%%%%%%%%%%%%%%%%%%%%%%%%%%%%%%%%%%
\end{tikzpicture}
\captionof{figure}{ Near-neighbor correlation functions as a function of density for $T=0.67$ and different values of interaction strength $U$. %, $C_{md}(r)$ and moment-holon correlations, $C_{mh}(r)$.
%\textbf{(a)} Moment-doublon correlations are reduced around half-filling where and moment-moment correlations are stronger. %At half-filling, it is suppressed as the system favors neighboring pairs of doublon-holon fluctuating into singlet pairs. In the particle doped side, it is favored as the system tries to have a local moment between two doublons to minimize doublon-doublon repulsion. 
%\textbf{(b)} Doublon-doublon correlation functions $C_{dd}(1)$. In the metallic regime, (below $U \sim 3.8$ ) there is finite repulsion between nearby doublon pairs at half-filling. In the correlated insulator regime, this goes to 0 due to the depletion of double occupancy.  \textbf{(c)} Nearest neighbor moment holon correlation function $C_{mh}(1)$. At half-filling, it is suppressed due to doublon-holon fluctuations, in the hole-doped side, it is positive due to hole-hole repulsion. \textbf{(d)} Nearest neighbor holon holon correlation function $C_{hh}(1)$. In the metallic regime, there is finite repulsion between neighboring holons at half-filling, in the correlated insulator regime, this goes to 0 at half-filling due to depletion of holes.
% I have removed because it would need rewriting and it is repeated.
\textbf{(a)} Moment-doublon, \textbf{(b)} doublon-doublon, \textbf{(c)} moment-holon, and \textbf{(d)} holon-holon.   
}  
\label{moment_doublon_holon_comparision}
\end{figure}
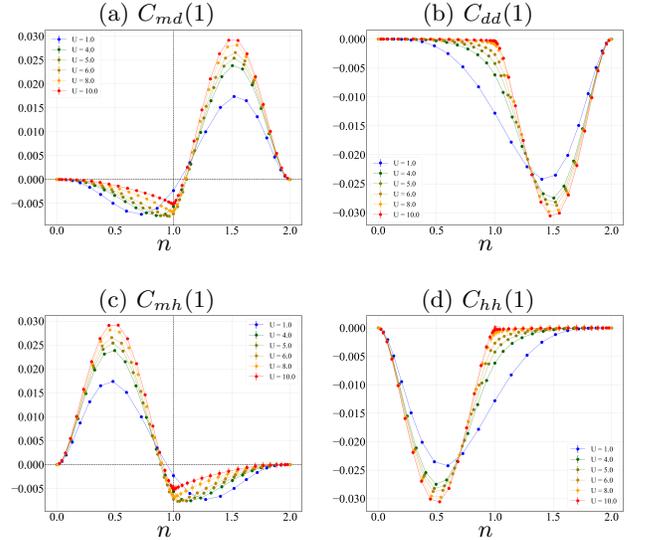
From the definition of doublons, holons and local moments, one can rewrite the moment-moment correlation function as

\begin{align}
    C_{mm}(\textbf{r}) &= \sum_{|\textbf{i}-\textbf{j}|=r}\langle m^{2}_{\textbf{i}} (1-h_{\textbf{j}}-d_{\textbf{j}})\rangle-\langle m^2_{\textbf{i}}\rangle\langle 1-h_{\textbf{j}}+d_{\textbf{j}}\rangle \nonumber \\
    &=\sum_{|\textbf{i}-\textbf{j}|=r}\langle m^2_{i}d_{\textbf{j}}\rangle - \langle m^2_{\textbf{i}} \rangle \langle d_{\textbf{j}} \rangle + \langle m^{2}_{\textbf{i}} h_{\textbf{j}} \rangle -\langle m^2_{\textbf{i}}\rangle \langle h_{\textbf{j}} \rangle \nonumber \\
    &= -\big [ C_{md}(\textbf{r})+C_{mh}(\textbf{r}) \big]
    \label{moment doublon holon correlations}
\end{align}

The moment-doublon and moment-holon correlations, as defined in Eq \ref{moment doublon holon correlations} are thus anticorrelated with moment-moment correlations. We note the similarity between the correlators defined in Eq \ref{moment doublon holon correlations} and the anti-moment correlation functions observed experimentally in Ref \cite{cheuk2016observation,chiu2019string} and simulated in Ref \cite{chen2021quantum}. The similarity stems from the constraint that on each site, there can be either a holon, a doublon or a local moment.

Nearest neighbor moment-holon (moment-doublon) correlations include fluctuations in which adjacent moments and holons (doublons) swap positions. At half-filling, strong onsite repulsion at large $U$ would force single occupancy (moment formation) on every site. On particle (hole) doping, one creates local puddles of doublons (holons), that can swap position with a nearby local moment. The doublon (holon) can then move through the entire lattice by these fluctuations, thus strongly reducing neighboring moment-moment correlations.

\begin{figure*}[t]
\begin{tikzpicture}
\node (img1) {\includegraphics[width=4.2cm,height=3.0cm]{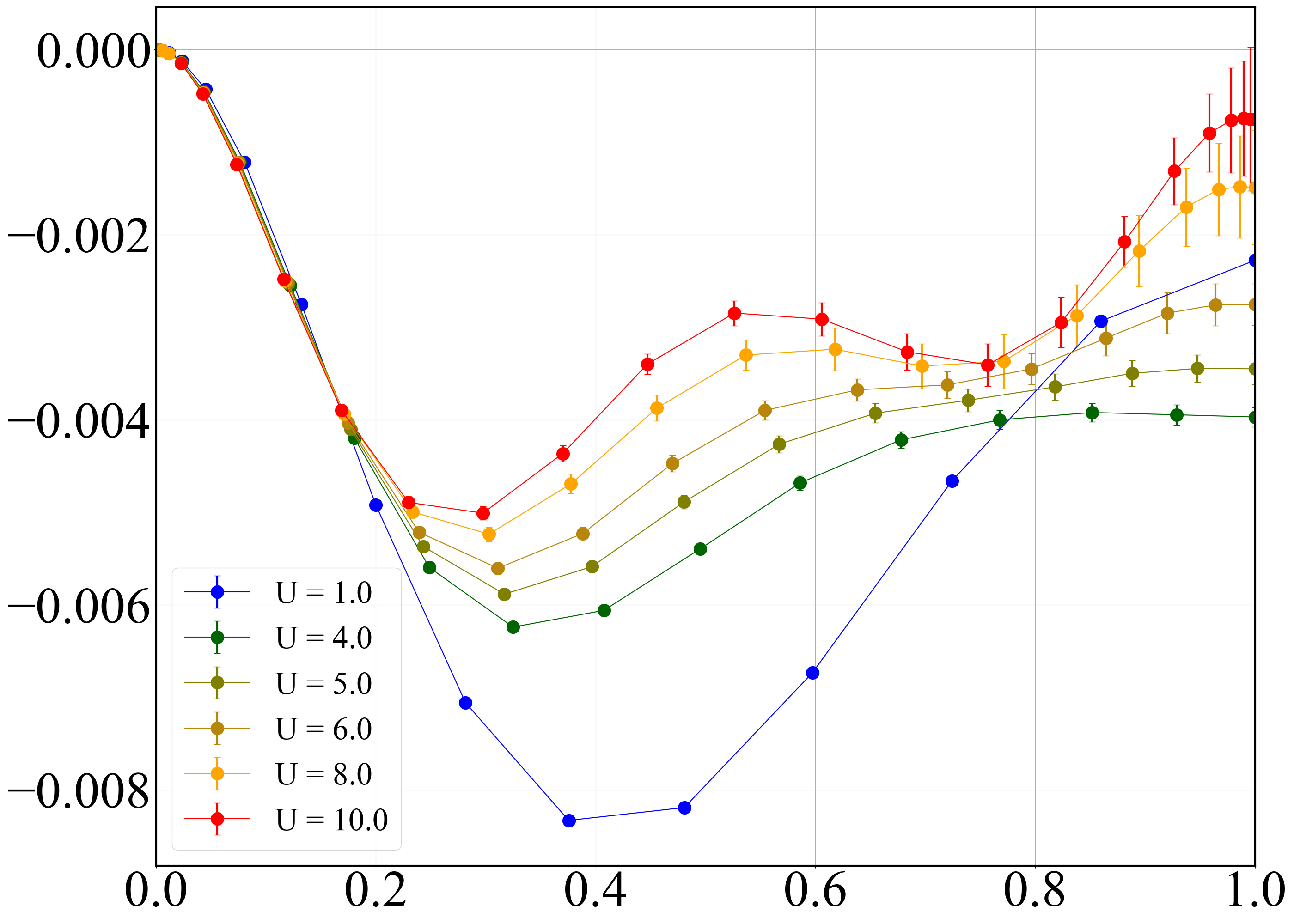}};
\node[above=of img1,node distance=0cm,yshift=-1.2cm,xshift=0.0cm]{\small{(a) $C_{nn}(\sqrt{2})$}};\node[below=of img1,node distance=0cm,yshift=1.2cm,xshift=0.1cm]{\small{$n$}};
%%%%%%%%%%%%%%%%%%%%%%%%%%%%%%%%%%%%%%%%%%%%
\node (img2) [right=of img1,xshift=-0.95cm]{\includegraphics[width=4.2cm,height=3cm]{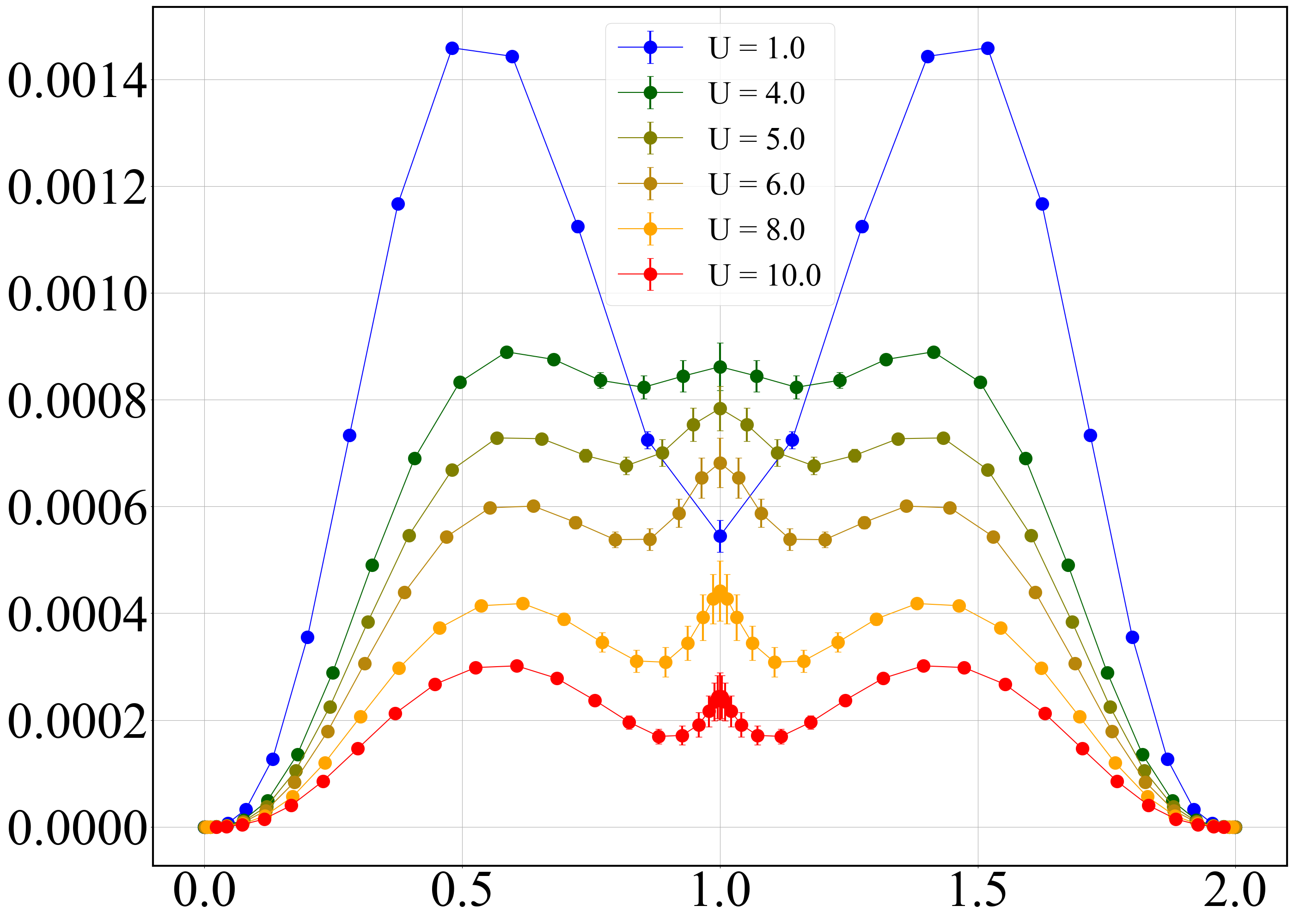}};
%\node[left=of img2,node distance=0cm,rotate=90,anchor=center,yshift=-1.0cm,xshift=0.0cm]{\small{ $C_{dd}(r)$}};
%\node[left=of img2,node distance=0cm,yshift=-0.84cm,xshift=4.6cm]{\scriptsize{(b)}};
\node[above=of img2,node distance=0cm,yshift=-1.2cm,xshift=0.0cm]{\small{(b) $C_{hd}(\sqrt{2})$}};
\node[below=of img2,node distance=0cm,yshift=1.2cm,xshift=0.1cm]{\small{$n$}};
%%%%%%%%%%%%%%%%%%%%%%%%%%%%%%%%%%%%%%%%%%%%
\node (img3)[right=of img2,yshift = 0.0cm,xshift=-0.95cm ]{\includegraphics[width=4.2cm,height=3.0cm]{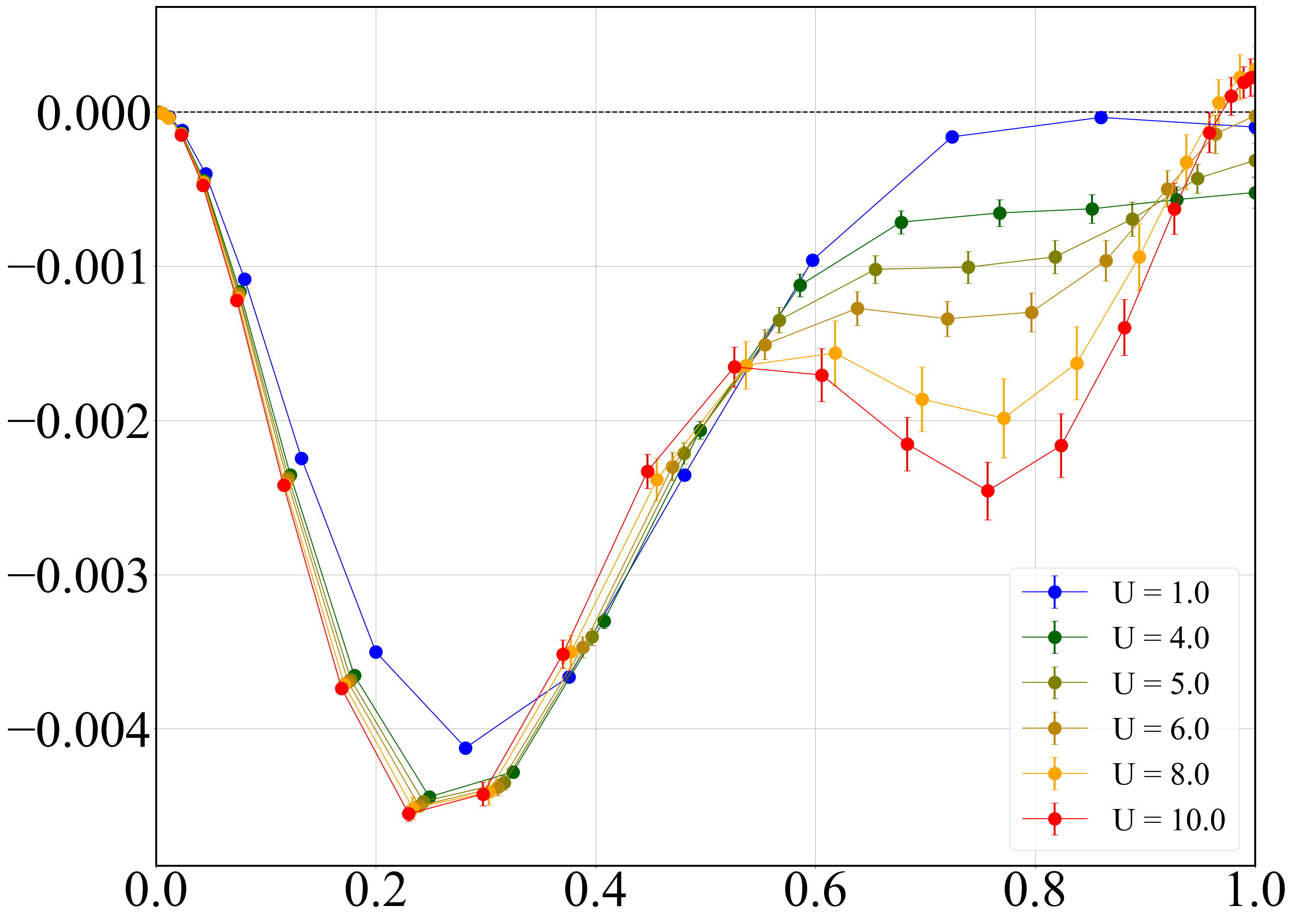}};
%\node[left=of img4,node distance=0cm,yshift=-0.84cm,xshift=2.1cm]{\scriptsize{(c)}};
%\node[left=of img4,node distance=0cm,rotate=90,anchor=center,yshift=-1.0cm,xshift=0.0cm]{\small{ $C_{mh}(r)$}};
\node[above=of img3,node distance=0cm,yshift=-1.2cm,xshift=0.0cm]{\small{(c) $C_{mm}(\sqrt{2})$}};
\node[below=of img3,node distance=0cm,yshift=1.2cm,xshift=0.1cm]{\small{$n$}};
%%%%%%%%%%%%%%%%%%%%%%%%%%%%%%%%%%%%%%%%%%%%%
\node (img4) [right=of img3,xshift=-0.95cm]{\includegraphics[width=4.2cm,height=3cm]{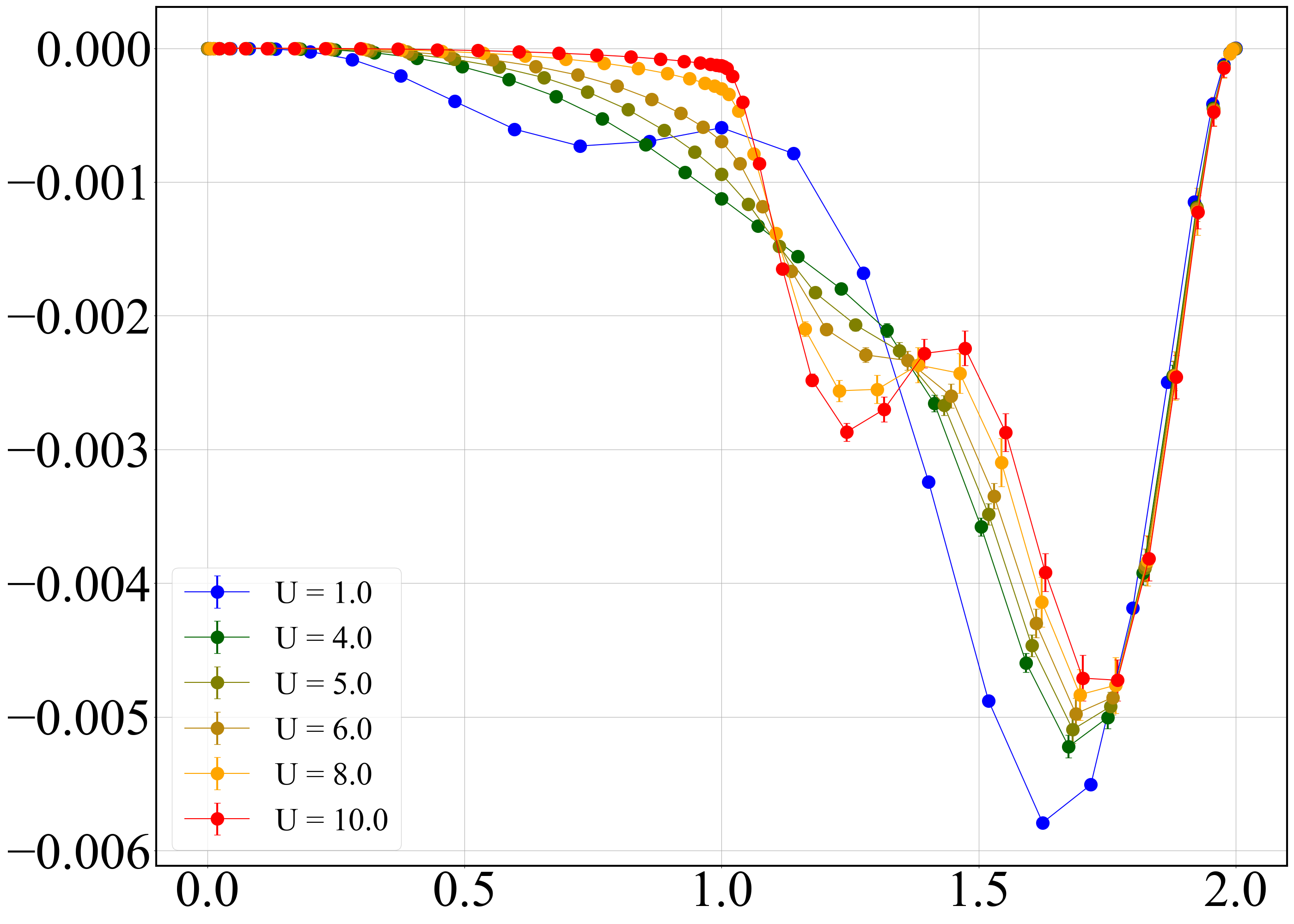}};
%\node[left=of img5,node distance=0cm,rotate=90,anchor=center,yshift=-1.0cm,xshift=0.0cm]{\small{ $C_{hh}(r)$}};
%\node[left=of img5,node distance=0cm,yshift=-0.84cm,xshift=2.1cm]{\scriptsize{(d)}};
\node[above=of img4,node distance=0cm,yshift=-1.2cm,xshift=0.0cm]{\small{(a) $C_{dd}(\sqrt{2})$}};
\node[below=of img4,node distance=0cm,yshift=1.2cm,xshift=0.1cm]{\small{$n$}};
%%%%%%%%%%%%%%%%%%%%%%%%%%%%%%%%%%%%%%%%%%%
\end{tikzpicture}
\captionof{figure}{Next-nearest neighbor correlations $C^{A-A}$ between same sublattice sites $A$ on a square lattice at $T = 0.67$ for different $U$. \textbf{(a)} Density-density correlator $C_{nn}(\sqrt{2})$, \textbf{(b)} holon-doublon correlator $C_{hd}(\sqrt{2})$, \textbf{(c)} moment-moment correlator $C_{mm}(\sqrt{2})$, and  \textbf{(d)} doublon-doublon correlator $C_{dd}(\sqrt{2})$. Note that the correlations show distinct behavior before and after opening of a gap in TDOS, marked by  the presence of additional minima/maxima at intermediate doping on the correlated insulator side.}  
\label{next_nearest_neighbor_correlator}
\end{figure*}
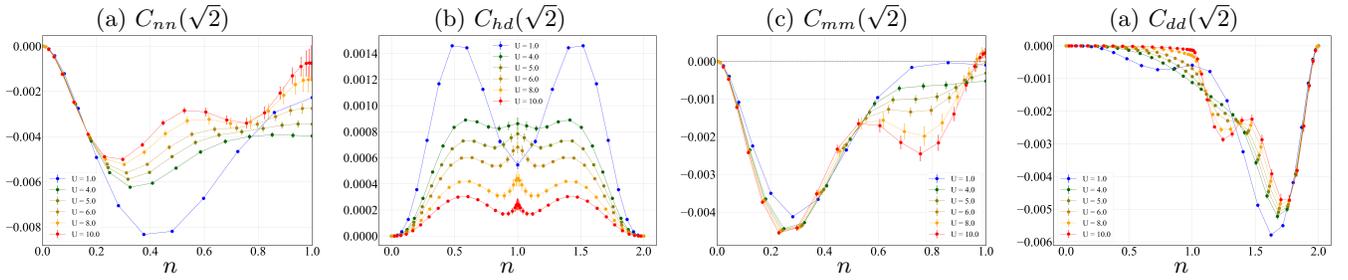

Fig \ref{moment_doublon_holon_comparision} shows the behavior of $C_{md}(1)$ and $C_{mh}(1)$ as a function of density for different values of the interaction strength at $T/t = 0.67$. Note that in the metallic regime $U<U_{cr}^{\rm LDOS}(T)$, the minima of $C_{md}(1)$ and $C_{mh}(1)$ are away from half-filling on the hole-doped and particle-doped side respectively. On entering the correlated insulator regime ($U>U_{cr}^{\rm LDOS}(T)$, the minima shifts towards half-filling. This is a consequence of the presence of a finite number of doublons and holons in the metallic limit, which decreases with increasing interaction strength in the correlated insulator regime. The presence of a finite number of doublons and holons at half-filling in the metallic regime can also be seen in the behavior of doublon-doublon $C_{dd}(1)$ and holon-holon $C_{hh}(1)$ correlators. 

To understand the non-trivial sign changes in the moment-doublon and moment-holon correlations at finite doping, shown in Fig \ref{moment doublon holon correlations}(a) and (c), we rewrite these correlations as:
\begin{align}
    \langle m^{2}_{i} d_{j}\rangle_C &= - \big [\langle h_{i} d_j \rangle_C + \langle d_i d_j \rangle_C \big ] \nonumber \\
    \langle m^{2}_{i} h_{j}\rangle_C &= -\big [ \langle h_{i} d_{j} \rangle_C + \langle h_i h_j \rangle_C \big ]
    \label{moment_doublon_holon_2}
\end{align}
which show that the moment-doublon (moment-holon) correlations derive from a competition between attractive holon-doublon pairs and repulsive doublon-doublon (holon-holon) pairs.  This is similar to the behavior noted in Ref \cite{chen2021quantum}. 

In the metallic regime, there is finite repulsion between doublon pairs $C_{dd}(1)$ and holon pairs $C_{hh}(1)$ at half-filling, which competes with the attraction between doublon and holon pairs, the latter being most dominant at half filling. Deep in the correlated insulator regime, the repulsion between doublon pairs and holon pairs are very small at half-filling, and the dominant contributions $C_{md}$ and $C_{mh}$ are due to $C_{hd}$. However, on the hole doped side, holon-holon repulsion forces the system to prefer singly occupied sites between holon pairs, to minimize holon-holon repulsion. This causes $C_{mh}(1)$ to be positive. A similar reasoning applies $C_{md}(1)$ on the particle doped side; the system tries to minimize doublon doublon repulsion, hence $C_{md}(1)$ is positive. At half filling, both $C_{mh}(1)$/$C_{md}(1)$) exhibit the same non monotonic nature as $C_{mm}(1)$ in Fig \ref{moment_correlations}(b).

\ 

\section{Next-nearest neighbor correlators and metal-insulator crossover}
\label{app_e}

With advances in cold atom experiments, measurements of non-local correlations beyond the first near neighbor are possible, although these might be orders of magnitude smaller than the nearest neighbor correlations\cite{mazurenko2017cold,chiu2019string,cheuk2016observation}. Since different correlation functions show different behavior as function of $U/t$ and doping, we look at all possible independent next nearest neighbor correlations (between same sublattice, $C^{A-A}$). % between density, holon and doublon pairs. 
We show density-density, holon-doublon, moment-moment and doublon-doublon correlators in Fig \ref{next_nearest_neighbor_correlator}, other correlators can be derived from these due to the constraint $m^2_i+h_i+d_i=1$ or particle-hole transformation.

As shown in Fig \ref{gap_scales}, increasing $U$ beyond a critical value opens a Mott gap. A common feature across all the correlators is the presence of a single minima/maxima in the metallic regime, and the appearance of multiple minima/maxima (on the appropriate hole/particle doped side) for interaction strengths larger than the ones where the gap in LDOS has opened up at half-filling. For the density-density, moment-moment and doublon-doublon correlations, a shoulder develops, that turns into secondary minima with increasing $U$, while pushing the primary minima away from the doping where it peaked in the metallic/gapless regime. For the holon-doublon correlations, minima at half-filling change into local maxima, with peaks decreasing with $U$. This is shown in Fig \ref{next_nearest_neighbor_correlator}, at $T = 0.67$. As these features are universally seen across all correlators in the charge sector, we can speculate that they are related to the appearance of  upper and lower Hubbard bands with strong interactions. % after a gap has opened up in LDOS; as opposed to a single band for the metallic regime. 
Tracking the evolution of multiple extrema in any of these correlation functions in experiments can qualitatively capture the crossover from metallic to correlated insulator regime. %, in addition to number density fluctuations.

%bibliography
\typeout{}
\bibliography{ref}

\providecommand{\noopsort}[1]{}\providecommand{\singleletter}[1]{#1}%
\begin{thebibliography}{73}%
\makeatletter
\providecommand \@ifxundefined [1]{%
 \@ifx{#1\undefined}
}%
\providecommand \@ifnum [1]{%
 \ifnum #1\expandafter \@firstoftwo
 \else \expandafter \@secondoftwo
 \fi
}%
\providecommand \@ifx [1]{%
 \ifx #1\expandafter \@firstoftwo
 \else \expandafter \@secondoftwo
 \fi
}%
\providecommand \natexlab [1]{#1}%
\providecommand \enquote  [1]{``#1''}%
\providecommand \bibnamefont  [1]{#1}%
\providecommand \bibfnamefont [1]{#1}%
\providecommand \citenamefont [1]{#1}%
\providecommand \href@noop [0]{\@secondoftwo}%
\providecommand \href [0]{\begingroup \@sanitize@url \@href}%
\providecommand \@href[1]{\@@startlink{#1}\@@href}%
\providecommand \@@href[1]{\endgroup#1\@@endlink}%
\providecommand \@sanitize@url [0]{\catcode `\\12\catcode `\$12\catcode `\&12\catcode `\#12\catcode `\^12\catcode `\_12\catcode `\%12\relax}%
\providecommand \@@startlink[1]{}%
\providecommand \@@endlink[0]{}%
\providecommand \url  [0]{\begingroup\@sanitize@url \@url }%
\providecommand \@url [1]{\endgroup\@href {#1}{\urlprefix }}%
\providecommand \urlprefix  [0]{URL }%
\providecommand \Eprint [0]{\href }%
\providecommand \doibase [0]{https://doi.org/}%
\providecommand \selectlanguage [0]{\@gobble}%
\providecommand \bibinfo  [0]{\@secondoftwo}%
\providecommand \bibfield  [0]{\@secondoftwo}%
\providecommand \translation [1]{[#1]}%
\providecommand \BibitemOpen [0]{}%
\providecommand \bibitemStop [0]{}%
\providecommand \bibitemNoStop [0]{.\EOS\space}%
\providecommand \EOS [0]{\spacefactor3000\relax}%
\providecommand \BibitemShut  [1]{\csname bibitem#1\endcsname}%
\let\auto@bib@innerbib\@empty
%</preamble>
\bibitem [{\citenamefont {Imada}\ \emph {et~al.}(1998)\citenamefont {Imada}, \citenamefont {Fujimori},\ and\ \citenamefont {Tokura}}]{imada1998metal}%
  \BibitemOpen
  \bibfield  {author} {\bibinfo {author} {\bibfnamefont {M.}~\bibnamefont {Imada}}, \bibinfo {author} {\bibfnamefont {A.}~\bibnamefont {Fujimori}},\ and\ \bibinfo {author} {\bibfnamefont {Y.}~\bibnamefont {Tokura}},\ }\bibfield  {title} {\bibinfo {title} {Metal-insulator transitions},\ }\href@noop {} {\bibfield  {journal} {\bibinfo  {journal} {Reviews of modern physics}\ }\textbf {\bibinfo {volume} {70}},\ \bibinfo {pages} {1039} (\bibinfo {year} {1998})}\BibitemShut {NoStop}%
\bibitem [{\citenamefont {Mott}(1961)}]{mott1961transition}%
  \BibitemOpen
  \bibfield  {author} {\bibinfo {author} {\bibfnamefont {N.~F.}\ \bibnamefont {Mott}},\ }\bibfield  {title} {\bibinfo {title} {The transition to the metallic state},\ }\href@noop {} {\bibfield  {journal} {\bibinfo  {journal} {Philosophical Magazine}\ }\textbf {\bibinfo {volume} {6}},\ \bibinfo {pages} {287} (\bibinfo {year} {1961})}\BibitemShut {NoStop}%
\bibitem [{\citenamefont {Gutzwiller}(1963)}]{gutzwiller1963effect}%
  \BibitemOpen
  \bibfield  {author} {\bibinfo {author} {\bibfnamefont {M.~C.}\ \bibnamefont {Gutzwiller}},\ }\bibfield  {title} {\bibinfo {title} {Effect of correlation on the ferromagnetism of transition metals},\ }\href@noop {} {\bibfield  {journal} {\bibinfo  {journal} {Physical Review Letters}\ }\textbf {\bibinfo {volume} {10}},\ \bibinfo {pages} {159} (\bibinfo {year} {1963})}\BibitemShut {NoStop}%
\bibitem [{\citenamefont {Kanamori}(1963)}]{kanamori1963electron}%
  \BibitemOpen
  \bibfield  {author} {\bibinfo {author} {\bibfnamefont {J.}~\bibnamefont {Kanamori}},\ }\bibfield  {title} {\bibinfo {title} {Electron correlation and ferromagnetism of transition metals},\ }\href@noop {} {\bibfield  {journal} {\bibinfo  {journal} {Progress of Theoretical Physics}\ }\textbf {\bibinfo {volume} {30}},\ \bibinfo {pages} {275} (\bibinfo {year} {1963})}\BibitemShut {NoStop}%
\bibitem [{\citenamefont {Hubbard}(1964)}]{hubbard1964electron}%
  \BibitemOpen
  \bibfield  {author} {\bibinfo {author} {\bibfnamefont {J.}~\bibnamefont {Hubbard}},\ }\bibfield  {title} {\bibinfo {title} {Electron correlations in narrow energy bands iii. an improved solution},\ }\href@noop {} {\bibfield  {journal} {\bibinfo  {journal} {Proceedings of the Royal Society of London. Series A. Mathematical and Physical Sciences}\ }\textbf {\bibinfo {volume} {281}},\ \bibinfo {pages} {401} (\bibinfo {year} {1964})}\BibitemShut {NoStop}%
\bibitem [{\citenamefont {Anderson}(1992)}]{anderson1992experimental}%
  \BibitemOpen
  \bibfield  {author} {\bibinfo {author} {\bibfnamefont {P.~.~W.}\ \bibnamefont {Anderson}},\ }\bibfield  {title} {\bibinfo {title} {Experimental constraints on the theory of high-t c superconductivity},\ }\href@noop {} {\bibfield  {journal} {\bibinfo  {journal} {Science}\ }\textbf {\bibinfo {volume} {256}},\ \bibinfo {pages} {1526} (\bibinfo {year} {1992})}\BibitemShut {NoStop}%
\bibitem [{\citenamefont {Sch{\"a}fer}\ \emph {et~al.}(2015)\citenamefont {Sch{\"a}fer}, \citenamefont {Geles}, \citenamefont {Rost}, \citenamefont {Rohringer}, \citenamefont {Arrigoni}, \citenamefont {Held}, \citenamefont {Bl{\"u}mer}, \citenamefont {Aichhorn},\ and\ \citenamefont {Toschi}}]{schafer2015fate}%
  \BibitemOpen
  \bibfield  {author} {\bibinfo {author} {\bibfnamefont {T.}~\bibnamefont {Sch{\"a}fer}}, \bibinfo {author} {\bibfnamefont {F.}~\bibnamefont {Geles}}, \bibinfo {author} {\bibfnamefont {D.}~\bibnamefont {Rost}}, \bibinfo {author} {\bibfnamefont {G.}~\bibnamefont {Rohringer}}, \bibinfo {author} {\bibfnamefont {E.}~\bibnamefont {Arrigoni}}, \bibinfo {author} {\bibfnamefont {K.}~\bibnamefont {Held}}, \bibinfo {author} {\bibfnamefont {N.}~\bibnamefont {Bl{\"u}mer}}, \bibinfo {author} {\bibfnamefont {M.}~\bibnamefont {Aichhorn}},\ and\ \bibinfo {author} {\bibfnamefont {A.}~\bibnamefont {Toschi}},\ }\bibfield  {title} {\bibinfo {title} {Fate of the false mott-hubbard transition in two dimensions},\ }\href@noop {} {\bibfield  {journal} {\bibinfo  {journal} {Physical Review B}\ }\textbf {\bibinfo {volume} {91}},\ \bibinfo {pages} {125109} (\bibinfo {year} {2015})}\BibitemShut {NoStop}%
\bibitem [{\citenamefont {Schrieffer}\ \emph {et~al.}(1989)\citenamefont {Schrieffer}, \citenamefont {Wen},\ and\ \citenamefont {Zhang}}]{schrieffer1989dynamic}%
  \BibitemOpen
  \bibfield  {author} {\bibinfo {author} {\bibfnamefont {J.}~\bibnamefont {Schrieffer}}, \bibinfo {author} {\bibfnamefont {X.}~\bibnamefont {Wen}},\ and\ \bibinfo {author} {\bibfnamefont {S.}~\bibnamefont {Zhang}},\ }\bibfield  {title} {\bibinfo {title} {Dynamic spin fluctuations and the bag mechanism of high-t c superconductivity},\ }\href@noop {} {\bibfield  {journal} {\bibinfo  {journal} {Physical Review B}\ }\textbf {\bibinfo {volume} {39}},\ \bibinfo {pages} {11663} (\bibinfo {year} {1989})}\BibitemShut {NoStop}%
\bibitem [{\citenamefont {Paiva}\ \emph {et~al.}(2010)\citenamefont {Paiva}, \citenamefont {Scalettar}, \citenamefont {Randeria},\ and\ \citenamefont {Trivedi}}]{paiva2010fermions}%
  \BibitemOpen
  \bibfield  {author} {\bibinfo {author} {\bibfnamefont {T.}~\bibnamefont {Paiva}}, \bibinfo {author} {\bibfnamefont {R.}~\bibnamefont {Scalettar}}, \bibinfo {author} {\bibfnamefont {M.}~\bibnamefont {Randeria}},\ and\ \bibinfo {author} {\bibfnamefont {N.}~\bibnamefont {Trivedi}},\ }\bibfield  {title} {\bibinfo {title} {Fermions in 2d optical lattices: temperature and entropy scales for observing antiferromagnetism and superfluidity},\ }\href@noop {} {\bibfield  {journal} {\bibinfo  {journal} {Physical Review Letters}\ }\textbf {\bibinfo {volume} {104}},\ \bibinfo {pages} {066406} (\bibinfo {year} {2010})}\BibitemShut {NoStop}%
\bibitem [{\citenamefont {Borejsza}\ and\ \citenamefont {Dupuis}(2004)}]{borejsza2004antiferromagnetism}%
  \BibitemOpen
  \bibfield  {author} {\bibinfo {author} {\bibfnamefont {K.}~\bibnamefont {Borejsza}}\ and\ \bibinfo {author} {\bibfnamefont {N.}~\bibnamefont {Dupuis}},\ }\bibfield  {title} {\bibinfo {title} {Antiferromagnetism and single-particle properties in the two-dimensional half-filled hubbard model: A nonlinear sigma model approach},\ }\href@noop {} {\bibfield  {journal} {\bibinfo  {journal} {Physical Review B}\ }\textbf {\bibinfo {volume} {69}},\ \bibinfo {pages} {085119} (\bibinfo {year} {2004})}\BibitemShut {NoStop}%
\bibitem [{\citenamefont {Bloch}\ \emph {et~al.}(2008)\citenamefont {Bloch}, \citenamefont {Dalibard},\ and\ \citenamefont {Zwerger}}]{bloch2008many}%
  \BibitemOpen
  \bibfield  {author} {\bibinfo {author} {\bibfnamefont {I.}~\bibnamefont {Bloch}}, \bibinfo {author} {\bibfnamefont {J.}~\bibnamefont {Dalibard}},\ and\ \bibinfo {author} {\bibfnamefont {W.}~\bibnamefont {Zwerger}},\ }\bibfield  {title} {\bibinfo {title} {Many-body physics with ultracold gases},\ }\href@noop {} {\bibfield  {journal} {\bibinfo  {journal} {Reviews of modern physics}\ }\textbf {\bibinfo {volume} {80}},\ \bibinfo {pages} {885} (\bibinfo {year} {2008})}\BibitemShut {NoStop}%
\bibitem [{\citenamefont {Bloch}(2005)}]{bloch2005ultracold}%
  \BibitemOpen
  \bibfield  {author} {\bibinfo {author} {\bibfnamefont {I.}~\bibnamefont {Bloch}},\ }\bibfield  {title} {\bibinfo {title} {Ultracold quantum gases in optical lattices},\ }\href@noop {} {\bibfield  {journal} {\bibinfo  {journal} {Nature physics}\ }\textbf {\bibinfo {volume} {1}},\ \bibinfo {pages} {23} (\bibinfo {year} {2005})}\BibitemShut {NoStop}%
\bibitem [{\citenamefont {Bloch}\ \emph {et~al.}(2012)\citenamefont {Bloch}, \citenamefont {Dalibard},\ and\ \citenamefont {Nascimbene}}]{bloch2012quantum}%
  \BibitemOpen
  \bibfield  {author} {\bibinfo {author} {\bibfnamefont {I.}~\bibnamefont {Bloch}}, \bibinfo {author} {\bibfnamefont {J.}~\bibnamefont {Dalibard}},\ and\ \bibinfo {author} {\bibfnamefont {S.}~\bibnamefont {Nascimbene}},\ }\bibfield  {title} {\bibinfo {title} {Quantum simulations with ultracold quantum gases},\ }\href@noop {} {\bibfield  {journal} {\bibinfo  {journal} {Nature Physics}\ }\textbf {\bibinfo {volume} {8}},\ \bibinfo {pages} {267} (\bibinfo {year} {2012})}\BibitemShut {NoStop}%
\bibitem [{\citenamefont {Bakr}\ \emph {et~al.}(2009)\citenamefont {Bakr}, \citenamefont {Gillen}, \citenamefont {Peng}, \citenamefont {F{\"o}lling},\ and\ \citenamefont {Greiner}}]{bakr2009quantum}%
  \BibitemOpen
  \bibfield  {author} {\bibinfo {author} {\bibfnamefont {W.~S.}\ \bibnamefont {Bakr}}, \bibinfo {author} {\bibfnamefont {J.~I.}\ \bibnamefont {Gillen}}, \bibinfo {author} {\bibfnamefont {A.}~\bibnamefont {Peng}}, \bibinfo {author} {\bibfnamefont {S.}~\bibnamefont {F{\"o}lling}},\ and\ \bibinfo {author} {\bibfnamefont {M.}~\bibnamefont {Greiner}},\ }\bibfield  {title} {\bibinfo {title} {A quantum gas microscope for detecting single atoms in a hubbard-regime optical lattice},\ }\href@noop {} {\bibfield  {journal} {\bibinfo  {journal} {Nature}\ }\textbf {\bibinfo {volume} {462}},\ \bibinfo {pages} {74} (\bibinfo {year} {2009})}\BibitemShut {NoStop}%
\bibitem [{\citenamefont {Aidelsburger}\ \emph {et~al.}(2013)\citenamefont {Aidelsburger}, \citenamefont {Atala}, \citenamefont {Lohse}, \citenamefont {Barreiro}, \citenamefont {Paredes},\ and\ \citenamefont {Bloch}}]{aidelsburger2013realization}%
  \BibitemOpen
  \bibfield  {author} {\bibinfo {author} {\bibfnamefont {M.}~\bibnamefont {Aidelsburger}}, \bibinfo {author} {\bibfnamefont {M.}~\bibnamefont {Atala}}, \bibinfo {author} {\bibfnamefont {M.}~\bibnamefont {Lohse}}, \bibinfo {author} {\bibfnamefont {J.~T.}\ \bibnamefont {Barreiro}}, \bibinfo {author} {\bibfnamefont {B.}~\bibnamefont {Paredes}},\ and\ \bibinfo {author} {\bibfnamefont {I.}~\bibnamefont {Bloch}},\ }\bibfield  {title} {\bibinfo {title} {Realization of the hofstadter hamiltonian with ultracold atoms in optical lattices},\ }\href@noop {} {\bibfield  {journal} {\bibinfo  {journal} {Physical review letters}\ }\textbf {\bibinfo {volume} {111}},\ \bibinfo {pages} {185301} (\bibinfo {year} {2013})}\BibitemShut {NoStop}%
\bibitem [{\citenamefont {Gross}\ and\ \citenamefont {Bloch}(2017)}]{gross2017quantum}%
  \BibitemOpen
  \bibfield  {author} {\bibinfo {author} {\bibfnamefont {C.}~\bibnamefont {Gross}}\ and\ \bibinfo {author} {\bibfnamefont {I.}~\bibnamefont {Bloch}},\ }\bibfield  {title} {\bibinfo {title} {Quantum simulations with ultracold atoms in optical lattices},\ }\href@noop {} {\bibfield  {journal} {\bibinfo  {journal} {Science}\ }\textbf {\bibinfo {volume} {357}},\ \bibinfo {pages} {995} (\bibinfo {year} {2017})}\BibitemShut {NoStop}%
\bibitem [{\citenamefont {Weitenberg}\ \emph {et~al.}(2011)\citenamefont {Weitenberg}, \citenamefont {Endres}, \citenamefont {Sherson}, \citenamefont {Cheneau}, \citenamefont {Schau{\ss}}, \citenamefont {Fukuhara}, \citenamefont {Bloch},\ and\ \citenamefont {Kuhr}}]{weitenberg2011single}%
  \BibitemOpen
  \bibfield  {author} {\bibinfo {author} {\bibfnamefont {C.}~\bibnamefont {Weitenberg}}, \bibinfo {author} {\bibfnamefont {M.}~\bibnamefont {Endres}}, \bibinfo {author} {\bibfnamefont {J.~F.}\ \bibnamefont {Sherson}}, \bibinfo {author} {\bibfnamefont {M.}~\bibnamefont {Cheneau}}, \bibinfo {author} {\bibfnamefont {P.}~\bibnamefont {Schau{\ss}}}, \bibinfo {author} {\bibfnamefont {T.}~\bibnamefont {Fukuhara}}, \bibinfo {author} {\bibfnamefont {I.}~\bibnamefont {Bloch}},\ and\ \bibinfo {author} {\bibfnamefont {S.}~\bibnamefont {Kuhr}},\ }\bibfield  {title} {\bibinfo {title} {Single-spin addressing in an atomic mott insulator},\ }\href@noop {} {\bibfield  {journal} {\bibinfo  {journal} {Nature}\ }\textbf {\bibinfo {volume} {471}},\ \bibinfo {pages} {319} (\bibinfo {year} {2011})}\BibitemShut {NoStop}%
\bibitem [{\citenamefont {Sherson}\ \emph {et~al.}(2010)\citenamefont {Sherson}, \citenamefont {Weitenberg}, \citenamefont {Endres}, \citenamefont {Cheneau}, \citenamefont {Bloch},\ and\ \citenamefont {Kuhr}}]{sherson2010single}%
  \BibitemOpen
  \bibfield  {author} {\bibinfo {author} {\bibfnamefont {J.~F.}\ \bibnamefont {Sherson}}, \bibinfo {author} {\bibfnamefont {C.}~\bibnamefont {Weitenberg}}, \bibinfo {author} {\bibfnamefont {M.}~\bibnamefont {Endres}}, \bibinfo {author} {\bibfnamefont {M.}~\bibnamefont {Cheneau}}, \bibinfo {author} {\bibfnamefont {I.}~\bibnamefont {Bloch}},\ and\ \bibinfo {author} {\bibfnamefont {S.}~\bibnamefont {Kuhr}},\ }\bibfield  {title} {\bibinfo {title} {Single-atom-resolved fluorescence imaging of an atomic mott insulator},\ }\href@noop {} {\bibfield  {journal} {\bibinfo  {journal} {Nature}\ }\textbf {\bibinfo {volume} {467}},\ \bibinfo {pages} {68} (\bibinfo {year} {2010})}\BibitemShut {NoStop}%
\bibitem [{\citenamefont {Greiner}\ \emph {et~al.}(2002)\citenamefont {Greiner}, \citenamefont {Mandel}, \citenamefont {Esslinger}, \citenamefont {H{\"a}nsch},\ and\ \citenamefont {Bloch}}]{greiner2002quantum}%
  \BibitemOpen
  \bibfield  {author} {\bibinfo {author} {\bibfnamefont {M.}~\bibnamefont {Greiner}}, \bibinfo {author} {\bibfnamefont {O.}~\bibnamefont {Mandel}}, \bibinfo {author} {\bibfnamefont {T.}~\bibnamefont {Esslinger}}, \bibinfo {author} {\bibfnamefont {T.~W.}\ \bibnamefont {H{\"a}nsch}},\ and\ \bibinfo {author} {\bibfnamefont {I.}~\bibnamefont {Bloch}},\ }\bibfield  {title} {\bibinfo {title} {Quantum phase transition from a superfluid to a mott insulator in a gas of ultracold atoms},\ }\href@noop {} {\bibfield  {journal} {\bibinfo  {journal} {nature}\ }\textbf {\bibinfo {volume} {415}},\ \bibinfo {pages} {39} (\bibinfo {year} {2002})}\BibitemShut {NoStop}%
\bibitem [{\citenamefont {Aidelsburger}\ \emph {et~al.}(2015)\citenamefont {Aidelsburger}, \citenamefont {Lohse}, \citenamefont {Schweizer}, \citenamefont {Atala}, \citenamefont {Barreiro}, \citenamefont {Nascimbene}, \citenamefont {Cooper}, \citenamefont {Bloch},\ and\ \citenamefont {Goldman}}]{aidelsburger2015measuring}%
  \BibitemOpen
  \bibfield  {author} {\bibinfo {author} {\bibfnamefont {M.}~\bibnamefont {Aidelsburger}}, \bibinfo {author} {\bibfnamefont {M.}~\bibnamefont {Lohse}}, \bibinfo {author} {\bibfnamefont {C.}~\bibnamefont {Schweizer}}, \bibinfo {author} {\bibfnamefont {M.}~\bibnamefont {Atala}}, \bibinfo {author} {\bibfnamefont {J.~T.}\ \bibnamefont {Barreiro}}, \bibinfo {author} {\bibfnamefont {S.}~\bibnamefont {Nascimbene}}, \bibinfo {author} {\bibfnamefont {N.}~\bibnamefont {Cooper}}, \bibinfo {author} {\bibfnamefont {I.}~\bibnamefont {Bloch}},\ and\ \bibinfo {author} {\bibfnamefont {N.}~\bibnamefont {Goldman}},\ }\bibfield  {title} {\bibinfo {title} {Measuring the chern number of hofstadter bands with ultracold bosonic atoms},\ }\href@noop {} {\bibfield  {journal} {\bibinfo  {journal} {Nature Physics}\ }\textbf {\bibinfo {volume} {11}},\ \bibinfo {pages} {162} (\bibinfo {year} {2015})}\BibitemShut {NoStop}%
\bibitem [{\citenamefont {Cocchi}\ \emph {et~al.}(2016)\citenamefont {Cocchi}, \citenamefont {Miller}, \citenamefont {Drewes}, \citenamefont {Koschorreck}, \citenamefont {Pertot}, \citenamefont {Brennecke},\ and\ \citenamefont {K{\"o}hl}}]{cocchi2016equation}%
  \BibitemOpen
  \bibfield  {author} {\bibinfo {author} {\bibfnamefont {E.}~\bibnamefont {Cocchi}}, \bibinfo {author} {\bibfnamefont {L.~A.}\ \bibnamefont {Miller}}, \bibinfo {author} {\bibfnamefont {J.~H.}\ \bibnamefont {Drewes}}, \bibinfo {author} {\bibfnamefont {M.}~\bibnamefont {Koschorreck}}, \bibinfo {author} {\bibfnamefont {D.}~\bibnamefont {Pertot}}, \bibinfo {author} {\bibfnamefont {F.}~\bibnamefont {Brennecke}},\ and\ \bibinfo {author} {\bibfnamefont {M.}~\bibnamefont {K{\"o}hl}},\ }\bibfield  {title} {\bibinfo {title} {Equation of state of the two-dimensional hubbard model},\ }\href@noop {} {\bibfield  {journal} {\bibinfo  {journal} {Physical review letters}\ }\textbf {\bibinfo {volume} {116}},\ \bibinfo {pages} {175301} (\bibinfo {year} {2016})}\BibitemShut {NoStop}%
\bibitem [{\citenamefont {Greif}\ \emph {et~al.}(2016)\citenamefont {Greif}, \citenamefont {Parsons}, \citenamefont {Mazurenko}, \citenamefont {Chiu}, \citenamefont {Blatt}, \citenamefont {Huber}, \citenamefont {Ji},\ and\ \citenamefont {Greiner}}]{greif2016site}%
  \BibitemOpen
  \bibfield  {author} {\bibinfo {author} {\bibfnamefont {D.}~\bibnamefont {Greif}}, \bibinfo {author} {\bibfnamefont {M.~F.}\ \bibnamefont {Parsons}}, \bibinfo {author} {\bibfnamefont {A.}~\bibnamefont {Mazurenko}}, \bibinfo {author} {\bibfnamefont {C.~S.}\ \bibnamefont {Chiu}}, \bibinfo {author} {\bibfnamefont {S.}~\bibnamefont {Blatt}}, \bibinfo {author} {\bibfnamefont {F.}~\bibnamefont {Huber}}, \bibinfo {author} {\bibfnamefont {G.}~\bibnamefont {Ji}},\ and\ \bibinfo {author} {\bibfnamefont {M.}~\bibnamefont {Greiner}},\ }\bibfield  {title} {\bibinfo {title} {Site-resolved imaging of a fermionic mott insulator},\ }\href@noop {} {\bibfield  {journal} {\bibinfo  {journal} {Science}\ }\textbf {\bibinfo {volume} {351}},\ \bibinfo {pages} {953} (\bibinfo {year} {2016})}\BibitemShut {NoStop}%
\bibitem [{\citenamefont {Cheuk}\ \emph {et~al.}(2016)\citenamefont {Cheuk}, \citenamefont {Nichols}, \citenamefont {Lawrence}, \citenamefont {Okan}, \citenamefont {Zhang},\ and\ \citenamefont {Zwierlein}}]{cheuk2016observation}%
  \BibitemOpen
  \bibfield  {author} {\bibinfo {author} {\bibfnamefont {L.~W.}\ \bibnamefont {Cheuk}}, \bibinfo {author} {\bibfnamefont {M.~A.}\ \bibnamefont {Nichols}}, \bibinfo {author} {\bibfnamefont {K.~R.}\ \bibnamefont {Lawrence}}, \bibinfo {author} {\bibfnamefont {M.}~\bibnamefont {Okan}}, \bibinfo {author} {\bibfnamefont {H.}~\bibnamefont {Zhang}},\ and\ \bibinfo {author} {\bibfnamefont {M.~W.}\ \bibnamefont {Zwierlein}},\ }\bibfield  {title} {\bibinfo {title} {Observation of 2d fermionic mott insulators of k 40 with single-site resolution},\ }\href@noop {} {\bibfield  {journal} {\bibinfo  {journal} {Physical review letters}\ }\textbf {\bibinfo {volume} {116}},\ \bibinfo {pages} {235301} (\bibinfo {year} {2016})}\BibitemShut {NoStop}%
\bibitem [{\citenamefont {Drewes}\ \emph {et~al.}(2016)\citenamefont {Drewes}, \citenamefont {Cocchi}, \citenamefont {Miller}, \citenamefont {Chan}, \citenamefont {Pertot}, \citenamefont {Brennecke},\ and\ \citenamefont {K{\"o}hl}}]{drewes2016thermodynamics}%
  \BibitemOpen
  \bibfield  {author} {\bibinfo {author} {\bibfnamefont {J.}~\bibnamefont {Drewes}}, \bibinfo {author} {\bibfnamefont {E.}~\bibnamefont {Cocchi}}, \bibinfo {author} {\bibfnamefont {L.}~\bibnamefont {Miller}}, \bibinfo {author} {\bibfnamefont {C.}~\bibnamefont {Chan}}, \bibinfo {author} {\bibfnamefont {D.}~\bibnamefont {Pertot}}, \bibinfo {author} {\bibfnamefont {F.}~\bibnamefont {Brennecke}},\ and\ \bibinfo {author} {\bibfnamefont {M.}~\bibnamefont {K{\"o}hl}},\ }\bibfield  {title} {\bibinfo {title} {Thermodynamics versus local density fluctuations in the metal--mott-insulator crossover},\ }\href@noop {} {\bibfield  {journal} {\bibinfo  {journal} {Physical review letters}\ }\textbf {\bibinfo {volume} {117}},\ \bibinfo {pages} {135301} (\bibinfo {year} {2016})}\BibitemShut {NoStop}%
\bibitem [{\citenamefont {Parsons}\ \emph {et~al.}(2016)\citenamefont {Parsons}, \citenamefont {Mazurenko}, \citenamefont {Chiu}, \citenamefont {Ji}, \citenamefont {Greif},\ and\ \citenamefont {Greiner}}]{parsons2016site}%
  \BibitemOpen
  \bibfield  {author} {\bibinfo {author} {\bibfnamefont {M.~F.}\ \bibnamefont {Parsons}}, \bibinfo {author} {\bibfnamefont {A.}~\bibnamefont {Mazurenko}}, \bibinfo {author} {\bibfnamefont {C.~S.}\ \bibnamefont {Chiu}}, \bibinfo {author} {\bibfnamefont {G.}~\bibnamefont {Ji}}, \bibinfo {author} {\bibfnamefont {D.}~\bibnamefont {Greif}},\ and\ \bibinfo {author} {\bibfnamefont {M.}~\bibnamefont {Greiner}},\ }\bibfield  {title} {\bibinfo {title} {Site-resolved measurement of the spin-correlation function in the fermi-hubbard model},\ }\href@noop {} {\bibfield  {journal} {\bibinfo  {journal} {Science}\ }\textbf {\bibinfo {volume} {353}},\ \bibinfo {pages} {1253} (\bibinfo {year} {2016})}\BibitemShut {NoStop}%
\bibitem [{\citenamefont {Drewes}\ \emph {et~al.}(2017)\citenamefont {Drewes}, \citenamefont {Miller}, \citenamefont {Cocchi}, \citenamefont {Chan}, \citenamefont {Wurz}, \citenamefont {Gall}, \citenamefont {Pertot}, \citenamefont {Brennecke},\ and\ \citenamefont {K{\"o}hl}}]{drewes2017antiferromagnetic}%
  \BibitemOpen
  \bibfield  {author} {\bibinfo {author} {\bibfnamefont {J.}~\bibnamefont {Drewes}}, \bibinfo {author} {\bibfnamefont {L.}~\bibnamefont {Miller}}, \bibinfo {author} {\bibfnamefont {E.}~\bibnamefont {Cocchi}}, \bibinfo {author} {\bibfnamefont {C.}~\bibnamefont {Chan}}, \bibinfo {author} {\bibfnamefont {N.}~\bibnamefont {Wurz}}, \bibinfo {author} {\bibfnamefont {M.}~\bibnamefont {Gall}}, \bibinfo {author} {\bibfnamefont {D.}~\bibnamefont {Pertot}}, \bibinfo {author} {\bibfnamefont {F.}~\bibnamefont {Brennecke}},\ and\ \bibinfo {author} {\bibfnamefont {M.}~\bibnamefont {K{\"o}hl}},\ }\bibfield  {title} {\bibinfo {title} {Antiferromagnetic correlations in two-dimensional fermionic mott-insulating and metallic phases},\ }\href@noop {} {\bibfield  {journal} {\bibinfo  {journal} {Physical review letters}\ }\textbf {\bibinfo {volume} {118}},\ \bibinfo {pages} {170401} (\bibinfo {year} {2017})}\BibitemShut {NoStop}%
\bibitem [{\citenamefont {Brown}\ \emph {et~al.}(2017)\citenamefont {Brown}, \citenamefont {Mitra}, \citenamefont {Guardado-Sanchez}, \citenamefont {Schau{\ss}}, \citenamefont {Kondov}, \citenamefont {Khatami}, \citenamefont {Paiva}, \citenamefont {Trivedi}, \citenamefont {Huse},\ and\ \citenamefont {Bakr}}]{brown2017spin}%
  \BibitemOpen
  \bibfield  {author} {\bibinfo {author} {\bibfnamefont {P.~T.}\ \bibnamefont {Brown}}, \bibinfo {author} {\bibfnamefont {D.}~\bibnamefont {Mitra}}, \bibinfo {author} {\bibfnamefont {E.}~\bibnamefont {Guardado-Sanchez}}, \bibinfo {author} {\bibfnamefont {P.}~\bibnamefont {Schau{\ss}}}, \bibinfo {author} {\bibfnamefont {S.~S.}\ \bibnamefont {Kondov}}, \bibinfo {author} {\bibfnamefont {E.}~\bibnamefont {Khatami}}, \bibinfo {author} {\bibfnamefont {T.}~\bibnamefont {Paiva}}, \bibinfo {author} {\bibfnamefont {N.}~\bibnamefont {Trivedi}}, \bibinfo {author} {\bibfnamefont {D.~A.}\ \bibnamefont {Huse}},\ and\ \bibinfo {author} {\bibfnamefont {W.~S.}\ \bibnamefont {Bakr}},\ }\bibfield  {title} {\bibinfo {title} {Spin-imbalance in a 2d fermi-hubbard system},\ }\href@noop {} {\bibfield  {journal} {\bibinfo  {journal} {Science}\ }\textbf {\bibinfo {volume} {357}},\ \bibinfo {pages} {1385} (\bibinfo {year} {2017})}\BibitemShut {NoStop}%
\bibitem [{\citenamefont {Nichols}\ \emph {et~al.}(2019)\citenamefont {Nichols}, \citenamefont {Cheuk}, \citenamefont {Okan}, \citenamefont {Hartke}, \citenamefont {Mendez}, \citenamefont {Senthil}, \citenamefont {Khatami}, \citenamefont {Zhang},\ and\ \citenamefont {Zwierlein}}]{nichols2019spin}%
  \BibitemOpen
  \bibfield  {author} {\bibinfo {author} {\bibfnamefont {M.~A.}\ \bibnamefont {Nichols}}, \bibinfo {author} {\bibfnamefont {L.~W.}\ \bibnamefont {Cheuk}}, \bibinfo {author} {\bibfnamefont {M.}~\bibnamefont {Okan}}, \bibinfo {author} {\bibfnamefont {T.~R.}\ \bibnamefont {Hartke}}, \bibinfo {author} {\bibfnamefont {E.}~\bibnamefont {Mendez}}, \bibinfo {author} {\bibfnamefont {T.}~\bibnamefont {Senthil}}, \bibinfo {author} {\bibfnamefont {E.}~\bibnamefont {Khatami}}, \bibinfo {author} {\bibfnamefont {H.}~\bibnamefont {Zhang}},\ and\ \bibinfo {author} {\bibfnamefont {M.~W.}\ \bibnamefont {Zwierlein}},\ }\bibfield  {title} {\bibinfo {title} {Spin transport in a mott insulator of ultracold fermions},\ }\href@noop {} {\bibfield  {journal} {\bibinfo  {journal} {Science}\ }\textbf {\bibinfo {volume} {363}},\ \bibinfo {pages} {383} (\bibinfo {year} {2019})}\BibitemShut {NoStop}%
\bibitem [{\citenamefont {Brown}\ \emph {et~al.}(2019)\citenamefont {Brown}, \citenamefont {Mitra}, \citenamefont {Guardado-Sanchez}, \citenamefont {Nourafkan}, \citenamefont {Reymbaut}, \citenamefont {H{\'e}bert}, \citenamefont {Bergeron}, \citenamefont {Tremblay}, \citenamefont {Kokalj}, \citenamefont {Huse} \emph {et~al.}}]{brown2019bad}%
  \BibitemOpen
  \bibfield  {author} {\bibinfo {author} {\bibfnamefont {P.~T.}\ \bibnamefont {Brown}}, \bibinfo {author} {\bibfnamefont {D.}~\bibnamefont {Mitra}}, \bibinfo {author} {\bibfnamefont {E.}~\bibnamefont {Guardado-Sanchez}}, \bibinfo {author} {\bibfnamefont {R.}~\bibnamefont {Nourafkan}}, \bibinfo {author} {\bibfnamefont {A.}~\bibnamefont {Reymbaut}}, \bibinfo {author} {\bibfnamefont {C.-D.}\ \bibnamefont {H{\'e}bert}}, \bibinfo {author} {\bibfnamefont {S.}~\bibnamefont {Bergeron}}, \bibinfo {author} {\bibfnamefont {A.-M.}\ \bibnamefont {Tremblay}}, \bibinfo {author} {\bibfnamefont {J.}~\bibnamefont {Kokalj}}, \bibinfo {author} {\bibfnamefont {D.~A.}\ \bibnamefont {Huse}}, \emph {et~al.},\ }\bibfield  {title} {\bibinfo {title} {Bad metallic transport in a cold atom fermi-hubbard system},\ }\href@noop {} {\bibfield  {journal} {\bibinfo  {journal} {Science}\ }\textbf {\bibinfo {volume} {363}},\ \bibinfo {pages} {379} (\bibinfo {year} {2019})}\BibitemShut {NoStop}%
\bibitem [{\citenamefont {Guardado-Sanchez}\ \emph {et~al.}(2020)\citenamefont {Guardado-Sanchez}, \citenamefont {Morningstar}, \citenamefont {Spar}, \citenamefont {Brown}, \citenamefont {Huse},\ and\ \citenamefont {Bakr}}]{guardado2020subdiffusion}%
  \BibitemOpen
  \bibfield  {author} {\bibinfo {author} {\bibfnamefont {E.}~\bibnamefont {Guardado-Sanchez}}, \bibinfo {author} {\bibfnamefont {A.}~\bibnamefont {Morningstar}}, \bibinfo {author} {\bibfnamefont {B.~M.}\ \bibnamefont {Spar}}, \bibinfo {author} {\bibfnamefont {P.~T.}\ \bibnamefont {Brown}}, \bibinfo {author} {\bibfnamefont {D.~A.}\ \bibnamefont {Huse}},\ and\ \bibinfo {author} {\bibfnamefont {W.~S.}\ \bibnamefont {Bakr}},\ }\bibfield  {title} {\bibinfo {title} {Subdiffusion and heat transport in a tilted two-dimensional fermi-hubbard system},\ }\href@noop {} {\bibfield  {journal} {\bibinfo  {journal} {Physical Review X}\ }\textbf {\bibinfo {volume} {10}},\ \bibinfo {pages} {011042} (\bibinfo {year} {2020})}\BibitemShut {NoStop}%
\bibitem [{\citenamefont {Mazurenko}\ \emph {et~al.}(2017)\citenamefont {Mazurenko}, \citenamefont {Chiu}, \citenamefont {Ji}, \citenamefont {Parsons}, \citenamefont {Kan{\'a}sz-Nagy}, \citenamefont {Schmidt}, \citenamefont {Grusdt}, \citenamefont {Demler}, \citenamefont {Greif},\ and\ \citenamefont {Greiner}}]{mazurenko2017cold}%
  \BibitemOpen
  \bibfield  {author} {\bibinfo {author} {\bibfnamefont {A.}~\bibnamefont {Mazurenko}}, \bibinfo {author} {\bibfnamefont {C.~S.}\ \bibnamefont {Chiu}}, \bibinfo {author} {\bibfnamefont {G.}~\bibnamefont {Ji}}, \bibinfo {author} {\bibfnamefont {M.~F.}\ \bibnamefont {Parsons}}, \bibinfo {author} {\bibfnamefont {M.}~\bibnamefont {Kan{\'a}sz-Nagy}}, \bibinfo {author} {\bibfnamefont {R.}~\bibnamefont {Schmidt}}, \bibinfo {author} {\bibfnamefont {F.}~\bibnamefont {Grusdt}}, \bibinfo {author} {\bibfnamefont {E.}~\bibnamefont {Demler}}, \bibinfo {author} {\bibfnamefont {D.}~\bibnamefont {Greif}},\ and\ \bibinfo {author} {\bibfnamefont {M.}~\bibnamefont {Greiner}},\ }\bibfield  {title} {\bibinfo {title} {A cold-atom fermi--hubbard antiferromagnet},\ }\href@noop {} {\bibfield  {journal} {\bibinfo  {journal} {Nature}\ }\textbf {\bibinfo {volume} {545}},\ \bibinfo {pages} {462} (\bibinfo {year} {2017})}\BibitemShut {NoStop}%
\bibitem [{\citenamefont {Koepsell}\ \emph {et~al.}(2019)\citenamefont {Koepsell}, \citenamefont {Vijayan}, \citenamefont {Sompet}, \citenamefont {Grusdt}, \citenamefont {Hilker}, \citenamefont {Demler}, \citenamefont {Salomon}, \citenamefont {Bloch},\ and\ \citenamefont {Gross}}]{koepsell2019imaging}%
  \BibitemOpen
  \bibfield  {author} {\bibinfo {author} {\bibfnamefont {J.}~\bibnamefont {Koepsell}}, \bibinfo {author} {\bibfnamefont {J.}~\bibnamefont {Vijayan}}, \bibinfo {author} {\bibfnamefont {P.}~\bibnamefont {Sompet}}, \bibinfo {author} {\bibfnamefont {F.}~\bibnamefont {Grusdt}}, \bibinfo {author} {\bibfnamefont {T.~A.}\ \bibnamefont {Hilker}}, \bibinfo {author} {\bibfnamefont {E.}~\bibnamefont {Demler}}, \bibinfo {author} {\bibfnamefont {G.}~\bibnamefont {Salomon}}, \bibinfo {author} {\bibfnamefont {I.}~\bibnamefont {Bloch}},\ and\ \bibinfo {author} {\bibfnamefont {C.}~\bibnamefont {Gross}},\ }\bibfield  {title} {\bibinfo {title} {Imaging magnetic polarons in the doped fermi--hubbard model},\ }\href@noop {} {\bibfield  {journal} {\bibinfo  {journal} {Nature}\ }\textbf {\bibinfo {volume} {572}},\ \bibinfo {pages} {358} (\bibinfo {year} {2019})}\BibitemShut {NoStop}%
\bibitem [{\citenamefont {Boll}\ \emph {et~al.}(2016)\citenamefont {Boll}, \citenamefont {Hilker}, \citenamefont {Salomon}, \citenamefont {Omran}, \citenamefont {Nespolo}, \citenamefont {Pollet}, \citenamefont {Bloch},\ and\ \citenamefont {Gross}}]{boll2016spin}%
  \BibitemOpen
  \bibfield  {author} {\bibinfo {author} {\bibfnamefont {M.}~\bibnamefont {Boll}}, \bibinfo {author} {\bibfnamefont {T.~A.}\ \bibnamefont {Hilker}}, \bibinfo {author} {\bibfnamefont {G.}~\bibnamefont {Salomon}}, \bibinfo {author} {\bibfnamefont {A.}~\bibnamefont {Omran}}, \bibinfo {author} {\bibfnamefont {J.}~\bibnamefont {Nespolo}}, \bibinfo {author} {\bibfnamefont {L.}~\bibnamefont {Pollet}}, \bibinfo {author} {\bibfnamefont {I.}~\bibnamefont {Bloch}},\ and\ \bibinfo {author} {\bibfnamefont {C.}~\bibnamefont {Gross}},\ }\bibfield  {title} {\bibinfo {title} {Spin-and density-resolved microscopy of antiferromagnetic correlations in fermi-hubbard chains},\ }\href@noop {} {\bibfield  {journal} {\bibinfo  {journal} {Science}\ }\textbf {\bibinfo {volume} {353}},\ \bibinfo {pages} {1257} (\bibinfo {year} {2016})}\BibitemShut {NoStop}%
\bibitem [{\citenamefont {Hartke}\ \emph {et~al.}(2020)\citenamefont {Hartke}, \citenamefont {Oreg}, \citenamefont {Jia},\ and\ \citenamefont {Zwierlein}}]{hartke2020doublon}%
  \BibitemOpen
  \bibfield  {author} {\bibinfo {author} {\bibfnamefont {T.}~\bibnamefont {Hartke}}, \bibinfo {author} {\bibfnamefont {B.}~\bibnamefont {Oreg}}, \bibinfo {author} {\bibfnamefont {N.}~\bibnamefont {Jia}},\ and\ \bibinfo {author} {\bibfnamefont {M.}~\bibnamefont {Zwierlein}},\ }\bibfield  {title} {\bibinfo {title} {Doublon-hole correlations and fluctuation thermometry in a fermi-hubbard gas},\ }\href@noop {} {\bibfield  {journal} {\bibinfo  {journal} {Physical Review Letters}\ }\textbf {\bibinfo {volume} {125}},\ \bibinfo {pages} {113601} (\bibinfo {year} {2020})}\BibitemShut {NoStop}%
\bibitem [{\citenamefont {Koepsell}\ \emph {et~al.}(2020)\citenamefont {Koepsell}, \citenamefont {Hirthe}, \citenamefont {Bourgund}, \citenamefont {Sompet}, \citenamefont {Vijayan}, \citenamefont {Salomon}, \citenamefont {Gross},\ and\ \citenamefont {Bloch}}]{koepsell2020robust}%
  \BibitemOpen
  \bibfield  {author} {\bibinfo {author} {\bibfnamefont {J.}~\bibnamefont {Koepsell}}, \bibinfo {author} {\bibfnamefont {S.}~\bibnamefont {Hirthe}}, \bibinfo {author} {\bibfnamefont {D.}~\bibnamefont {Bourgund}}, \bibinfo {author} {\bibfnamefont {P.}~\bibnamefont {Sompet}}, \bibinfo {author} {\bibfnamefont {J.}~\bibnamefont {Vijayan}}, \bibinfo {author} {\bibfnamefont {G.}~\bibnamefont {Salomon}}, \bibinfo {author} {\bibfnamefont {C.}~\bibnamefont {Gross}},\ and\ \bibinfo {author} {\bibfnamefont {I.}~\bibnamefont {Bloch}},\ }\bibfield  {title} {\bibinfo {title} {Robust bilayer charge pumping for spin-and density-resolved quantum gas microscopy},\ }\href@noop {} {\bibfield  {journal} {\bibinfo  {journal} {Physical Review Letters}\ }\textbf {\bibinfo {volume} {125}},\ \bibinfo {pages} {010403} (\bibinfo {year} {2020})}\BibitemShut {NoStop}%
\bibitem [{\citenamefont {Hirthe}\ \emph {et~al.}(2023)\citenamefont {Hirthe}, \citenamefont {Chalopin}, \citenamefont {Bourgund}, \citenamefont {Bojovi{\'c}}, \citenamefont {Bohrdt}, \citenamefont {Demler}, \citenamefont {Grusdt}, \citenamefont {Bloch},\ and\ \citenamefont {Hilker}}]{hirthe2023magnetically}%
  \BibitemOpen
  \bibfield  {author} {\bibinfo {author} {\bibfnamefont {S.}~\bibnamefont {Hirthe}}, \bibinfo {author} {\bibfnamefont {T.}~\bibnamefont {Chalopin}}, \bibinfo {author} {\bibfnamefont {D.}~\bibnamefont {Bourgund}}, \bibinfo {author} {\bibfnamefont {P.}~\bibnamefont {Bojovi{\'c}}}, \bibinfo {author} {\bibfnamefont {A.}~\bibnamefont {Bohrdt}}, \bibinfo {author} {\bibfnamefont {E.}~\bibnamefont {Demler}}, \bibinfo {author} {\bibfnamefont {F.}~\bibnamefont {Grusdt}}, \bibinfo {author} {\bibfnamefont {I.}~\bibnamefont {Bloch}},\ and\ \bibinfo {author} {\bibfnamefont {T.~A.}\ \bibnamefont {Hilker}},\ }\bibfield  {title} {\bibinfo {title} {Magnetically mediated hole pairing in fermionic ladders of ultracold atoms},\ }\href@noop {} {\bibfield  {journal} {\bibinfo  {journal} {Nature}\ }\textbf {\bibinfo {volume} {613}},\ \bibinfo {pages} {463} (\bibinfo {year} {2023})}\BibitemShut {NoStop}%
\bibitem [{\citenamefont {Salomon}\ \emph {et~al.}(2019)\citenamefont {Salomon}, \citenamefont {Koepsell}, \citenamefont {Vijayan}, \citenamefont {Hilker}, \citenamefont {Nespolo}, \citenamefont {Pollet}, \citenamefont {Bloch},\ and\ \citenamefont {Gross}}]{salomon2019direct}%
  \BibitemOpen
  \bibfield  {author} {\bibinfo {author} {\bibfnamefont {G.}~\bibnamefont {Salomon}}, \bibinfo {author} {\bibfnamefont {J.}~\bibnamefont {Koepsell}}, \bibinfo {author} {\bibfnamefont {J.}~\bibnamefont {Vijayan}}, \bibinfo {author} {\bibfnamefont {T.~A.}\ \bibnamefont {Hilker}}, \bibinfo {author} {\bibfnamefont {J.}~\bibnamefont {Nespolo}}, \bibinfo {author} {\bibfnamefont {L.}~\bibnamefont {Pollet}}, \bibinfo {author} {\bibfnamefont {I.}~\bibnamefont {Bloch}},\ and\ \bibinfo {author} {\bibfnamefont {C.}~\bibnamefont {Gross}},\ }\bibfield  {title} {\bibinfo {title} {Direct observation of incommensurate magnetism in hubbard chains},\ }\href@noop {} {\bibfield  {journal} {\bibinfo  {journal} {Nature}\ }\textbf {\bibinfo {volume} {565}},\ \bibinfo {pages} {56} (\bibinfo {year} {2019})}\BibitemShut {NoStop}%
\bibitem [{\citenamefont {Vijayan}\ \emph {et~al.}(2020)\citenamefont {Vijayan}, \citenamefont {Sompet}, \citenamefont {Salomon}, \citenamefont {Koepsell}, \citenamefont {Hirthe}, \citenamefont {Bohrdt}, \citenamefont {Grusdt}, \citenamefont {Bloch},\ and\ \citenamefont {Gross}}]{vijayan2020time}%
  \BibitemOpen
  \bibfield  {author} {\bibinfo {author} {\bibfnamefont {J.}~\bibnamefont {Vijayan}}, \bibinfo {author} {\bibfnamefont {P.}~\bibnamefont {Sompet}}, \bibinfo {author} {\bibfnamefont {G.}~\bibnamefont {Salomon}}, \bibinfo {author} {\bibfnamefont {J.}~\bibnamefont {Koepsell}}, \bibinfo {author} {\bibfnamefont {S.}~\bibnamefont {Hirthe}}, \bibinfo {author} {\bibfnamefont {A.}~\bibnamefont {Bohrdt}}, \bibinfo {author} {\bibfnamefont {F.}~\bibnamefont {Grusdt}}, \bibinfo {author} {\bibfnamefont {I.}~\bibnamefont {Bloch}},\ and\ \bibinfo {author} {\bibfnamefont {C.}~\bibnamefont {Gross}},\ }\bibfield  {title} {\bibinfo {title} {Time-resolved observation of spin-charge deconfinement in fermionic hubbard chains},\ }\href@noop {} {\bibfield  {journal} {\bibinfo  {journal} {Science}\ }\textbf {\bibinfo {volume} {367}},\ \bibinfo {pages} {186} (\bibinfo {year} {2020})}\BibitemShut {NoStop}%
\bibitem [{\citenamefont {Koepsell}\ \emph {et~al.}(2021)\citenamefont {Koepsell}, \citenamefont {Bourgund}, \citenamefont {Sompet}, \citenamefont {Hirthe}, \citenamefont {Bohrdt}, \citenamefont {Wang}, \citenamefont {Grusdt}, \citenamefont {Demler}, \citenamefont {Salomon}, \citenamefont {Gross} \emph {et~al.}}]{koepsell2021microscopic}%
  \BibitemOpen
  \bibfield  {author} {\bibinfo {author} {\bibfnamefont {J.}~\bibnamefont {Koepsell}}, \bibinfo {author} {\bibfnamefont {D.}~\bibnamefont {Bourgund}}, \bibinfo {author} {\bibfnamefont {P.}~\bibnamefont {Sompet}}, \bibinfo {author} {\bibfnamefont {S.}~\bibnamefont {Hirthe}}, \bibinfo {author} {\bibfnamefont {A.}~\bibnamefont {Bohrdt}}, \bibinfo {author} {\bibfnamefont {Y.}~\bibnamefont {Wang}}, \bibinfo {author} {\bibfnamefont {F.}~\bibnamefont {Grusdt}}, \bibinfo {author} {\bibfnamefont {E.}~\bibnamefont {Demler}}, \bibinfo {author} {\bibfnamefont {G.}~\bibnamefont {Salomon}}, \bibinfo {author} {\bibfnamefont {C.}~\bibnamefont {Gross}}, \emph {et~al.},\ }\bibfield  {title} {\bibinfo {title} {Microscopic evolution of doped mott insulators from polaronic metal to fermi liquid},\ }\href@noop {} {\bibfield  {journal} {\bibinfo  {journal} {Science}\ }\textbf {\bibinfo {volume} {374}},\ \bibinfo {pages} {82} (\bibinfo {year} {2021})}\BibitemShut {NoStop}%
\bibitem [{\citenamefont {Sompet}\ \emph {et~al.}(2022)\citenamefont {Sompet}, \citenamefont {Hirthe}, \citenamefont {Bourgund}, \citenamefont {Chalopin}, \citenamefont {Bibo}, \citenamefont {Koepsell}, \citenamefont {Bojovi{\'c}}, \citenamefont {Verresen}, \citenamefont {Pollmann}, \citenamefont {Salomon} \emph {et~al.}}]{sompet2022realizing}%
  \BibitemOpen
  \bibfield  {author} {\bibinfo {author} {\bibfnamefont {P.}~\bibnamefont {Sompet}}, \bibinfo {author} {\bibfnamefont {S.}~\bibnamefont {Hirthe}}, \bibinfo {author} {\bibfnamefont {D.}~\bibnamefont {Bourgund}}, \bibinfo {author} {\bibfnamefont {T.}~\bibnamefont {Chalopin}}, \bibinfo {author} {\bibfnamefont {J.}~\bibnamefont {Bibo}}, \bibinfo {author} {\bibfnamefont {J.}~\bibnamefont {Koepsell}}, \bibinfo {author} {\bibfnamefont {P.}~\bibnamefont {Bojovi{\'c}}}, \bibinfo {author} {\bibfnamefont {R.}~\bibnamefont {Verresen}}, \bibinfo {author} {\bibfnamefont {F.}~\bibnamefont {Pollmann}}, \bibinfo {author} {\bibfnamefont {G.}~\bibnamefont {Salomon}}, \emph {et~al.},\ }\bibfield  {title} {\bibinfo {title} {Realizing the symmetry-protected haldane phase in fermi--hubbard ladders},\ }\href@noop {} {\bibfield  {journal} {\bibinfo  {journal} {Nature}\ }\textbf {\bibinfo {volume} {606}},\ \bibinfo {pages} {484} (\bibinfo {year} {2022})}\BibitemShut {NoStop}%
\bibitem [{\citenamefont {Chiu}\ \emph {et~al.}(2018)\citenamefont {Chiu}, \citenamefont {Ji}, \citenamefont {Mazurenko}, \citenamefont {Greif},\ and\ \citenamefont {Greiner}}]{chiu2018quantum}%
  \BibitemOpen
  \bibfield  {author} {\bibinfo {author} {\bibfnamefont {C.~S.}\ \bibnamefont {Chiu}}, \bibinfo {author} {\bibfnamefont {G.}~\bibnamefont {Ji}}, \bibinfo {author} {\bibfnamefont {A.}~\bibnamefont {Mazurenko}}, \bibinfo {author} {\bibfnamefont {D.}~\bibnamefont {Greif}},\ and\ \bibinfo {author} {\bibfnamefont {M.}~\bibnamefont {Greiner}},\ }\bibfield  {title} {\bibinfo {title} {Quantum state engineering of a hubbard system with ultracold fermions},\ }\href@noop {} {\bibfield  {journal} {\bibinfo  {journal} {Physical review letters}\ }\textbf {\bibinfo {volume} {120}},\ \bibinfo {pages} {243201} (\bibinfo {year} {2018})}\BibitemShut {NoStop}%
\bibitem [{\citenamefont {Kale}\ \emph {et~al.}(2022)\citenamefont {Kale}, \citenamefont {Huhn}, \citenamefont {Xu}, \citenamefont {Kendrick}, \citenamefont {Lebrat}, \citenamefont {Chiu}, \citenamefont {Ji}, \citenamefont {Grusdt}, \citenamefont {Bohrdt},\ and\ \citenamefont {Greiner}}]{kale2022schrieffer}%
  \BibitemOpen
  \bibfield  {author} {\bibinfo {author} {\bibfnamefont {A.}~\bibnamefont {Kale}}, \bibinfo {author} {\bibfnamefont {J.~H.}\ \bibnamefont {Huhn}}, \bibinfo {author} {\bibfnamefont {M.}~\bibnamefont {Xu}}, \bibinfo {author} {\bibfnamefont {L.~H.}\ \bibnamefont {Kendrick}}, \bibinfo {author} {\bibfnamefont {M.}~\bibnamefont {Lebrat}}, \bibinfo {author} {\bibfnamefont {C.}~\bibnamefont {Chiu}}, \bibinfo {author} {\bibfnamefont {G.}~\bibnamefont {Ji}}, \bibinfo {author} {\bibfnamefont {F.}~\bibnamefont {Grusdt}}, \bibinfo {author} {\bibfnamefont {A.}~\bibnamefont {Bohrdt}},\ and\ \bibinfo {author} {\bibfnamefont {M.}~\bibnamefont {Greiner}},\ }\bibfield  {title} {\bibinfo {title} {Schrieffer-wolff transformations for experiments: Dynamically suppressing virtual doublon-hole excitations in a fermi-hubbard simulator},\ }\href@noop {} {\bibfield  {journal} {\bibinfo  {journal} {Physical Review A}\ }\textbf {\bibinfo {volume} {106}},\ \bibinfo {pages} {012428} (\bibinfo {year} {2022})}\BibitemShut {NoStop}%
\bibitem [{\citenamefont {Ji}\ \emph {et~al.}(2021)\citenamefont {Ji}, \citenamefont {Xu}, \citenamefont {Kendrick}, \citenamefont {Chiu}, \citenamefont {Br{\"u}ggenj{\"u}rgen}, \citenamefont {Greif}, \citenamefont {Bohrdt}, \citenamefont {Grusdt}, \citenamefont {Demler}, \citenamefont {Lebrat} \emph {et~al.}}]{ji2021coupling}%
  \BibitemOpen
  \bibfield  {author} {\bibinfo {author} {\bibfnamefont {G.}~\bibnamefont {Ji}}, \bibinfo {author} {\bibfnamefont {M.}~\bibnamefont {Xu}}, \bibinfo {author} {\bibfnamefont {L.~H.}\ \bibnamefont {Kendrick}}, \bibinfo {author} {\bibfnamefont {C.~S.}\ \bibnamefont {Chiu}}, \bibinfo {author} {\bibfnamefont {J.~C.}\ \bibnamefont {Br{\"u}ggenj{\"u}rgen}}, \bibinfo {author} {\bibfnamefont {D.}~\bibnamefont {Greif}}, \bibinfo {author} {\bibfnamefont {A.}~\bibnamefont {Bohrdt}}, \bibinfo {author} {\bibfnamefont {F.}~\bibnamefont {Grusdt}}, \bibinfo {author} {\bibfnamefont {E.}~\bibnamefont {Demler}}, \bibinfo {author} {\bibfnamefont {M.}~\bibnamefont {Lebrat}}, \emph {et~al.},\ }\bibfield  {title} {\bibinfo {title} {Coupling a mobile hole to an antiferromagnetic spin background: Transient dynamics of a magnetic polaron},\ }\href@noop {} {\bibfield  {journal} {\bibinfo  {journal} {Physical Review X}\ }\textbf {\bibinfo {volume} {11}},\ \bibinfo {pages} {021022} (\bibinfo {year} {2021})}\BibitemShut {NoStop}%
\bibitem [{\citenamefont {Chiu}\ \emph {et~al.}(2019)\citenamefont {Chiu}, \citenamefont {Ji}, \citenamefont {Bohrdt}, \citenamefont {Xu}, \citenamefont {Knap}, \citenamefont {Demler}, \citenamefont {Grusdt}, \citenamefont {Greiner},\ and\ \citenamefont {Greif}}]{chiu2019string}%
  \BibitemOpen
  \bibfield  {author} {\bibinfo {author} {\bibfnamefont {C.~S.}\ \bibnamefont {Chiu}}, \bibinfo {author} {\bibfnamefont {G.}~\bibnamefont {Ji}}, \bibinfo {author} {\bibfnamefont {A.}~\bibnamefont {Bohrdt}}, \bibinfo {author} {\bibfnamefont {M.}~\bibnamefont {Xu}}, \bibinfo {author} {\bibfnamefont {M.}~\bibnamefont {Knap}}, \bibinfo {author} {\bibfnamefont {E.}~\bibnamefont {Demler}}, \bibinfo {author} {\bibfnamefont {F.}~\bibnamefont {Grusdt}}, \bibinfo {author} {\bibfnamefont {M.}~\bibnamefont {Greiner}},\ and\ \bibinfo {author} {\bibfnamefont {D.}~\bibnamefont {Greif}},\ }\bibfield  {title} {\bibinfo {title} {String patterns in the doped hubbard model},\ }\href@noop {} {\bibfield  {journal} {\bibinfo  {journal} {Science}\ }\textbf {\bibinfo {volume} {365}},\ \bibinfo {pages} {251} (\bibinfo {year} {2019})}\BibitemShut {NoStop}%
\bibitem [{\citenamefont {Parsons}\ \emph {et~al.}(2015)\citenamefont {Parsons}, \citenamefont {Huber}, \citenamefont {Mazurenko}, \citenamefont {Chiu}, \citenamefont {Setiawan}, \citenamefont {Wooley-Brown}, \citenamefont {Blatt},\ and\ \citenamefont {Greiner}}]{parsons2015site}%
  \BibitemOpen
  \bibfield  {author} {\bibinfo {author} {\bibfnamefont {M.~F.}\ \bibnamefont {Parsons}}, \bibinfo {author} {\bibfnamefont {F.}~\bibnamefont {Huber}}, \bibinfo {author} {\bibfnamefont {A.}~\bibnamefont {Mazurenko}}, \bibinfo {author} {\bibfnamefont {C.~S.}\ \bibnamefont {Chiu}}, \bibinfo {author} {\bibfnamefont {W.}~\bibnamefont {Setiawan}}, \bibinfo {author} {\bibfnamefont {K.}~\bibnamefont {Wooley-Brown}}, \bibinfo {author} {\bibfnamefont {S.}~\bibnamefont {Blatt}},\ and\ \bibinfo {author} {\bibfnamefont {M.}~\bibnamefont {Greiner}},\ }\bibfield  {title} {\bibinfo {title} {Site-resolved imaging of fermionic li 6 in an optical lattice},\ }\href@noop {} {\bibfield  {journal} {\bibinfo  {journal} {Physical review letters}\ }\textbf {\bibinfo {volume} {114}},\ \bibinfo {pages} {213002} (\bibinfo {year} {2015})}\BibitemShut {NoStop}%
\bibitem [{\citenamefont {Simon}\ \emph {et~al.}(2011)\citenamefont {Simon}, \citenamefont {Bakr}, \citenamefont {Ma}, \citenamefont {Tai}, \citenamefont {Preiss},\ and\ \citenamefont {Greiner}}]{simon2011quantum}%
  \BibitemOpen
  \bibfield  {author} {\bibinfo {author} {\bibfnamefont {J.}~\bibnamefont {Simon}}, \bibinfo {author} {\bibfnamefont {W.~S.}\ \bibnamefont {Bakr}}, \bibinfo {author} {\bibfnamefont {R.}~\bibnamefont {Ma}}, \bibinfo {author} {\bibfnamefont {M.~E.}\ \bibnamefont {Tai}}, \bibinfo {author} {\bibfnamefont {P.~M.}\ \bibnamefont {Preiss}},\ and\ \bibinfo {author} {\bibfnamefont {M.}~\bibnamefont {Greiner}},\ }\bibfield  {title} {\bibinfo {title} {Quantum simulation of antiferromagnetic spin chains in an optical lattice},\ }\href@noop {} {\bibfield  {journal} {\bibinfo  {journal} {Nature}\ }\textbf {\bibinfo {volume} {472}},\ \bibinfo {pages} {307} (\bibinfo {year} {2011})}\BibitemShut {NoStop}%
\bibitem [{\citenamefont {Greiner}\ \emph {et~al.}(2004)\citenamefont {Greiner}, \citenamefont {Regal}, \citenamefont {Ticknor}, \citenamefont {Bohn},\ and\ \citenamefont {Jin}}]{greiner2004detection}%
  \BibitemOpen
  \bibfield  {author} {\bibinfo {author} {\bibfnamefont {M.}~\bibnamefont {Greiner}}, \bibinfo {author} {\bibfnamefont {C.}~\bibnamefont {Regal}}, \bibinfo {author} {\bibfnamefont {C.}~\bibnamefont {Ticknor}}, \bibinfo {author} {\bibfnamefont {J.}~\bibnamefont {Bohn}},\ and\ \bibinfo {author} {\bibfnamefont {D.}~\bibnamefont {Jin}},\ }\bibfield  {title} {\bibinfo {title} {Detection of spatial correlations in an ultracold gas of fermions},\ }\href@noop {} {\bibfield  {journal} {\bibinfo  {journal} {Physical review letters}\ }\textbf {\bibinfo {volume} {92}},\ \bibinfo {pages} {150405} (\bibinfo {year} {2004})}\BibitemShut {NoStop}%
\bibitem [{\citenamefont {Cheuk}\ \emph {et~al.}(2015)\citenamefont {Cheuk}, \citenamefont {Nichols}, \citenamefont {Okan}, \citenamefont {Gersdorf}, \citenamefont {Ramasesh}, \citenamefont {Bakr}, \citenamefont {Lompe},\ and\ \citenamefont {Zwierlein}}]{cheuk2015quantum}%
  \BibitemOpen
  \bibfield  {author} {\bibinfo {author} {\bibfnamefont {L.~W.}\ \bibnamefont {Cheuk}}, \bibinfo {author} {\bibfnamefont {M.~A.}\ \bibnamefont {Nichols}}, \bibinfo {author} {\bibfnamefont {M.}~\bibnamefont {Okan}}, \bibinfo {author} {\bibfnamefont {T.}~\bibnamefont {Gersdorf}}, \bibinfo {author} {\bibfnamefont {V.~V.}\ \bibnamefont {Ramasesh}}, \bibinfo {author} {\bibfnamefont {W.~S.}\ \bibnamefont {Bakr}}, \bibinfo {author} {\bibfnamefont {T.}~\bibnamefont {Lompe}},\ and\ \bibinfo {author} {\bibfnamefont {M.~W.}\ \bibnamefont {Zwierlein}},\ }\bibfield  {title} {\bibinfo {title} {Quantum-gas microscope for fermionic atoms},\ }\href@noop {} {\bibfield  {journal} {\bibinfo  {journal} {Physical review letters}\ }\textbf {\bibinfo {volume} {114}},\ \bibinfo {pages} {193001} (\bibinfo {year} {2015})}\BibitemShut {NoStop}%
\bibitem [{\citenamefont {Sommer}\ \emph {et~al.}(2011)\citenamefont {Sommer}, \citenamefont {Ku}, \citenamefont {Roati},\ and\ \citenamefont {Zwierlein}}]{sommer2011universal}%
  \BibitemOpen
  \bibfield  {author} {\bibinfo {author} {\bibfnamefont {A.}~\bibnamefont {Sommer}}, \bibinfo {author} {\bibfnamefont {M.}~\bibnamefont {Ku}}, \bibinfo {author} {\bibfnamefont {G.}~\bibnamefont {Roati}},\ and\ \bibinfo {author} {\bibfnamefont {M.~W.}\ \bibnamefont {Zwierlein}},\ }\bibfield  {title} {\bibinfo {title} {Universal spin transport in a strongly interacting fermi gas},\ }\href@noop {} {\bibfield  {journal} {\bibinfo  {journal} {Nature}\ }\textbf {\bibinfo {volume} {472}},\ \bibinfo {pages} {201} (\bibinfo {year} {2011})}\BibitemShut {NoStop}%
\bibitem [{\citenamefont {Zwierlein}\ \emph {et~al.}(2005)\citenamefont {Zwierlein}, \citenamefont {Schunck}, \citenamefont {Stan}, \citenamefont {Raupach},\ and\ \citenamefont {Ketterle}}]{zwierlein2005formation}%
  \BibitemOpen
  \bibfield  {author} {\bibinfo {author} {\bibfnamefont {M.}~\bibnamefont {Zwierlein}}, \bibinfo {author} {\bibfnamefont {C.}~\bibnamefont {Schunck}}, \bibinfo {author} {\bibfnamefont {C.}~\bibnamefont {Stan}}, \bibinfo {author} {\bibfnamefont {S.}~\bibnamefont {Raupach}},\ and\ \bibinfo {author} {\bibfnamefont {W.}~\bibnamefont {Ketterle}},\ }\bibfield  {title} {\bibinfo {title} {Formation dynamics of a fermion pair condensate},\ }\href@noop {} {\bibfield  {journal} {\bibinfo  {journal} {Physical review letters}\ }\textbf {\bibinfo {volume} {94}},\ \bibinfo {pages} {180401} (\bibinfo {year} {2005})}\BibitemShut {NoStop}%
\bibitem [{\citenamefont {Hertkorn}\ \emph {et~al.}(2021)\citenamefont {Hertkorn}, \citenamefont {Schmidt}, \citenamefont {B{\"o}ttcher}, \citenamefont {Guo}, \citenamefont {Schmidt}, \citenamefont {Ng}, \citenamefont {Graham}, \citenamefont {B{\"u}chler}, \citenamefont {Langen}, \citenamefont {Zwierlein} \emph {et~al.}}]{hertkorn2021density}%
  \BibitemOpen
  \bibfield  {author} {\bibinfo {author} {\bibfnamefont {J.}~\bibnamefont {Hertkorn}}, \bibinfo {author} {\bibfnamefont {J.-N.}\ \bibnamefont {Schmidt}}, \bibinfo {author} {\bibfnamefont {F.}~\bibnamefont {B{\"o}ttcher}}, \bibinfo {author} {\bibfnamefont {M.}~\bibnamefont {Guo}}, \bibinfo {author} {\bibfnamefont {M.}~\bibnamefont {Schmidt}}, \bibinfo {author} {\bibfnamefont {K.}~\bibnamefont {Ng}}, \bibinfo {author} {\bibfnamefont {S.}~\bibnamefont {Graham}}, \bibinfo {author} {\bibfnamefont {H.}~\bibnamefont {B{\"u}chler}}, \bibinfo {author} {\bibfnamefont {T.}~\bibnamefont {Langen}}, \bibinfo {author} {\bibfnamefont {M.}~\bibnamefont {Zwierlein}}, \emph {et~al.},\ }\bibfield  {title} {\bibinfo {title} {Density fluctuations across the superfluid-supersolid phase transition in a dipolar quantum gas},\ }\href@noop {} {\bibfield  {journal} {\bibinfo  {journal} {Physical Review X}\ }\textbf {\bibinfo {volume} {11}},\ \bibinfo {pages} {011037} (\bibinfo {year} {2021})}\BibitemShut {NoStop}%
\bibitem [{\citenamefont {Hartke}\ \emph {et~al.}(2023)\citenamefont {Hartke}, \citenamefont {Oreg}, \citenamefont {Turnbaugh}, \citenamefont {Jia},\ and\ \citenamefont {Zwierlein}}]{hartke2023direct}%
  \BibitemOpen
  \bibfield  {author} {\bibinfo {author} {\bibfnamefont {T.}~\bibnamefont {Hartke}}, \bibinfo {author} {\bibfnamefont {B.}~\bibnamefont {Oreg}}, \bibinfo {author} {\bibfnamefont {C.}~\bibnamefont {Turnbaugh}}, \bibinfo {author} {\bibfnamefont {N.}~\bibnamefont {Jia}},\ and\ \bibinfo {author} {\bibfnamefont {M.}~\bibnamefont {Zwierlein}},\ }\bibfield  {title} {\bibinfo {title} {Direct observation of nonlocal fermion pairing in an attractive fermi-hubbard gas},\ }\href@noop {} {\bibfield  {journal} {\bibinfo  {journal} {Science}\ }\textbf {\bibinfo {volume} {381}},\ \bibinfo {pages} {82} (\bibinfo {year} {2023})}\BibitemShut {NoStop}%
\bibitem [{\citenamefont {Blankenbecler}\ \emph {et~al.}(1981)\citenamefont {Blankenbecler}, \citenamefont {Scalapino},\ and\ \citenamefont {Sugar}}]{blankenbecler1981monte}%
  \BibitemOpen
  \bibfield  {author} {\bibinfo {author} {\bibfnamefont {R.}~\bibnamefont {Blankenbecler}}, \bibinfo {author} {\bibfnamefont {D.}~\bibnamefont {Scalapino}},\ and\ \bibinfo {author} {\bibfnamefont {R.}~\bibnamefont {Sugar}},\ }\bibfield  {title} {\bibinfo {title} {Monte carlo calculations of coupled boson-fermion systems. i},\ }\href@noop {} {\bibfield  {journal} {\bibinfo  {journal} {Physical Review D}\ }\textbf {\bibinfo {volume} {24}},\ \bibinfo {pages} {2278} (\bibinfo {year} {1981})}\BibitemShut {NoStop}%
\bibitem [{\citenamefont {White}\ \emph {et~al.}(1989)\citenamefont {White}, \citenamefont {Scalapino}, \citenamefont {Sugar}, \citenamefont {Loh}, \citenamefont {Gubernatis},\ and\ \citenamefont {Scalettar}}]{white1989numerical}%
  \BibitemOpen
  \bibfield  {author} {\bibinfo {author} {\bibfnamefont {S.~R.}\ \bibnamefont {White}}, \bibinfo {author} {\bibfnamefont {D.~J.}\ \bibnamefont {Scalapino}}, \bibinfo {author} {\bibfnamefont {R.~L.}\ \bibnamefont {Sugar}}, \bibinfo {author} {\bibfnamefont {E.}~\bibnamefont {Loh}}, \bibinfo {author} {\bibfnamefont {J.~E.}\ \bibnamefont {Gubernatis}},\ and\ \bibinfo {author} {\bibfnamefont {R.~T.}\ \bibnamefont {Scalettar}},\ }\bibfield  {title} {\bibinfo {title} {Numerical study of the two-dimensional hubbard model},\ }\href@noop {} {\bibfield  {journal} {\bibinfo  {journal} {Physical Review B}\ }\textbf {\bibinfo {volume} {40}},\ \bibinfo {pages} {506} (\bibinfo {year} {1989})}\BibitemShut {NoStop}%
\bibitem [{\citenamefont {Hirsch}(1985)}]{hirsch1985two}%
  \BibitemOpen
  \bibfield  {author} {\bibinfo {author} {\bibfnamefont {J.~E.}\ \bibnamefont {Hirsch}},\ }\bibfield  {title} {\bibinfo {title} {Two-dimensional hubbard model: Numerical simulation study},\ }\href@noop {} {\bibfield  {journal} {\bibinfo  {journal} {Physical Review B}\ }\textbf {\bibinfo {volume} {31}},\ \bibinfo {pages} {4403} (\bibinfo {year} {1985})}\BibitemShut {NoStop}%
\bibitem [{\citenamefont {Kim}\ \emph {et~al.}(2020)\citenamefont {Kim}, \citenamefont {Simkovic~IV},\ and\ \citenamefont {Kozik}}]{kim2020spin}%
  \BibitemOpen
  \bibfield  {author} {\bibinfo {author} {\bibfnamefont {A.~J.}\ \bibnamefont {Kim}}, \bibinfo {author} {\bibfnamefont {F.}~\bibnamefont {Simkovic~IV}},\ and\ \bibinfo {author} {\bibfnamefont {E.}~\bibnamefont {Kozik}},\ }\bibfield  {title} {\bibinfo {title} {Spin and charge correlations across the metal-to-insulator crossover in the half-filled 2d hubbard model},\ }\href@noop {} {\bibfield  {journal} {\bibinfo  {journal} {Physical Review Letters}\ }\textbf {\bibinfo {volume} {124}},\ \bibinfo {pages} {117602} (\bibinfo {year} {2020})}\BibitemShut {NoStop}%
\bibitem [{\citenamefont {Park}\ \emph {et~al.}(2008)\citenamefont {Park}, \citenamefont {Haule},\ and\ \citenamefont {Kotliar}}]{park2008cluster}%
  \BibitemOpen
  \bibfield  {author} {\bibinfo {author} {\bibfnamefont {H.}~\bibnamefont {Park}}, \bibinfo {author} {\bibfnamefont {K.}~\bibnamefont {Haule}},\ and\ \bibinfo {author} {\bibfnamefont {G.}~\bibnamefont {Kotliar}},\ }\bibfield  {title} {\bibinfo {title} {Cluster dynamical mean field theory of the mott transition},\ }\href@noop {} {\bibfield  {journal} {\bibinfo  {journal} {Physical review letters}\ }\textbf {\bibinfo {volume} {101}},\ \bibinfo {pages} {186403} (\bibinfo {year} {2008})}\BibitemShut {NoStop}%
\bibitem [{\citenamefont {Macridin}\ \emph {et~al.}(2006)\citenamefont {Macridin}, \citenamefont {Jarrell}, \citenamefont {Maier}, \citenamefont {Kent},\ and\ \citenamefont {D’Azevedo}}]{macridin2006pseudogap}%
  \BibitemOpen
  \bibfield  {author} {\bibinfo {author} {\bibfnamefont {A.}~\bibnamefont {Macridin}}, \bibinfo {author} {\bibfnamefont {M.}~\bibnamefont {Jarrell}}, \bibinfo {author} {\bibfnamefont {T.}~\bibnamefont {Maier}}, \bibinfo {author} {\bibfnamefont {P.}~\bibnamefont {Kent}},\ and\ \bibinfo {author} {\bibfnamefont {E.}~\bibnamefont {D’Azevedo}},\ }\bibfield  {title} {\bibinfo {title} {Pseudogap and antiferromagnetic correlations in the hubbard model},\ }\href@noop {} {\bibfield  {journal} {\bibinfo  {journal} {Physical review letters}\ }\textbf {\bibinfo {volume} {97}},\ \bibinfo {pages} {036401} (\bibinfo {year} {2006})}\BibitemShut {NoStop}%
\bibitem [{\citenamefont {Kyung}\ \emph {et~al.}(2006)\citenamefont {Kyung}, \citenamefont {Kancharla}, \citenamefont {S{\'e}n{\'e}chal}, \citenamefont {Tremblay}, \citenamefont {Civelli},\ and\ \citenamefont {Kotliar}}]{kyung2006pseudogap}%
  \BibitemOpen
  \bibfield  {author} {\bibinfo {author} {\bibfnamefont {B.}~\bibnamefont {Kyung}}, \bibinfo {author} {\bibfnamefont {S.}~\bibnamefont {Kancharla}}, \bibinfo {author} {\bibfnamefont {D.}~\bibnamefont {S{\'e}n{\'e}chal}}, \bibinfo {author} {\bibfnamefont {A.-M.}\ \bibnamefont {Tremblay}}, \bibinfo {author} {\bibfnamefont {M.}~\bibnamefont {Civelli}},\ and\ \bibinfo {author} {\bibfnamefont {G.}~\bibnamefont {Kotliar}},\ }\bibfield  {title} {\bibinfo {title} {Pseudogap induced by short-range spin correlations in a doped mott insulator},\ }\href@noop {} {\bibfield  {journal} {\bibinfo  {journal} {Physical Review B}\ }\textbf {\bibinfo {volume} {73}},\ \bibinfo {pages} {165114} (\bibinfo {year} {2006})}\BibitemShut {NoStop}%
\bibitem [{\citenamefont {Boschini}\ \emph {et~al.}(2020)\citenamefont {Boschini}, \citenamefont {Zonno}, \citenamefont {Razzoli}, \citenamefont {Day}, \citenamefont {Michiardi}, \citenamefont {Zwartsenberg}, \citenamefont {Nigge}, \citenamefont {Schneider}, \citenamefont {da~Silva~Neto}, \citenamefont {Erb} \emph {et~al.}}]{boschini2020emergence}%
  \BibitemOpen
  \bibfield  {author} {\bibinfo {author} {\bibfnamefont {F.}~\bibnamefont {Boschini}}, \bibinfo {author} {\bibfnamefont {M.}~\bibnamefont {Zonno}}, \bibinfo {author} {\bibfnamefont {E.}~\bibnamefont {Razzoli}}, \bibinfo {author} {\bibfnamefont {R.~P.}\ \bibnamefont {Day}}, \bibinfo {author} {\bibfnamefont {M.}~\bibnamefont {Michiardi}}, \bibinfo {author} {\bibfnamefont {B.}~\bibnamefont {Zwartsenberg}}, \bibinfo {author} {\bibfnamefont {P.}~\bibnamefont {Nigge}}, \bibinfo {author} {\bibfnamefont {M.}~\bibnamefont {Schneider}}, \bibinfo {author} {\bibfnamefont {E.~H.}\ \bibnamefont {da~Silva~Neto}}, \bibinfo {author} {\bibfnamefont {A.}~\bibnamefont {Erb}}, \emph {et~al.},\ }\bibfield  {title} {\bibinfo {title} {Emergence of pseudogap from short-range spin-correlations in electron-doped cuprates},\ }\href@noop {} {\bibfield  {journal} {\bibinfo  {journal} {npj Quantum Materials}\ }\textbf {\bibinfo {volume} {5}},\ \bibinfo {pages} {6} (\bibinfo {year} {2020})}\BibitemShut {NoStop}%
\bibitem [{\citenamefont {Krien}\ \emph {et~al.}(2022)\citenamefont {Krien}, \citenamefont {Worm}, \citenamefont {Chalupa-Gantner}, \citenamefont {Toschi},\ and\ \citenamefont {Held}}]{krien2022explaining}%
  \BibitemOpen
  \bibfield  {author} {\bibinfo {author} {\bibfnamefont {F.}~\bibnamefont {Krien}}, \bibinfo {author} {\bibfnamefont {P.}~\bibnamefont {Worm}}, \bibinfo {author} {\bibfnamefont {P.}~\bibnamefont {Chalupa-Gantner}}, \bibinfo {author} {\bibfnamefont {A.}~\bibnamefont {Toschi}},\ and\ \bibinfo {author} {\bibfnamefont {K.}~\bibnamefont {Held}},\ }\bibfield  {title} {\bibinfo {title} {Explaining the pseudogap through damping and antidamping on the fermi surface by imaginary spin scattering},\ }\href@noop {} {\bibfield  {journal} {\bibinfo  {journal} {Communications Physics}\ }\textbf {\bibinfo {volume} {5}},\ \bibinfo {pages} {336} (\bibinfo {year} {2022})}\BibitemShut {NoStop}%
\bibitem [{\citenamefont {Paiva}\ \emph {et~al.}(2001)\citenamefont {Paiva}, \citenamefont {Scalettar}, \citenamefont {Huscroft},\ and\ \citenamefont {McMahan}}]{paiva2001signatures}%
  \BibitemOpen
  \bibfield  {author} {\bibinfo {author} {\bibfnamefont {T.}~\bibnamefont {Paiva}}, \bibinfo {author} {\bibfnamefont {R.}~\bibnamefont {Scalettar}}, \bibinfo {author} {\bibfnamefont {C.}~\bibnamefont {Huscroft}},\ and\ \bibinfo {author} {\bibfnamefont {A.}~\bibnamefont {McMahan}},\ }\bibfield  {title} {\bibinfo {title} {Signatures of spin and charge energy scales in the local moment and specific heat of the half-filled two-dimensional hubbard model},\ }\href@noop {} {\bibfield  {journal} {\bibinfo  {journal} {Physical Review B}\ }\textbf {\bibinfo {volume} {63}},\ \bibinfo {pages} {125116} (\bibinfo {year} {2001})}\BibitemShut {NoStop}%
\bibitem [{\citenamefont {Endres}\ \emph {et~al.}(2013)\citenamefont {Endres}, \citenamefont {Cheneau}, \citenamefont {Fukuhara}, \citenamefont {Weitenberg}, \citenamefont {Schau{\ss}}, \citenamefont {Gross}, \citenamefont {Mazza}, \citenamefont {Ba{\~n}uls}, \citenamefont {Pollet}, \citenamefont {Bloch} \emph {et~al.}}]{endres2013single}%
  \BibitemOpen
  \bibfield  {author} {\bibinfo {author} {\bibfnamefont {M.}~\bibnamefont {Endres}}, \bibinfo {author} {\bibfnamefont {M.}~\bibnamefont {Cheneau}}, \bibinfo {author} {\bibfnamefont {T.}~\bibnamefont {Fukuhara}}, \bibinfo {author} {\bibfnamefont {C.}~\bibnamefont {Weitenberg}}, \bibinfo {author} {\bibfnamefont {P.}~\bibnamefont {Schau{\ss}}}, \bibinfo {author} {\bibfnamefont {C.}~\bibnamefont {Gross}}, \bibinfo {author} {\bibfnamefont {L.}~\bibnamefont {Mazza}}, \bibinfo {author} {\bibfnamefont {M.~C.}\ \bibnamefont {Ba{\~n}uls}}, \bibinfo {author} {\bibfnamefont {L.}~\bibnamefont {Pollet}}, \bibinfo {author} {\bibfnamefont {I.}~\bibnamefont {Bloch}}, \emph {et~al.},\ }\bibfield  {title} {\bibinfo {title} {Single-site-and single-atom-resolved measurement of correlation functions},\ }\href@noop {} {\bibfield  {journal} {\bibinfo  {journal} {Applied Physics B}\ }\textbf {\bibinfo {volume} {113}},\ \bibinfo {pages} {27} (\bibinfo {year} {2013})}\BibitemShut {NoStop}%
\bibitem [{\citenamefont {Walsh}\ \emph {et~al.}(2019)\citenamefont {Walsh}, \citenamefont {S{\'e}mon}, \citenamefont {Sordi},\ and\ \citenamefont {Tremblay}}]{walsh2019critical}%
  \BibitemOpen
  \bibfield  {author} {\bibinfo {author} {\bibfnamefont {C.}~\bibnamefont {Walsh}}, \bibinfo {author} {\bibfnamefont {P.}~\bibnamefont {S{\'e}mon}}, \bibinfo {author} {\bibfnamefont {G.}~\bibnamefont {Sordi}},\ and\ \bibinfo {author} {\bibfnamefont {A.-M.}\ \bibnamefont {Tremblay}},\ }\bibfield  {title} {\bibinfo {title} {Critical opalescence across the doping-driven mott transition in optical lattices of ultracold atoms},\ }\href@noop {} {\bibfield  {journal} {\bibinfo  {journal} {Physical Review B}\ }\textbf {\bibinfo {volume} {99}},\ \bibinfo {pages} {165151} (\bibinfo {year} {2019})}\BibitemShut {NoStop}%
\bibitem [{\citenamefont {Mahan}(2013)}]{mahan2013many}%
  \BibitemOpen
  \bibfield  {author} {\bibinfo {author} {\bibfnamefont {G.~D.}\ \bibnamefont {Mahan}},\ }\href@noop {} {\emph {\bibinfo {title} {Many-particle physics}}}\ (\bibinfo  {publisher} {Springer Science \& Business Media},\ \bibinfo {year} {2013})\BibitemShut {NoStop}%
\bibitem [{\citenamefont {Dar{\'e}}\ \emph {et~al.}(2007)\citenamefont {Dar{\'e}}, \citenamefont {Raymond}, \citenamefont {Albinet},\ and\ \citenamefont {Tremblay}}]{dare2007interaction}%
  \BibitemOpen
  \bibfield  {author} {\bibinfo {author} {\bibfnamefont {A.-M.}\ \bibnamefont {Dar{\'e}}}, \bibinfo {author} {\bibfnamefont {L.}~\bibnamefont {Raymond}}, \bibinfo {author} {\bibfnamefont {G.}~\bibnamefont {Albinet}},\ and\ \bibinfo {author} {\bibfnamefont {A.-M.}\ \bibnamefont {Tremblay}},\ }\bibfield  {title} {\bibinfo {title} {Interaction-induced adiabatic cooling for antiferromagnetism in optical lattices},\ }\href@noop {} {\bibfield  {journal} {\bibinfo  {journal} {Physical Review B}\ }\textbf {\bibinfo {volume} {76}},\ \bibinfo {pages} {064402} (\bibinfo {year} {2007})}\BibitemShut {NoStop}%
\bibitem [{\citenamefont {Roy}\ and\ \citenamefont {Trivedi}()}]{roylieb}%
  \BibitemOpen
  \bibfield  {author} {\bibinfo {author} {\bibfnamefont {S.}~\bibnamefont {Roy}}\ and\ \bibinfo {author} {\bibfnamefont {N.}~\bibnamefont {Trivedi}},\ }\bibfield  {title} {\bibinfo {title} {in preparation},\ }\href@noop {} {\bibfield  {journal} {\bibinfo  {journal} {""}\ } (\bibinfo {year} {""})}\BibitemShut {NoStop}%
\bibitem [{\citenamefont {Slot}\ \emph {et~al.}(2017)\citenamefont {Slot}, \citenamefont {Gardenier}, \citenamefont {Jacobse}, \citenamefont {Van~Miert}, \citenamefont {Kempkes}, \citenamefont {Zevenhuizen}, \citenamefont {Smith}, \citenamefont {Vanmaekelbergh},\ and\ \citenamefont {Swart}}]{slot2017experimental}%
  \BibitemOpen
  \bibfield  {author} {\bibinfo {author} {\bibfnamefont {M.~R.}\ \bibnamefont {Slot}}, \bibinfo {author} {\bibfnamefont {T.~S.}\ \bibnamefont {Gardenier}}, \bibinfo {author} {\bibfnamefont {P.~H.}\ \bibnamefont {Jacobse}}, \bibinfo {author} {\bibfnamefont {G.~C.}\ \bibnamefont {Van~Miert}}, \bibinfo {author} {\bibfnamefont {S.~N.}\ \bibnamefont {Kempkes}}, \bibinfo {author} {\bibfnamefont {S.~J.}\ \bibnamefont {Zevenhuizen}}, \bibinfo {author} {\bibfnamefont {C.~M.}\ \bibnamefont {Smith}}, \bibinfo {author} {\bibfnamefont {D.}~\bibnamefont {Vanmaekelbergh}},\ and\ \bibinfo {author} {\bibfnamefont {I.}~\bibnamefont {Swart}},\ }\bibfield  {title} {\bibinfo {title} {Experimental realization and characterization of an electronic lieb lattice},\ }\href@noop {} {\bibfield  {journal} {\bibinfo  {journal} {Nature physics}\ }\textbf {\bibinfo {volume} {13}},\ \bibinfo {pages} {672} (\bibinfo {year} {2017})}\BibitemShut {NoStop}%
\bibitem [{\citenamefont {Mukherjee}\ \emph {et~al.}(2015)\citenamefont {Mukherjee}, \citenamefont {Spracklen}, \citenamefont {Choudhury}, \citenamefont {Goldman}, \citenamefont {{\"O}hberg}, \citenamefont {Andersson},\ and\ \citenamefont {Thomson}}]{mukherjee2015observation}%
  \BibitemOpen
  \bibfield  {author} {\bibinfo {author} {\bibfnamefont {S.}~\bibnamefont {Mukherjee}}, \bibinfo {author} {\bibfnamefont {A.}~\bibnamefont {Spracklen}}, \bibinfo {author} {\bibfnamefont {D.}~\bibnamefont {Choudhury}}, \bibinfo {author} {\bibfnamefont {N.}~\bibnamefont {Goldman}}, \bibinfo {author} {\bibfnamefont {P.}~\bibnamefont {{\"O}hberg}}, \bibinfo {author} {\bibfnamefont {E.}~\bibnamefont {Andersson}},\ and\ \bibinfo {author} {\bibfnamefont {R.~R.}\ \bibnamefont {Thomson}},\ }\bibfield  {title} {\bibinfo {title} {Observation of a localized flat-band state in a photonic lieb lattice},\ }\href@noop {} {\bibfield  {journal} {\bibinfo  {journal} {Physical review letters}\ }\textbf {\bibinfo {volume} {114}},\ \bibinfo {pages} {245504} (\bibinfo {year} {2015})}\BibitemShut {NoStop}%
\bibitem [{\citenamefont {Lenihan}\ \emph {et~al.}(2021)\citenamefont {Lenihan}, \citenamefont {Kim}, \citenamefont {{\v{S}}imkovic~IV},\ and\ \citenamefont {Kozik}}]{lenihan2021entropy}%
  \BibitemOpen
  \bibfield  {author} {\bibinfo {author} {\bibfnamefont {C.}~\bibnamefont {Lenihan}}, \bibinfo {author} {\bibfnamefont {A.~J.}\ \bibnamefont {Kim}}, \bibinfo {author} {\bibfnamefont {F.}~\bibnamefont {{\v{S}}imkovic~IV}},\ and\ \bibinfo {author} {\bibfnamefont {E.}~\bibnamefont {Kozik}},\ }\bibfield  {title} {\bibinfo {title} {Entropy in the non-fermi-liquid regime of the doped 2 d hubbard model},\ }\href@noop {} {\bibfield  {journal} {\bibinfo  {journal} {Physical Review Letters}\ }\textbf {\bibinfo {volume} {126}},\ \bibinfo {pages} {105701} (\bibinfo {year} {2021})}\BibitemShut {NoStop}%
\bibitem [{\citenamefont {{\v{S}}imkovic~IV}\ \emph {et~al.}(2020)\citenamefont {{\v{S}}imkovic~IV}, \citenamefont {LeBlanc}, \citenamefont {Kim}, \citenamefont {Deng}, \citenamefont {Prokof’ev}, \citenamefont {Svistunov},\ and\ \citenamefont {Kozik}}]{vsimkovic2020extended}%
  \BibitemOpen
  \bibfield  {author} {\bibinfo {author} {\bibfnamefont {F.}~\bibnamefont {{\v{S}}imkovic~IV}}, \bibinfo {author} {\bibfnamefont {J.}~\bibnamefont {LeBlanc}}, \bibinfo {author} {\bibfnamefont {A.~J.}\ \bibnamefont {Kim}}, \bibinfo {author} {\bibfnamefont {Y.}~\bibnamefont {Deng}}, \bibinfo {author} {\bibfnamefont {N.}~\bibnamefont {Prokof’ev}}, \bibinfo {author} {\bibfnamefont {B.}~\bibnamefont {Svistunov}},\ and\ \bibinfo {author} {\bibfnamefont {E.}~\bibnamefont {Kozik}},\ }\bibfield  {title} {\bibinfo {title} {Extended crossover from a fermi liquid to a quasiantiferromagnet in the half-filled 2d hubbard model},\ }\href@noop {} {\bibfield  {journal} {\bibinfo  {journal} {Physical Review Letters}\ }\textbf {\bibinfo {volume} {124}},\ \bibinfo {pages} {017003} (\bibinfo {year} {2020})}\BibitemShut {NoStop}%
\bibitem [{\citenamefont {Werner}\ \emph {et~al.}(2005)\citenamefont {Werner}, \citenamefont {Parcollet}, \citenamefont {Georges},\ and\ \citenamefont {Hassan}}]{werner2005interaction}%
  \BibitemOpen
  \bibfield  {author} {\bibinfo {author} {\bibfnamefont {F.}~\bibnamefont {Werner}}, \bibinfo {author} {\bibfnamefont {O.}~\bibnamefont {Parcollet}}, \bibinfo {author} {\bibfnamefont {A.}~\bibnamefont {Georges}},\ and\ \bibinfo {author} {\bibfnamefont {S.}~\bibnamefont {Hassan}},\ }\bibfield  {title} {\bibinfo {title} {Interaction-induced adiabatic cooling and antiferromagnetism of cold fermions in optical lattices},\ }\href@noop {} {\bibfield  {journal} {\bibinfo  {journal} {Physical review letters}\ }\textbf {\bibinfo {volume} {95}},\ \bibinfo {pages} {056401} (\bibinfo {year} {2005})}\BibitemShut {NoStop}%
\bibitem [{\citenamefont {Chen}\ \emph {et~al.}(2021)\citenamefont {Chen}, \citenamefont {Chen}, \citenamefont {Chen}, \citenamefont {Cui}, \citenamefont {Zhai}, \citenamefont {Weichselbaum}, \citenamefont {von Delft}, \citenamefont {Meng},\ and\ \citenamefont {Li}}]{chen2021quantum}%
  \BibitemOpen
  \bibfield  {author} {\bibinfo {author} {\bibfnamefont {B.-B.}\ \bibnamefont {Chen}}, \bibinfo {author} {\bibfnamefont {C.}~\bibnamefont {Chen}}, \bibinfo {author} {\bibfnamefont {Z.}~\bibnamefont {Chen}}, \bibinfo {author} {\bibfnamefont {J.}~\bibnamefont {Cui}}, \bibinfo {author} {\bibfnamefont {Y.}~\bibnamefont {Zhai}}, \bibinfo {author} {\bibfnamefont {A.}~\bibnamefont {Weichselbaum}}, \bibinfo {author} {\bibfnamefont {J.}~\bibnamefont {von Delft}}, \bibinfo {author} {\bibfnamefont {Z.~Y.}\ \bibnamefont {Meng}},\ and\ \bibinfo {author} {\bibfnamefont {W.}~\bibnamefont {Li}},\ }\bibfield  {title} {\bibinfo {title} {Quantum many-body simulations of the two-dimensional fermi-hubbard model in ultracold optical lattices},\ }\href@noop {} {\bibfield  {journal} {\bibinfo  {journal} {Physical Review B}\ }\textbf {\bibinfo {volume} {103}},\ \bibinfo {pages} {L041107} (\bibinfo {year} {2021})}\BibitemShut {NoStop}%
\end{thebibliography}%

\end{document}